\documentclass[traditabstract]{aa} 
\usepackage{graphicx}
\usepackage{txfonts}
\usepackage{array}
\usepackage{courier}
\begin{document}
   \title{The VIMOS Ultra-Deep Survey: $\sim$10\,000 galaxies
with spectroscopic redshifts to study galaxy assembly at early epochs $2<z\simeq6$ 
          \thanks{Based on data obtained with the European Southern Observatory Very Large
Telescope, Paranal, Chile, under Large Program 185.A-0791. }
}

\author{O. Le F\`evre\inst{1}
\and L.A.M. Tasca\inst{1}
\and P. Cassata\inst{1}
\and B. Garilli\inst{3}
\and V. Le Brun\inst{1}
\and D. Maccagni\inst{3}
\and L. Pentericci\inst{4}
\and R. Thomas\inst{1}
\and E. Vanzella\inst{2}
\and G. Zamorani \inst{2}
\and E. Zucca\inst{2}
\and R. Amorin\inst{4}
\and S. Bardelli\inst{2}
\and P. Capak\inst{12}
\and L. Cassar\`a\inst{3}
\and M. Castellano\inst{4}
\and A. Cimatti\inst{5}
\and J.G. Cuby\inst{1}
\and O. Cucciati\inst{5,2}
\and S. de la Torre\inst{1}
\and A. Durkalec\inst{1}
\and A. Fontana\inst{4}
\and M. Giavalisco\inst{13}
\and A. Grazian\inst{4}
\and N. P. Hathi\inst{1}
\and O. Ilbert\inst{1}
\and B. C. Lemaux \inst{1}
\and C. Moreau\inst{1}
\and S. Paltani\inst{9}
\and B. Ribeiro\inst{1}
\and M. Salvato\inst{14}
\and D. Schaerer\inst{10,8}
\and M. Scodeggio\inst{3}
\and V. Sommariva\inst{5,4}
\and M. Talia\inst{5}
\and Y. Taniguchi\inst{15}
\and L. Tresse\inst{1}
\and D. Vergani\inst{6,2}
\and P.W. Wang\inst{1}
\and S. Charlot\inst{7}
\and T. Contini\inst{8}
\and S. Fotopoulou\inst{9}
\and C. L\'opez-Sanjuan\inst{11}
\and Y. Mellier\inst{7}
\and N. Scoville\inst{12}
}

\institute{Aix Marseille Universit\'e, CNRS, LAM (Laboratoire d'Astrophysique  de Marseille) UMR 7326, 13388, Marseille, France
\and
INAF--Osservatorio Astronomico di Bologna, via Ranzani,1, I-40127, Bologna, Italy
\and
INAF--IASF, via Bassini 15, I-20133,  Milano, Italy
\and
INAF--Osservatorio Astronomico di Roma, via di Frascati 33, I-00040,  Monte Porzio Catone, Italy
\and
University of Bologna, Department of Physics and Astronomy (DIFA), V.le Berti Pichat, 6/2 - 40127, Bologna
\and
INAF--IASF Bologna, via Gobetti 101, I--40129,  Bologna, Italy
\and
Institut d'Astrophysique de Paris, UMR7095 CNRS,
Universit\'e Pierre et Marie Curie, 98 bis Boulevard Arago, 75014
Paris, France
\and
Institut de Recherche en Astrophysique et Plan\'etologie - IRAP, CNRS, Université de Toulouse, UPS-OMP, 14, avenue E. Belin, F31400
Toulouse, France
\and
Department of Astronomy, University of Geneva
ch. d'Écogia 16, CH-1290 Versoix
\and
Geneva Observatory, University of Geneva, ch. des Maillettes 51, CH-1290 Versoix, Switzerland
\and
Centro de Estudios de F\'isica del Cosmos de Arag\'on, Teruel, Spain
\and
Department of Astronomy, California Institute of Technology, 1200 E. California Blvd., MC 249--17, Pasadena, CA 91125, USA 
\and
Astronomy Department, University of Massachusetts, Amherst, MA 01003, USA 
\and
Max-Planck-Institut f\"ur Extraterrestrische Physik, Postfach 1312, D-85741, Garching bei M\"unchen, Germany
\and
Research Center for Space and Cosmic Evolution, Ehime University, Bunkyo-cho 2-5, Matsuyama 790-8577, Japan 
 \\ \\
             \email{olivier.lefevre@lam.fr}
             }

   \date{Received...; accepted...}

 
  \abstract
  {We present the VIMOS Ultra Deep Survey (VUDS), a spectroscopic redshift survey of $\sim$10\,000 very faint galaxies to study the 
major phase of galaxy assembly $2<z\simeq6$. The survey covers 1 deg$^2$ in 3 separate fields: COSMOS, ECDFS and VVDS-02h, with
targets selection based on an {\it inclusive} combination of photometric redshifts and color properties. 
Spectra covering $3650<\lambda<9350$\AA ~are obtained with VIMOS on the ESO-VLT with integration times of 14h.
Here we present the survey strategy, the target selection, the data processing, as well as the redshift measurement process, emphasizing the
specific methods adapted to this high redshift range. The spectra quality and redshift reliability are discussed, and
we derive a completeness in redshift measurement of 91\%, or 74\% for the most reliable measurements, down to $i_{AB}=25$, and
measurements are performed all the way down to $i_{AB}=27$.
The redshift distribution of the main sample peaks at $z=3-4$ and extends over a 
large redshift range mainly in $2 < z < 6$. At $3<z<5$, the galaxies cover a large range 
of luminosities $-23< M_{NUV} < -20.5$, stellar mass $10^9$M$_{\sun}< M_* < 10^{11}$M$_{\sun}$, 
and star formation rates $1$ M$_{\sun}$/yr$< SFR < 10^3$M$_{\sun}$/yr. 
We discuss the spectral properties of galaxies using individual as well as stacked spectra. 
The comparison between spectroscopic and photometric redshifts as well as color selection
demonstrate the effectiveness of our selection scheme.
With $\sim6000$ galaxies with reliable spectroscopic redshifts in $2<z<6$ expected when complete, this survey is the largest at these 
redshifts and offers the opportunity for unprecedented studies of the star-forming galaxy population and its 
distribution in large scale structures during the major phase of galaxy assembly.
}


\authorrunning{Le F\`evre, O., and VUDS team 
}
\titlerunning{VUDS: $\sim10000$ galaxies with spectroscopic redshifts to study galaxy assembly at $2<z<6$ }

   \maketitle
%

\section{Introduction}

The study of the first billion years of 
galaxy evolution is one of the key frontiers of modern cosmology. 
The current theoretical paradigm rests on the hierarchical build-up of
dark matter halos in a $\Lambda$CDM cosmology (see e.g. Mo, van den Bosch \& White 2010). 
Galaxies formed in these deep potential
wells, are expected to transform primordial gas into stars from the initial
reservoir and to be fed from new accreted gas. As dark matter halos merge, galaxies in
them are also expected to merge, these events deeply transforming the dynamics
and overall star, gas, and dark matter content of the merging galaxies into the newly
formed one. Along with these processes super-massive black holes are expected to form at the bottom
of the potential wells. Complex processes are invoked to regulate the 
growth of galaxies, including supernovae or AGN feedback possibly quenching star formation,
or the role of different environments impacting the way galaxies are nurtured.
All these processes are combined in increasingly sophisticated numerical models producing
galaxy simulations in representative volumes (e.g. Springel et al. 2008) coupled to semi-analytic
description of galaxy evolution (e.g. Guo et al. 2011). These simulations
face the challenge to reproduce both the internal physics in complex galaxy systems
and the general volume-averaged properties of large galaxy populations as a function
of cosmic time, and need better constraints from observations to be thoroughly tested. 

Impressive progress has been made on the observational front over the past two decades in
an attempt to test and detail a galaxy formation and evolution scenario from robust measurements. A key element
driving observational progress is the need to cover all major phases of galaxy
evolution from the early formation and galaxy assembly to today, a formidable endeavor. 
Deep galaxy surveys have florished to conduct this exploration. The latest few billion years 
have been extensively mapped by surveys like the 2dFGRS (Colless et al. 2001) and then the
various stages of the SDSS (Abazajian et al. 2009), setting a firm observational reference for 
status of galaxies after more than 13 billion years of evolution. At larger redshifts deep
surveys are providing a complex picture with strong evolutionary
features like the build-up along cosmic time of stellar mass in galaxies of different types
or the star formation history. Different types of imaging and spectroscopic surveys are playing a complementary role, 
the deepest studies being performed in photometry and augmented with photometric redshifts (e.g. Ilbert et al. 2006), and with spectroscopic
surveys bringing accurate redshifts and spectro-photometry, spectral features properties, as well as internal velocity information.  
The contribution of space observations in combination with ground-based surveys has been key to provide morphological information (Rix et al. 2004,
Koekemoer et al. 2007, Koekemoer et al. 2011)
and access to photometric bands invisible from the ground either in the far UV or in the mid to far infrared.  
Despite this progress, the exploration of early phases of galaxy formation and evolution
is still largely incomplete. We do not know which objects ignited first at the end of the dark ages, when and 
how the universe was reionised, when and how the first massive galaxies 
formed, and the importance of any physical connection between 
galaxy and black-hole formation and growth. Star-forming
galaxies are providing key information to understand how galaxies grow with time, 
in particular enabling to measure fundamental quantities such as the cosmic star formation
history (e.g. Lilly et al. 1996, Madau et al. 1996, Tresse
et al. 2007, Bouwens et al. 2009, Cucciati et al. 2012, Madau \& Dickinson 2014), and the
history of stellar mass assembly (e.g. Arnouts et al. 2007, Ilbert et al. 2013). 

At redshifts $z\sim1$ the pioneering CFRS survey (Lilly et al. 1995, Le F\`evre et al. 1995) was followed
by more extensive galaxy redshift surveys covering larger volumes like the DEEP2 (Davis et al. 2003), the VVDS (Le F\`evre et al. 2005a, 
Le F\`evre et al. 2013), zCOSMOS (Lilly et al., 2007), now reaching the 100\,000 redshift mark at $z\sim1$
with VIPERS (Guzzo et al. 2014). These surveys have brought a wealth of quantitative and accurate
measurements now reaching large enough areas of a few tens of square degrees and volumes of $\sim 5 \times 10^7 h^{-3}$Mpc$^3$
so that the most fundamental statistical quantities
describing the galaxy population like the luminosity function (LF), the mass function (MF) or the
correlation function (CF) are becoming very accurate and less affected by the cosmic variance related to the small fields of earlier studies.  
At higher redshifts ($z >\sim 2$), the rapid progress has been driven by the effectiveness in selecting high
redshift galaxies and, most importantly, by the impressive gains
in sensitivity and efficiency provided by high multiplex multi-slit spectrographs like 
LRIS (Oke et al. 1995) and DEIMOS (Faber et al. 2003) on the Keck telescope,  FORS (Appenzeller et al. 1998) and VIMOS (Le F\`evre et al. 2003) on the VLT. 
The effectiveness of the Lyman-break galaxies (LBG) selection has provided the capability
to find large numbers of galaxies at $z>2.5$, and is continuing to be the single-most used technique 
to select galaxies at the highest possible redshifts (e.g. Steidel et al. 2003,  Bouwens et al. 2009, Ellis et al. 2013).
It is supplemented by narrow band imaging techniques isolating Lyman-$\alpha$ emitters 
(LAE; Taniguchi et al. 2005, Shimasaku et al. 2006; Ouchi et al. 2008), 
highly efficiency when reaching sufficiently deep to dig deeper into the LAE luminosity function at increasingly high redshifts. 
In addition to these pre-selection techniques, deep purely magnitude selected surveys were conducted in order to probe a 
large population mostly free of pre-selection biases. The largest to date probing redshifts $z>1.5$ is the $i-band$ magnitude
selected VVDS survey covering up to $z\sim6.5$ with $>1000$ galaxies with redshifts $z\geq2$ (Le F\`evre et al. 2005a, 2013). 
Other magnitude selected surveys have attempted using redder bands to alleviate selecting only galaxies with strong rest-frame
UV continuum. The K20 survey used K-band magnitude selection down to $K=20$ to identify extremely red objects (EROs) at $z\sim1.5-2$
which turned out to be dust obscured star forming galaxie or old passive early-type galaxies (Cimatti et al. 2002). In a following
work the GMASS survey selected objects on the basis of Spitzer NIR photometry with $m_{4.5\mu m} < 23.0$ (AB) coupled to photometric redshift 
$z_{phot}>1.4$ to identify a few hundred galaxies with $1.5<z<3$, including 13 red passive galaxies (Cimatti et al. 2008).
However, and despite these attempts, the approach using pure magnitude selection is costly in observing time
when going much beyond redshift $z\sim2$ or so.

Performing a complete galaxy census is a basic {\it Astronomy} input necessary for any subsequent {\it astrophysical} analysis.
While at redshifts $\sim1$ this census is mostly complete down to stellar mass $10^8$M$_{\sun}$, it is not
yet the case at redshifts $z>1$ for several reasons. First, the color selection schemes applied to photometric samples
to extract the high redshift populations, while efficient to identify galaxies, are affected by significant incompleteness,
loosing some fraction of the population at the selected redshift, or by contamination from galaxies at other redshifts.
While the latter can be dealt with by obtaining spectroscopic redshifts, the former remains a serious difficulty
especially at faint magnitude and at the highest redshifts.
Unfortunately the level of incompleteness strongly depends on the photometric filters used for imaging, the 
depth of the observations, as well as the image quality, which requires a case by case study involving 
source simulations complicated by the need to make apriori hypotheses on the properties of galaxies one is trying to establish. 
It was realised that color-color selection like the LBG technique at $z>2.5$ or
like the BzK working at $z\simeq2$ (Daddi et al. 2004) would miss a part of the general galaxy population in their
selection process (Le F\`evre et al. 2005b), therefore making the galaxy census incomplete (Le F\`evre et al. 2014).
The consequences of this maybe far-reaching, as incompleteness in counts leads to underestimates in
luminosity density, star formation rates, as well as mass density, just to cite these few.
An important aspect of on-going and future studies is to revisit galaxy counts as a function
of redshift making sure that no significant population is missing and that no significant bias is introduced in deriving astrophysical quantities.
  
A key element is then the availability of large samples of galaxies with a well defined and well
controlled selection function. Spectroscopic redshift surveys play a key role as they provide
samples with confirmed redshifts. Photometric redshift surveys are widely used and have now reached
an impressive accuracy. However the level of 'catastrophic failures' when photometric
redshifts disagree with their training set of spectroscopic redshifts, even if low at a 
few percent (Ilbert et al. 2013), could still produce large unknowns because of the shape
of the N(z) of flux limited samples. An error of 1\% at the peak, $z\simeq1$, of the N(z) 
of a $i_{AB}=25$ sample could spread galaxies with wrong photometric redshifts to higher
redshifts, e.g. at $z\sim3$ where the projected galaxy density is less than 10 times the N(z) at peak,
which could then represent several tens of percent of uncertainty. One recent example 
is the difficulty to distinguish $z\sim5$ very massive objects from lower redshifts $z\sim2$ galaxies of lower mass (Wiklind et al. 2008, Caputi et al. 2012).
Obtaining a spectroscopic redshift therefore remains a fundamental measurement.

Unfortunately, the total number of galaxies spectroscopically confirmed today by all these surveys at $z>2$ is still limited. 
Published LBG samples reach $\sim$2000 redshifts at $z\sim3$, only $\sim150$ at $z\sim4$, and a few tens
of galaxies beyond that (e.g. Steidel et al., 2003, Vanzella et al., 2009, Bielby et al. 2013).
Samples of LAE galaxies selected with narrow band techniques and confirmed in spectroscopy reach a few hundred objects beyond $z=3$ 
(Ouchi et al. 2008, Kashikawa, 2011). 
The VVDS has assembled $\sim$35\,000 galaxies with spectroscopic redshifts down to $i_{AB}=24.75$, but the high redshift 
tail at $z > 2$ contains about 1000 galaxies (Le F\`evre et al. 2014). 
In front of the difficulty of obtaining large samples of spectroscopicaly confirmed galaxies,
many surveys use samples defined solely on the basis of photometric color selection techniques like LBG or other simple color cuts,
relying on completeness and contamination estimates difficult to control. To overcome the uncertainties linked to
small existing spectroscopic samples, and to understand the biases and limitations of photometry-based studies 
in their ability to provide a complete census of the galaxy population, extremely deep 
spectroscopic surveys over large volumes are needed.

Here we present VUDS, the VIMOS Ultra Deep Survey of $\sim$10\,000 galaxies specifically designed 
to study the early phases of galaxy evolution $2 < z < 6+$. 
The VUDS sample contains an unprecedented number of galaxies with secure spectroscopic redshifts
at this epoch, obtained in three different fields: COSMOS, ECDFS and VVDS-02h. 
The survey design including the target selection is based mainly on photometric redshifts, as 
presented in Section \ref{overview}. The VIMOS multi-slit spectroscopic observations, the data reduction 
and redshift measurement scheme are discussed in Section \ref{observations}.
Properties of the VUDS sample are presented in Section \ref{sample}, including the redshift distribution of the sample,
the distribution of intrinsic properties like stellar mass and star formation rate, and the average spectra properties
based on high signal-to-noise stacked spectra. After comparing to other surveys in Section \ref{comparison},
we summarize our results in Section \ref{summary} and conclude on the usefulness of the VUDS to study the early phases of galaxy assembly with 
unprecedented accuracy.  \\ \\

All magnitudes are given in the AB system unless specified,
and we use a Cosmology with $\Omega_M=0.3$,
$\Omega_{\Lambda}=0.7$ and $h=0.7$.


\section{Survey design}
\label{overview}

\subsection{Science drivers}

When the VUDS survey observing time proposal was approved at the end of 2009, spectroscopic redshift surveys were still
of limited scope at the epoch of galaxy assembly significantly beyond redshifts $z\sim2-3$. 
The VUDS survey therefore aimed to address several core science goals at the epoch $2< z < 6+$, including:
(i) the early history of the global star formation rate,
(ii) the build up of the mass function for different galaxy populations, and the
contribution to mass assembly and star formation of merging, feedback and accretion,
(iii) the study of very young galaxies in their early stages of assembly, including the earliest stellar populations like Pop-III.
(iv) the identification of proto-structures and the effects of environment at early stages of galaxy evolution,
(v) the study of the clustering of galaxies to infer the mass growth of underlying  dark matter halos, 

These broad science goals still remain hot science topics today and are the main focus of VUDS.

\subsection{Survey strategy overview}

The VUDS survey is designed to study the galaxy population beyond redshift $z\simeq2$ and up
to the highest redshifts possible in the wavelength range up to $\lambda = 9350$\AA ~accessible
with the VIMOS spectrograph on the VLT (see below), corresponding to Ly$\alpha$ at a redshift up to
$z\simeq6.6$. A total of one square degree is observed in three well separated fields, the 
COSMOS field, the extended Chandra Deep Field South (ECDFS) and the VVDS-02h field, in order to mitigate cosmic
variance effects. This program was awarded Large Program status with 640h of VLT-VIMOS observing time.

A key feature of the survey is the target selection. It is primarily based on photometric redshifts
and their probability distribution function (PDF), complemented by color-color selections 
and analysis of the SED shape when not already in the photo-z selection.
When the geometry of the observed slit-masks allows it after placing the previous priority targets, 
a purely magnitude-selected random sample
of galaxies with $23 \leq i_{AB} \leq 25$ has been added to the target list.
This is further described in Section \ref{select}.

The second key feature is the extended wavelength coverage of the VIMOS spectroscopic observations covering
$3650 \leq \lambda \leq 9350$\AA, which minimizes any instrument-induced redshift desert and strongly
reduces the degeneracies in redshift measurements (Le F\`evre et al. 2014).

The last key point is the integration time of $\simeq14$h per target, which allows reaching
a S/N on the continuum at $8500$\AA ~of $S/N=5$ for $i_{AB}=25$, and $S/N=5$ for an emission line with a flux
$F=1.5 \times 10^{-18}$ erg/s/cm$^{2}$/\AA.

\subsection{Survey fields}

\begin{figure*}[h]
  \includegraphics[width=\hsize]{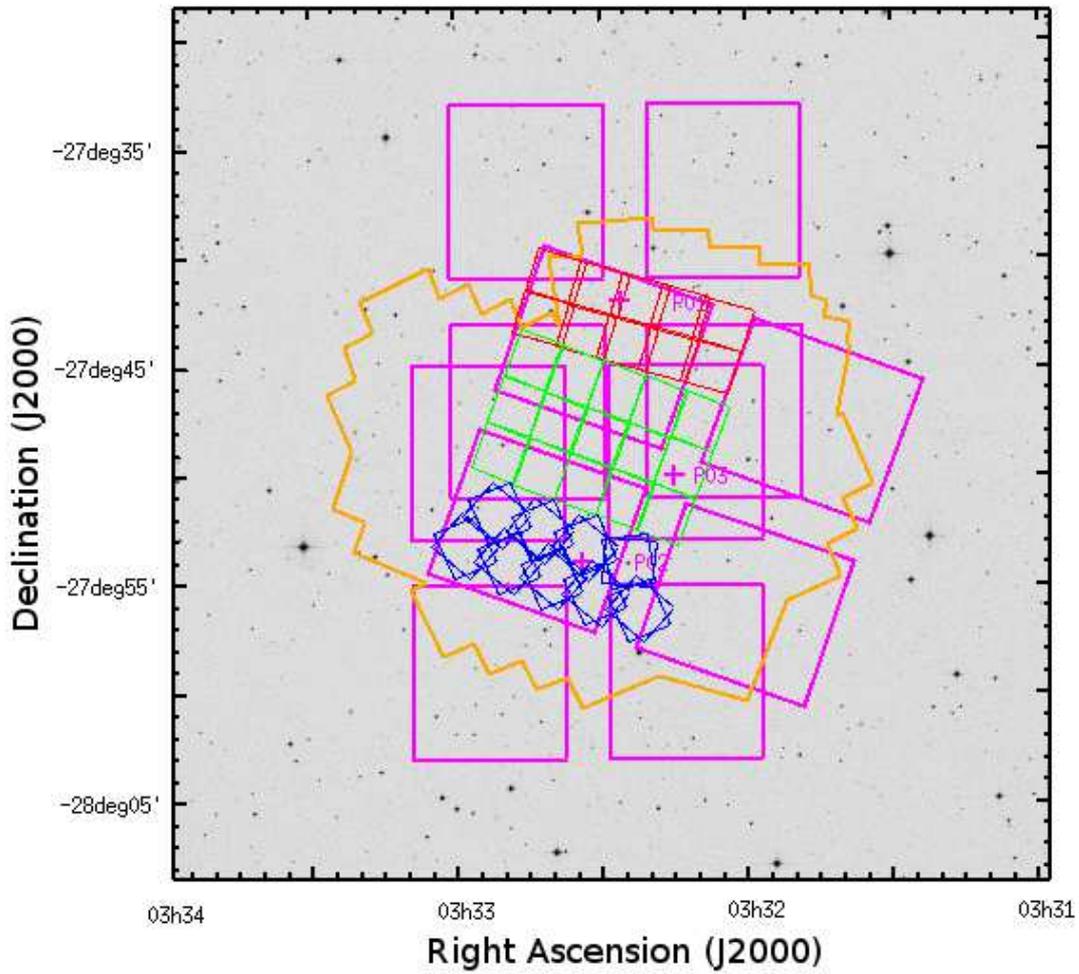}
      \caption{Layout of the observed 3 VUDS VIMOS pointings in the ECDFS field. The center of each VIMOS pointing is identified
               by a cross with the pointing number (see Table \ref{obs}), while the imprint of the 4 quadrants is in magenta.
               The positions of the CANDELS WFC3 deep area is indicated in green, and the CANDELS wide area in blue.
               The red region indicates the WFC3 coverage of the ERS.
               The outline of the existing ACS-F814W imaging is identified in orange. The size of the image is $0.67 \times 0.67$ deg$^2$.
              }
         \label{fecdfs}
\end{figure*}

\begin{figure*}[h]
  \includegraphics[width=\hsize]{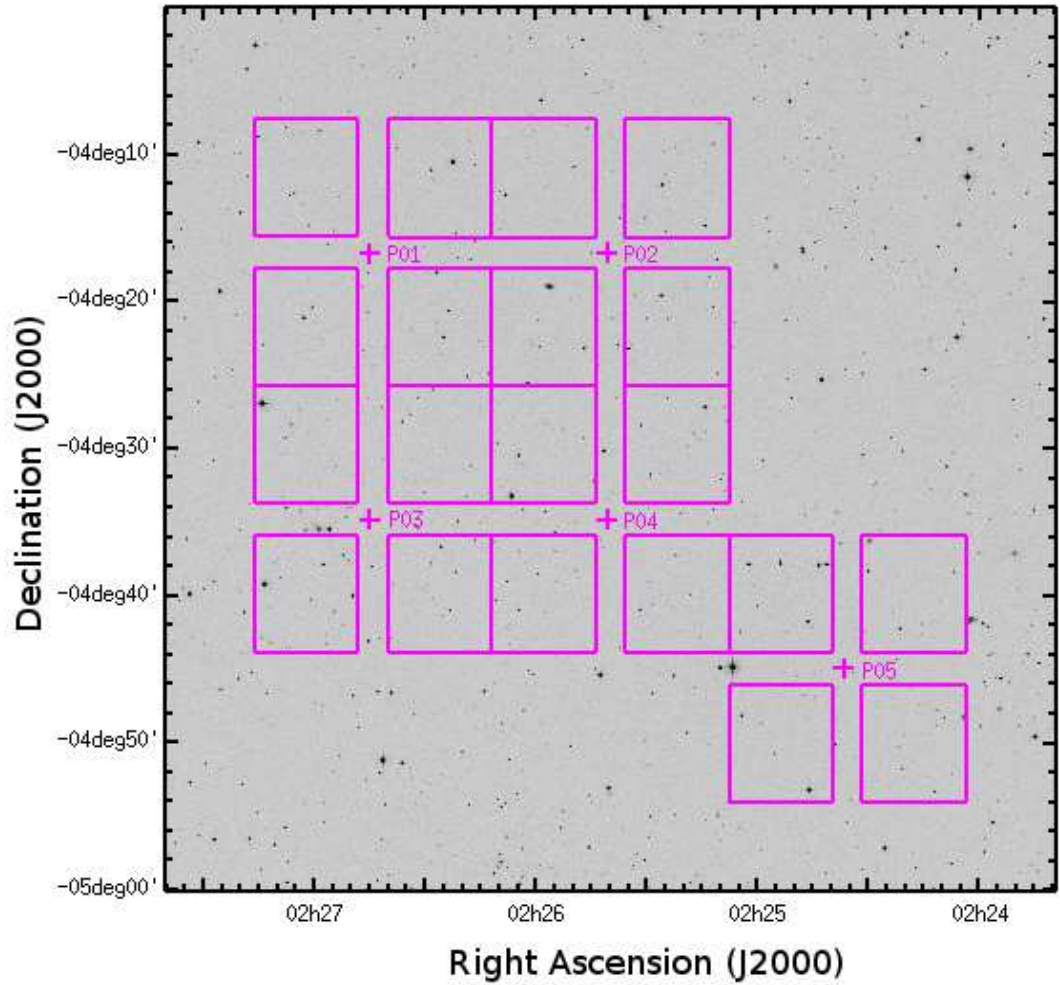}
      \caption{Layout of the observed 5 VUDS VIMOS pointings in the VVDS-02h field. The center of each VIMOS pointing is identified
               by a cross with the pointing number (see Table \ref{obs}), while the imprint of the 4 quadrants is in magenta.
               The size of the image is $1 \times 1$ deg$^2$. All of the VIMOS pointings are covered by CFHTLS visible photometry
               and WIRDS near infrared photometry.
              }
         \label{fvvds}
\end{figure*}

VUDS covers three of the most observed extragalactic fields: the COSMOS field, the ECDFS and the 
VVDS-02h (also known as the CFHTLS-D1/XMM-LSS field). A total of 0.5 square degrees are observed in the COSMOS
field, 0.31 square degrees in the VVDS-02h field, and 0.14 square degrees in the ECDFS. 
The fields location and covered area for each are summarized in Table \ref{fields}, and identified in
Figures \ref{fcosmos} to \ref{fvvds}.

\subsection{Multi-wavelength data and photometric catalogues}

By design the three survey fields cumulate a wealth of deep multi-wavelength data and spectroscopy.
This is an important part of VUDS as multi-wavelength coverage of the spectral energy distribution (SED)
is essential in combination with accurate spectroscopic redshifts, in particular to derive key physical quantities
like absolute magnitudes,  stellar masses, star formation rates or ages.
We describe for each field the most relevant existing data.

\begin{figure*}[h]
  \includegraphics[width=\hsize]{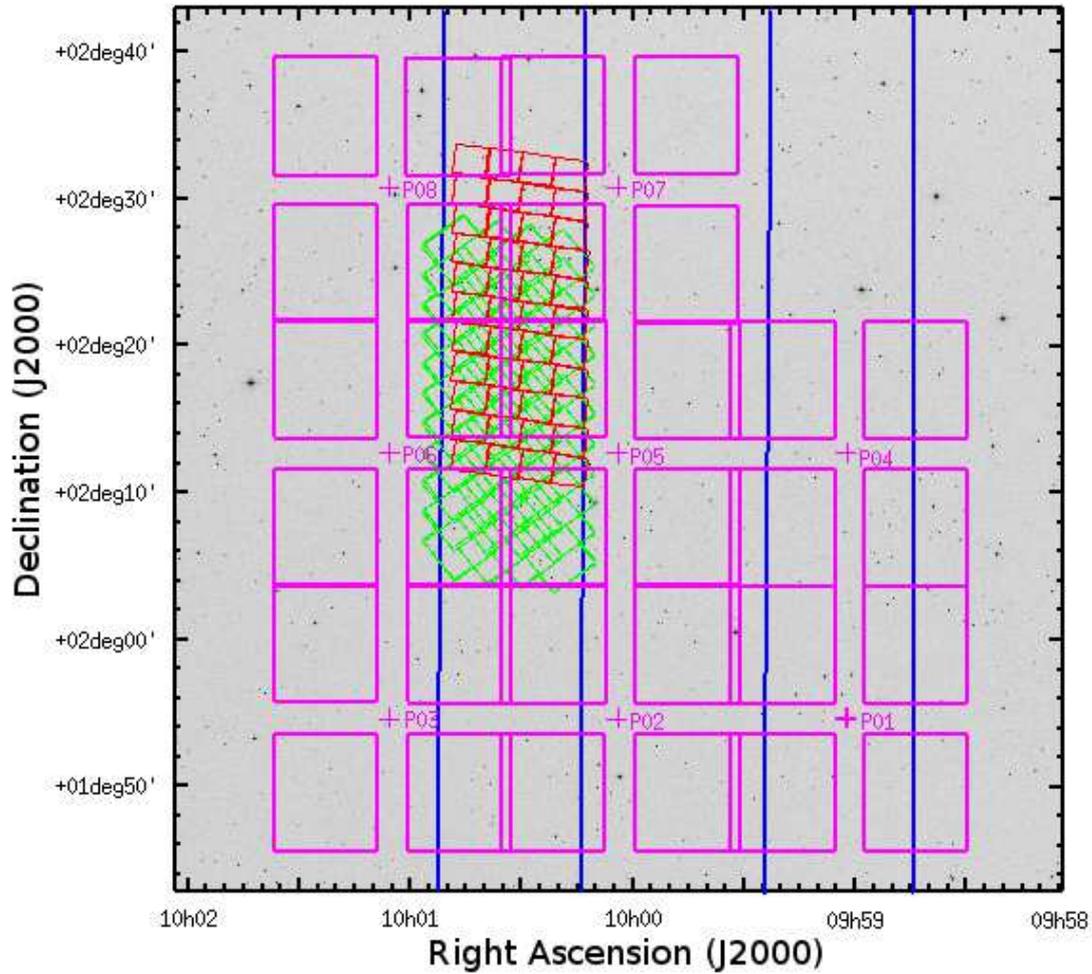}
      \caption{Layout of the observed 8 VUDS VIMOS pointings in the COSMOS field. The center of each VIMOS pointing is identified
               by a cross with the pointing number (see Table \ref{obs}), while the imprint of the 4 quadrants is in magenta.
               The positions of the UltraVista Deep stripes overlapping with the VUDS area are identified by the blue regions.
               The CANDELS ACS-F814W (in green) and WFC3-F160W (in red) areas are indicated. The size of the image is $1 \times 1$ deg$^2$.
              }
         \label{fcosmos}
\end{figure*}

The COSMOS field (Scoville et al. 2007) is extensively observed at all wavelengths.
Its location was setup close to one of the original VVDS fields, recentered to a region with
less galactic extinction when better extinction maps became available. The field was observed with
HST-ACS in the F814W filter (Koekemoer et al. 2007) providing high spatial resolution imaging
down to a depth $AB(F814W)=27.2$ ($5 \sigma$). Extensive imaging observations were obtained in $g, r, i,$ and $z$ bands 
from the Subaru SuprimeCam (Taniguchi et al. 2007), as well as from the CFHT Megacam from the CFHT-Legacy Survey including
the u-band, and photometry is available from 12 medium bands (references available in Ilbert et al. 2009). 
The UltraVista survey is acquiring very deep near-infrared imaging in the Y, J, H and K
bands using the VIRCAM camera on the VISTA telescope (McCracken et al. 2012). The UltraVista DR2 release 
reaches a depth $K_{AB}=24.8$ at $5 \sigma$ (in 2 arcsec apertures), and it is planned to increase the depth
down to $K_{AB}=25.5$ in half the area by the end of the survey. The VUDS VIMOS pointings are placed
in a way to optimize the overlap with the deeper UltraVista strips (Figure \ref{fcosmos}).
Following on the initial Spitzer observations (Sanders et al. 2007), deeper Spitzer-warm observations are obtained by 
the Splash program, reaching $AB=25.4$ ($5 \sigma$ at $3.6\mu$m) (Capak et al. in prep.). The COSMOS field has deep Herschel imaging
reaching 8 $mJy$ at $250$ $\mu$m (Oliver et al. 2012). The CANDELS survey (Grogin et al. 2011) in this field 
includes WFC3 imaging in the F125W (J) and F160W (H) filters down to $H_{AB}=27.1$ ($5 \sigma$); 
the CANDELS area is almost entirely covered by the VUDS pointings (Figure \ref{fcosmos}).
Extensive spectroscopy was obtained in the COSMOS field prior to VUDS. The zCOSMOS survey (Lilly et al. 2007)
contains $~20\,000$ galaxies with $0<z<1.2$ selected with $i_{AB} \leq 22.5$ and $\sim6000$ galaxies selected
with $g_{AB}\simeq 25-25.5$ with $1.4 < z < 3$ and a median redshift $z=2.15$ (Lilly et al. in preparation).
Deep spectroscopy from Keck has identified several hundred galaxies in the same redshift range as VUDS (e.g. Capak et al. 2011).
VUDS targeting avoided to reobserve those galaxies with existing reliable redshifts, when known. 
A wealth of observations at other wavelengths are available in this field,
we refer the reader to the COSMOS web site (\texttt{http://cosmos.astro.caltech.edu/}) for the detailed list and properties.

The ECDFS field is the focus of a number of deep multi-wavelength surveys.
Built around the field with deep 1Ms Chandra observations (Giacconi et al. 2002),
the field was extended and is covered with deep UBVRI imaging down to $R_{AB}=25.3$ ($5 \sigma$, Cardamone et al. 2010 and references
therein). Several spectroscopic surveys have been conducted (Le F\`evre et al. 2004, Cardamone et al. 2010, Cooper et al. 2012), with more than 5000
galaxies with redshifts $0<z<2$, but the number of spectroscopically confirmed galaxies at $z>2$ remains small. 
The central part of this field cumulates a number of deep HST
imaging with the HDF-South (Williams et al. 2000), GEMS (Rix et al. 2004), and ERS (Windhorst et al. 2011) surveys, 
and more recently with the CANDELS survey (Grogin et al. 2012) adding WFC3 near-IR
imaging reaching as deep as $H_{AB}=27.3-27.6$. The SERVS  Spitzer-warm obtained 3.6 and 4.5 $\mu$m imaging data down to $AB=23.1$ (Mauduit et al. 2012).

The VVDS-02h field was orginally defined as the 'deep field' of the VVDS survey. It was observed in BVRI at 
CFHT (Le F\`{e}vre et al. 2004), followed by deeper $u', g, r, i$ observations as the D1 deep field in the 
CFHTLS survey reaching $i_{AB}=25.44$ at 50\% completeness in the latest DR7 (Cuillandre et al. 2012). Deep infrared imaging has been
obtained with the WIRCAM at CFHT in YJHK bands down to $Ks_{AB}=24.8$ also at 50\% completness (Bielby et al. 2012).
Extensive multi-slit spectroscopy has been obtained with VIMOS as the 'Deep' and 'Ultra-Deep' surveys
of the VVDS, with magnitude-selected samples down to $I_{AB}=24$ and $i_{AB}=24.75$ respectively.
A total of 11\,139 objects have spectroscopic redshift measurements from the VVDS, including 774 with $z>2$.
This field was observed in all Spitzer bands as part of the SWIRE survey (Lonsdale et al. 2003),
reaching a magnitude in the 3.6 microns band of $AB=21.5$. More recently the SERVS survey obtained deeper data 
with Spitzer in the 3.6 $\mu$m and 4.5 $\mu$m bands down to $AB=23.1$ at 3.6 $\mu$m (Mauduit et al. 2012).
It is one of the fields of the HERMES survey (Oliver et al. 2012), matched to CFHTLS and VVDS data (Lemaux et al. 2014). 
A number of other deep data are available including X-ray (Pierre et al. 2004) and radio (Bondi et al. 2003) observations.  

In each field photometric catalogues including at least from u to 4.5 $\mu$m bands have been assembled, matching different imaging sources,
and extracting photometry in 2 arcsecond apertures in dual mode using SExtractor (Bertin \& Arnouts 1996). 

\subsection{Spectroscopic target selection}
\label{select}

As shown from the VVDS, the redshift distribution of faint galaxies, e.g. down to $i_{AB} \simeq 25$ is peaked 
at $z\sim1.3-1.5$, with a rapidly decreasing high redshift tail (Le F\`evre et al. 2014). 
Magnitude selection is then inefficient in picking-up $z>2$ galaxies among the far more numerous foreground.
This prompted the use of the Lyman-break selection technique, as pioneered by Steidel and collaborators (Steidel et al., 1996).
This technique makes simple use of the color properties of galaxies as a function of redshift, identifying a 
locus in color-color space associated to specific redshift ranges. 

The development of the photometric
redshift technique and its demonstrated success  in terms of the redshift measurement accuracy with typical errors
of less 5\% on $\Delta z / (1+z)$ and a low rate of catastrophic errors when compared to spectroscopic redshifts (e.g. Ilbert et al. 2010,
Cardamone et al. 2010)  
has opened new perspectives in pre-selecting galaxies. In essence, photometric redshift techniques are a 
generalisation of color selection techniques such as LBG or BzK to the complete spectral
energy distribution (SED) which must verify a color distribution accross wavelengths compatible
with reference galaxy templates.
We elected to use the best possible photometric redshifts as our main selection criterion. 
Photometric redshifts have been computed with the code Le Phare (Arnouts et al. 1999, Ilbert et al. 2006) 
using the multi-band photometric catalogues covering from
the u-band to the Spitzer 4.5$\mu$m band. The techniques are described in Ilbert et al. (2009), and more recently in Ilbert et al. (2013).  
The observed photometric data are correlated against 33 templates covering a range from early to late type
galaxies of different ages, star formation histories and metallicities, leaving the E(B-V)
extinction as a free parameter.  Emission lines with a flux empirically computed from the UV continuum flux 
are added to the templates. The 'best' redshift is assigned from the median of the 
marginalized redshift probability distribution function (PDF). 

To select the spectroscopic targets in VUDS we followed an {\it inclusive} rather than
exclusive strategy, adding samples pre-selected from several different criteria, as described below:

\begin{itemize}
\item The spectroscopic targets must verify $z_{phot} + 1 \sigma \geq 2.4$,
and $i_{AB} \leq 25$. 
As degeneracies are known to occur, we selected sources for which either the first or 
the second peak in the photometric redshift probability distribution function satisfy this criterion.
\item Sources which were not selected from the primary $z_{phot}$ criterion but satisfy one of the ugr, gri or riz
LBG color-color selection criteria have been added to the target list.
\item The full SED over all filters available in ugrizYJHK is used to identify galaxies with a break in the continuum
and not identified by any of the $z_{phot}$ or LBG criteria, extending the {\it dropout} technique.
These targest are allowed to have $K_{AB} \leq 24$ and hence may not necessarily be brighter than
$i_{AB}=25$.
\item Finally, when space on the slit mask was still available after the above selections, a random
sample of galaxies with $23 \leq i_{AB} \leq 25$ has been selected.
\end{itemize}
 For the last half of the survey observations, priority for slit placement was given
to the targets with $z_{estimate} \geq 4$ to increase the number of objects at these redshifts in the final sample.  
 
In adding these different selection criteria we aim at maximizing the pre-selection of objects above $z=2.4$.
Given the dispersion of photometric redshift errors, we expect the redshift distribution of VUDS sources
to start rising at $z\simeq2$.
While the selection strategy described above could lead to an increase in the contamination from objects at lower redshifts,
this contamination remains quite small as described below.
For each of the targeted samples, slits are placed at random using the \texttt{vmmps} slit mask
design software (Bottini et al. 2005), providing a fair and representative sample of 
the general population. 

These various selection criteria are summarized in Table \ref{selection}.
A large fraction $\sim 75$\% of the sample has been selected on the basis of the first peak
of photometric redshifts, $\sim 15$\% on the second peak of photometric redshifts, and the rest $\sim 10$\%
about equally split between the LBG color-color selection and purely magnitude selected samples.

\section{Observations, data processing, and redshift measurement}
\label{observations}

\subsection{VIMOS on the VLT}
\label{vimos}

The VIsible Multi-Object Spectrograph (VIMOS) is installed on the European Southern Observatory
Very Large Telescope unit 3 Melipal. VIMOS is a wide field imaging multi-slit spectrograph (Le F\`evre et al. 2003), 
offering broad band imaging capabilities, as well as multi-slit spectroscopy. 
VIMOS is a high performance Multi-Object Spectrograph (MOS) with 4 parallel channels, each
a complete imaging-spectrograph with a field of view $8 \times 7$ arcmin$^2$, or a total field
of 224 arcmin$^2$. The key features of VIMOS are a high multiplex (number of slits) and 
the excellent sky subtraction accuracy reaching $\sigma_{sky ~residual}\simeq0.1$\% of the sky signal (Le F\`evre et al. 2013). 

For VUDS we use the low resolution multi-slit mode of VIMOS; with the 4 channels, this offers the largest
multiplex for multi-slit spectroscopy. Following Scodeggio et al. (2009) we have optimized the slit length allowing 
lengths as small as 6 arcseconds, maximizing the  number of observed slits.

\subsection{VIMOS observations}

We observe a total of 16 VIMOS pointings, at the coordinates of the pointings identified in Table \ref{obs}.
Fifteen pointings are observed with both the LRBLUE grism 
covering $3650 \leq \lambda \leq 6800$\AA ~and the LRRED grism covering $5500 \leq \lambda \leq 9350$\AA,
leading to a full wavelength coverage of $3650 \leq \lambda \leq 9350$\AA.
With slits one arc-second wide, these grisms provide a spectral resolution $R=230$
quite uniform over the wavelength range. At this resolution, each of the four $2048\times4096$ pixels detectors can accommodate
3--4 full length spectra along the dispersion direction, and given the projected space density of VUDS targets
we therefore observe on average $N_{slits}\simeq600$ individual slits simultaneously. Each slit may contain
not only the VUDS pre-selected target but also serendipitous objects falling in the slit by chance. 

One pointing on the ECDFS (\#3) has been observed with the MR grating. This setup covers $5000 \leq \lambda \leq 9500$\AA
~with a spectral resolution $R=580$. With this resolution about 2 full length spectra can be placed along the 
dispersion direction, and a total of $\sim220$ objects have been observed in this pointing.
The targets for this pointing have been optimized towards the highest redshifts $z>4.5$; the improved resolution compared
to the LR grisms making it somewhat easier, in principle, to identify emission lines like Ly$\alpha$ between
the OH sky emission features (but possibly at the expense of sensitivity on continuum measurements). 

Most VUDS observations were obtained after the CCD detectors were upgraded in summer 2010. 
At that time, the original blue--sensitive thinned E2V CCDs were changed to red--optimized thick E2V CCDs
in 2010 (Hammersley et al. 2010). The global VIMOS sensitivity at 9000\AA ~increased by $\sim \times 2$, making it comparable
to the FORS2 sensitivity in the red with a field of view $4.8\times$ larger, and significantly reducing the fringing
above 8000\AA thanks to the thicker substrate. 

To reach a total integration of 14h, each of the LRBLUE or LRRED grism observations consist on average of 13 observing blocks (OBs)
executed at the telescope. Each OB includes three spectroscopic exposures of 1250 to 1350 seconds obtained 
dithering $-0.75,0,+0.75$ arcseconds along the slit. The OBs specify the observing conditions that must be met,
including a seeing better than 1 arcsecond, sky transparency set to 'clear', airmass less than 1.5, dark time
with constraints on the moon phase and distance from the field (lunar illumination 0.3 to 0.5, and distance
to the observed field of more than 60-90 degrees). Arc lamp and flat field calibrations are obtained after each
set of 3 OBs, and flux calibration on standard stars are performed in the standard ESO procedure. 
When ready, OBs are sent to the VLT-VIMOS service observing queue via the P2PP tool
and get executed when the atmospheric and moon conditions are met.

\subsection{Spectroscopic data processing, redshift measurement, and reliability flags}
\label{zmeas}

The spectroscopic data processing followed the same general principles as
defined for the VVDS (Le F\`evre et al. 2005a, 2013).
We summarize this process below and emphasize the specific data processing steps 
that we follow for VUDS.

The general outline of the VUDS data processing is now a standard for multi-slit spectroscopy.
We use the VIPGI environment to process the spectra (Scodeggio et al. 2005).
First the 40 individual 2D spectrograms coming from the 13 OBs for one of the 
LRBLUE or LRRED observations are extracted finding the location of the
slit projection on the detector using the expectation from the slit mask design.
Sky subtraction is performed with a low order spline fit along the slit
for each wavelength sampled. The sky subtracted 2D spectrograms are combined
with sigma clipping  to produce a single stacked 2D spectrogram calibrated in
wavelength and flux. The 2D spectrogram is collapsed along the dispersion direction
to produce a slit profile in which objects are identified.
The spectral trace of the target and other detected objects in a given slit 
are linked to the astrometric frame to identify the corresponding target
in the parent photometric catalogue. 
This processing is performed separately for each of the LRBLUE and LRRED observations.
The 2D and 1D fully calibrated spectra are then cross matched per observed slit,
and joined to form spectra with full $3650 \leq \lambda \leq 9350$\AA ~wavelength coverage.
It is to be noted that, in the wavelength range of overlap $5500 \leq \lambda \leq 6800$\AA, 
the end spectra are the average of the LRBLUE and LRRED, cumulating about 28h of observation.
At the end of this process 1D sky-corrected stacked spectra are extracted. They
are fully calibrated both in wavelength and in flux using spectrophotometric standard stars.
In this process additional objects can be detected in the slit and have subsequently 
their 1D spectra extracted. These objects are called "secondary detections", and 
a photometric counterpart is searched for in the photometric catalogue. If an object is found within
less than one arcsecond from the spectral trace, the secondary spectrum identification is assigned
the identifier of the object in the photometric catalogue. If no object is found in the photometric
catalogue, the object is extracted and a new entry is produced in the catalogue with coordinates at the location 
corresponding to the trace of the object. It might happen that a single emission line is identified
upon visual examination of the 2D spectrogram of a given slit, but not extracted by the automated procedure: these objects are
flagged and then manually extracted. This  procedure is particularly important for
objects with single emission lines and no or very little detected continuum, which often turn out to be Ly$\alpha$ emitters 
at high redshifts (e.g. Cassata et al. 2011).  

Upon comparison between the photometric magnitudes and the magnitudes derived from the
calibrated spectra, we realized that the $u-$band part of the spectra (and to
a lesser extent the $g-$band) was lacking photons at the $\sim40$\% ($\sim15$\%) level. We proceeded to add three well-defined 
corrections to the spectra: (1) atmospheric absorption, (2) atmospheric refraction, and (3) Galactic extinction.
The atmosphere absorbs photons depending on the airmass along the light path; this is corrected
using the prescription defined  for the Paranal observatory in Patat et al. (2011). In addition atmospheric refraction
acts as a small prism before entering the telescope, spreading the incoming light into a spectrum
with length depending on the airmass and paralactic angle, the angle of the slit to the zenithal angle. With slit-masks  placed on sky objects using an $r-$band 
filter prior to spectroscopic observations, and slits one arcsecond wide, this may introduce a significant
loss of uv-blue photons falling out of the slit. The Galactic extinction on distant sources also produces a chromatic correction, which
has been applied using the E(B-V) maps of Schlegel et al. (1998). In this process we also proceed to produce spectra with continuum
flux which are calibrated on the $i-$band photometric flux, therefore correcting for any slit losses occuring
in that band if the object extension is larger than the slit width.
Adding these corrections we are able to
correct the uv-blue spectroscopic flux in such a way that there is excellent agreement between the spectroscopic flux and the photometric flux measurements 
at better than the 5\% level in all wavelengths $3650 \leq \lambda \leq 9350$\AA. This is of particular importance when fitting the spectra to reference 
templates in order to derive internal galaxy properties (Thomas et al., in prep.).
One of the last observations obtained on this program includes a number of repeated observations
on a sub-sample of VUDS galaxies. When processed it will allow to estimate redshift (velocity) measurements
accuracy, as well as to have an independent check of the reliability of each of the redshift reliability flags (see below).
Based on the previous VVDS, zCOSMOS and VIPERS surveys with the same VIMOS instrument, we expect 
the redshift accuracy to be in the range $dz/(1+z)=0.0005 - 0.0007$, or an absolute velocity accuracy $150-200$ km/s (Le F\`evre et al. 2013).
Relative velocities in the same slit (along the slit profile) can be measured to a better precision using e.g. accurate spectral line fitting.

The 2D and 1D spectra are then available for spectroscopic redshift 
measurements using the EZ environment (Garilli et al. 2010). The core algorithm to find a redshift 
is the cross-correlation with templates, confronted to a separate estimate of an emission line redshift when applicable. 
A key element for the cross-correlation engine to deliver a robust measurement
is the availability of reference templates covering a large range of galaxy
and star types, as well as a large range of rest wavelengths. This last point has 
to be carefully dealt with when measuring the highest redshift galaxies, as it is
necessary that the templates go far enough in the UV, bluer than the Lyman-912\AA ~limit,
for the wavelength overlap between the observed galaxy and the redshifted template
to be large enough to provide a robust correlation signal. We have used
templates built over the years from VIMOS observations for the VVDS (Le F\`evre et al. 2005a, 2013)
and the zCOSMOS survey (Lilly et al. 2007). As is relevant for $z>2$ we have used templates with
and without Lyman$-\alpha$ emission. 

The redshift measurements are first obtained from an
automated run with EZ. This serves as a basis for a visual examination of each spectrum,
with an iteration on the redshift measurement using EZ in manual mode, if necessary. We 
find that at these high redshifts more than half of the spectra need manual intervention
to properly measure a redshift. This is mainly due to residual defects in the spectra like
sky features residuals after sky subtraction or second order spectra superimposition.  
As the manual intervention remains an important feature in this process, 
we have implemented a method to minimize measurement biases linked to one single person. 
One spectrum is measured by two team members separately, and
these measurements are then confronted to produce a single measurement agreed on
by the two measurers. With this scheme, we have implemented the same redshift reliability estimator as
developed for the CFRS (Le F\`evre et al. 1995) and refined for the VVDS (Le F\`evre
et al. 2005, 2013), zCOSMOS (Lilly et al. 2007) or the VIPERS (Guzzo et al. 2013) surveys.
The reliability of a redshift measurement is expressed with a flag  giving the
range of probability for a redshift to be right. The reliability flag may take the following values:
\begin{itemize}
\item 0:  No redshift could be assigned (the redshifts are then set to 9.9999).
\item 1:  50--75\% probability to be correct
\item 2:  75--85\% probability to be correct
\item 3:  95--100\% probability to be correct
\item 4: 100\% probability to be correct
\item 9:  spectrum with a single emission line. The redshift given is the most probable
          given the observed continuum, it has a $\sim80$\% probability to be correct. 
\end{itemize}
The probabilities associated to these reliability flags are remarkably stable
because of the process involving several independant people, smoothing out
individual biases (Le F\`evre et al. 2013). 

\begin{figure*}[h]
   \includegraphics[width=14cm]{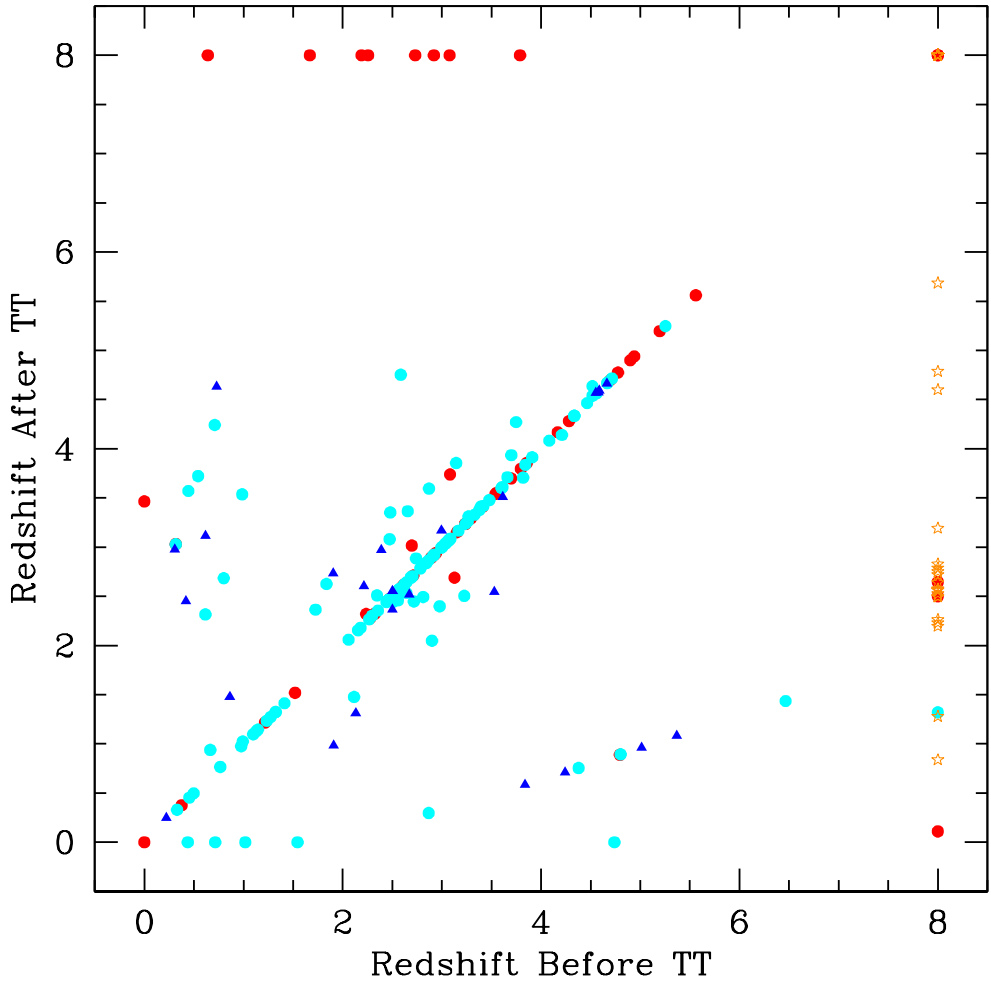}
      \caption{Comparison between VUDS spectroscopic redshifts before and after the {\it Tiger Team} (TT)
               check of redshifts measured after a first measurement pass (see text). 
               About 10\% of all objects had a redshift change (6\%)
               or a flag change (4\%). 
               Red circles represent objects for which the reliability flag was downgraded after the Tiger Team work;
               filled cyan circles are those objects with upgraded flags;
               and blue triangles are objects keeping the same flag.
               Objects which had undetermined redshifts (flag=0) before the TT work are placed at $z_{before}=8$
               and represented with orange starred symbols. Objects which have undetermined redshifts
               after the TT work (either keeping their original undetermined status or the TT work decided to
               downgrade them) are the red circles at $z_{after}=8$.
              }
         \label{tiger}
\end{figure*}

VUDS enters a redshift domain which  has never been probed by spectroscopic redshift
surveys on this scale. The expertise of the VUDS team members grew steadily as
more and more of the data was being processed. Upon examination of the first measurements
we realised that redshifts were probably wrongly assigned for a small but sizeable ($\sim10$\%)
fraction of the objects. Several standard cases for erroneous measurements 
were identified, including the possible confusion between early M stars (M0-M3)
and $z \sim 5$ absorption-line only galaxies (going both ways), the assignment
of Ly$\alpha$-1215\AA ~instead of OII-3727\AA ~(or vice-versa), the setting of
a continuum break to the Balmer-D4000 break rather than the Ly$\alpha$ break (or vice-versa).
In addition, the reliability flag as defined above was sometimes either too cautious
when a measurer found a spectrum he/she was not yet familiar with, or too
optimistic. In view of this, we opted to create a 'Tiger Team' (OLF, PC, EV, BG, DM, VLB, OLF, LP, LTa) 
in charge to conduct an additional redshift check,
provided with a set of well defined reference cases and their treatment.
This redshift check was done again by two independent people going through all the measured spectra,
separately identifying which ones needed to have their redshift and/or reliability flag modified,
and agreeing on the modifications. The check of all pointings was done by four pairs of 2 people,
each couple proposing a list of modifications to be examined and agreed upon by
the other 'Tiger team' members. 
We compare in Figure \ref{tiger} the {\it old redshifts} with the {\it Tiger team redshifts} which summarizes this process.
At the end of this process, about 10\% of the 
objects had either a redshift or a flag change. 
While this does not garantee that there are no more obvious 'catastrophic failures' in spectroscopic
redshift meassurements, this process ensures a homogeneous treatment of all spectra and the reduction
of the main degeneracies present in measuring the redshifts of high redshift galaxies. Obviously, for 
the fainter objects where the information content of the spectra is not sufficient to solve a possible degeneracy,
the reliability flag is assigned to the 'flag 1' category. 

\begin{figure*}[h]
\centering
   \includegraphics[width=\hsize]{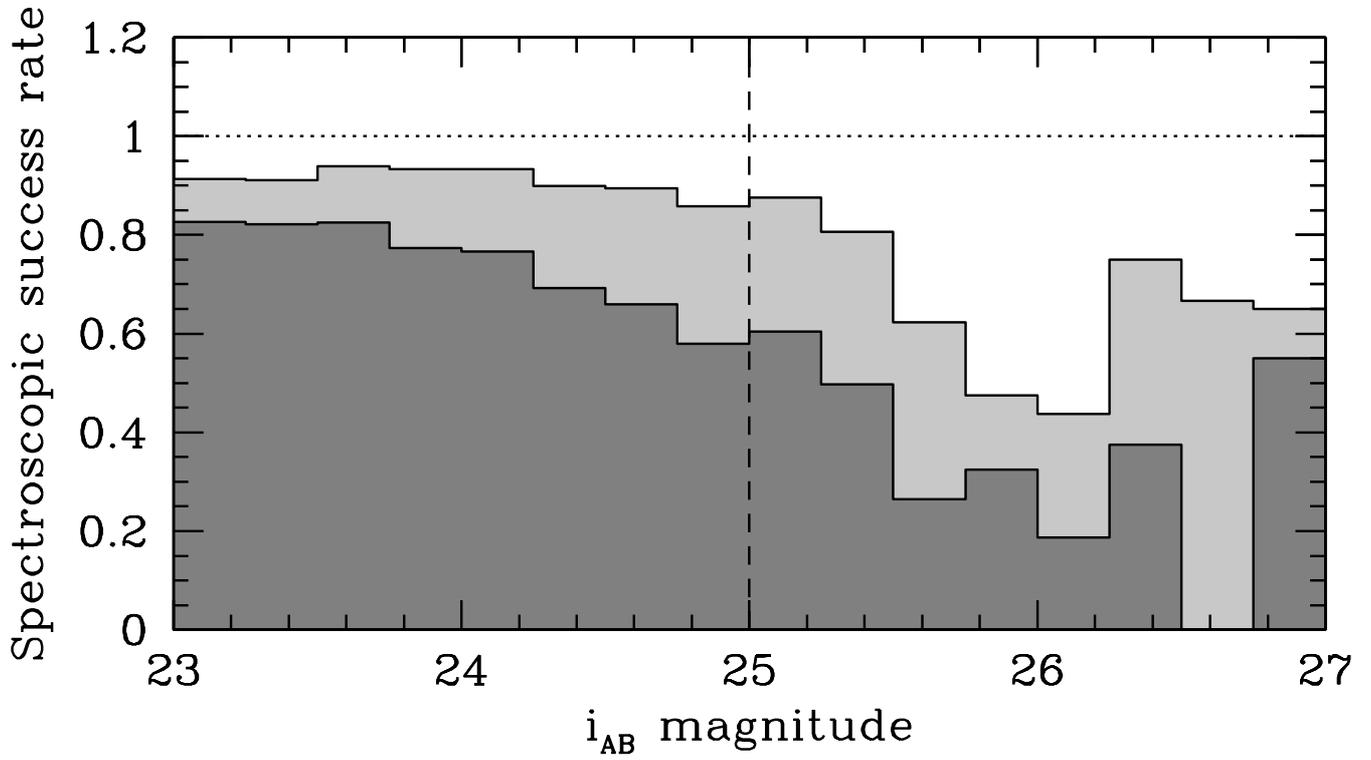}
      \caption{Spectroscopic success rate versus $i_{AB}$ magnitude, for all objects with a redshift measurement (light grey)
               and all objects with a $>75$\% reliable redshift measurement (flags 2, 3, 4, and 9; dark grey). 
              }
         \label{ssr_magi}
\end{figure*}

One key element of the selection function of the VUDS sample is the  
target sampling rate (TSR) defined as the ratio of the observed galaxies (all reliability flags)
to the underlying parent photometric populations from which the spectroscopic targets
have been selected. We find a global TSR of $\sim30$\% for the VUDS survey, similar for the three observed fields. 
With respect to the total population with $i_{AB} \leq 25$, the parent sample of galaxies satisfying the
VUDS selection criteria represents 10\%, and hence the observed VUDS sample represents 3.3\% of all galaxies
with $i_{AB} \leq 25$.
The most reliable redshifts including flag 2, 3, 4 and 9 account for 70.2\%
of the sample, flags 1 represent 21.4\% and 8.4\% could not be measured (flag 0). The spectroscopic
success rate (SSR) as a function of magnitude is shown in Figure \ref{ssr_magi}. Down to $i_{AB}=25$
91\% of targets have a redshift measurement (flags 1, 2, 3, 4, 9), and  74.3\% have a reliable 
measurement (flags 2, 3, 4, 9). This fraction is decreasing to 58\% in the last 0.25 magnitude bin before $i_{AB}=25$. The 
spectroscopic success rate then goes down with magnitude with a reliable redshift obtained for $\simeq 30$\% 
of the galaxies targeted at $i_{AB}=26$.
A complete description of the survey selection function including the analysis of 
the spectroscopic success rate at different redshifts will be provided elsewhere
(Tasca et al. in prep.).

The experience gained in this process is invaluable for future massive
high redshift surveys. As in Le F\`evre et al. (2013), we emphasize that
redshift measurement {\it at these high redshifts} is a complex process
which deserves dedicated and expert care beyond a simplistic 'good' vs. 'bad' redshift measurement
scheme to fully exploit the information content of faint object spectroscopy at the instrumental limit.
This is further discussed in Section \ref{sec_select} in view of the
aposteriori comparison between spectroscopic and photometric redshifts.

\subsection{VUDS spectra}

\begin{figure*}[h]
   \centering
   \includegraphics[width=\hsize]{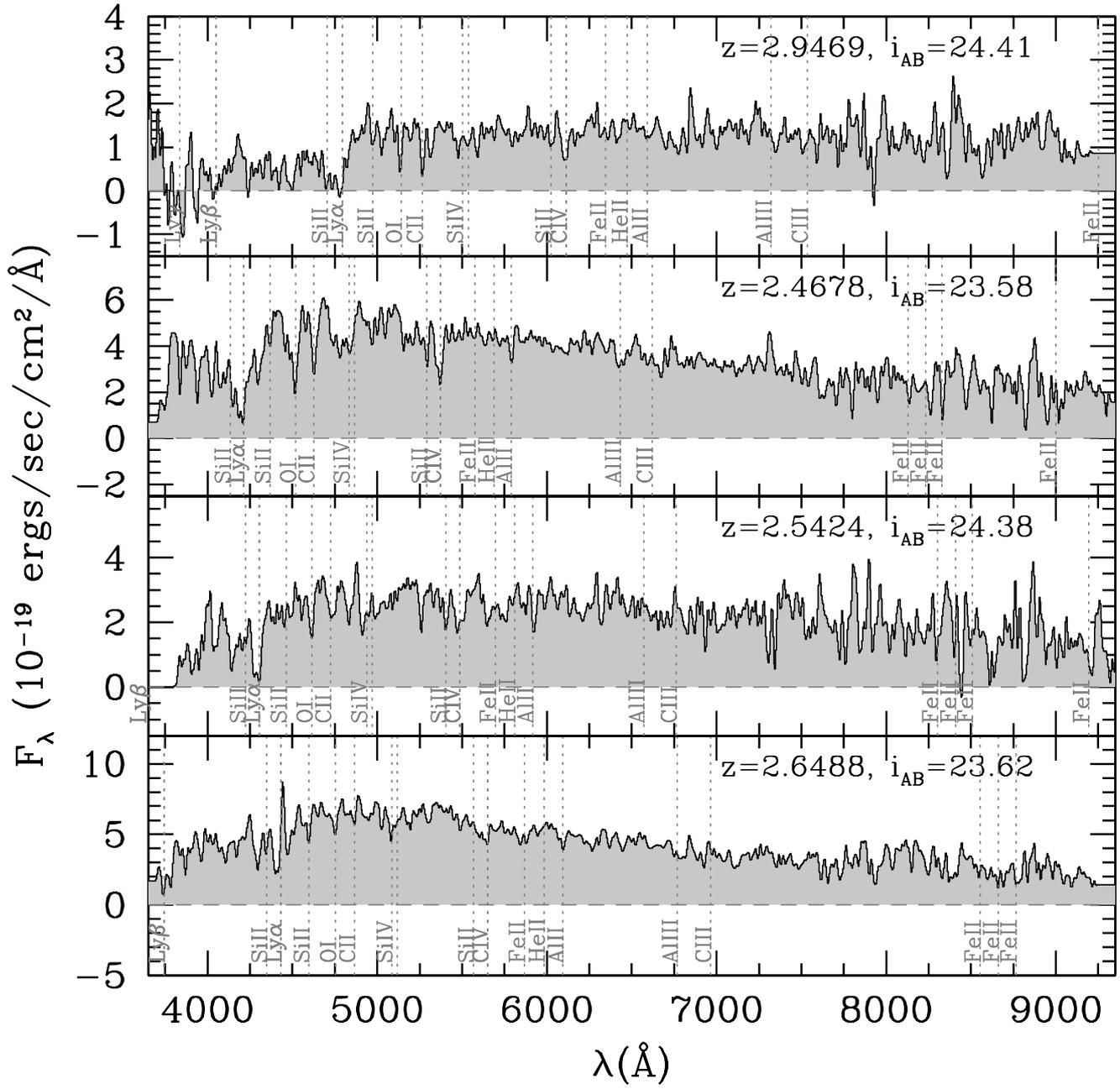}
      \caption{Sample VUDS spectra in the range $2<z<3$.
              }
         \label{spec_z2}
\end{figure*}

\begin{figure*}[h]
   \centering
   \includegraphics[width=\hsize]{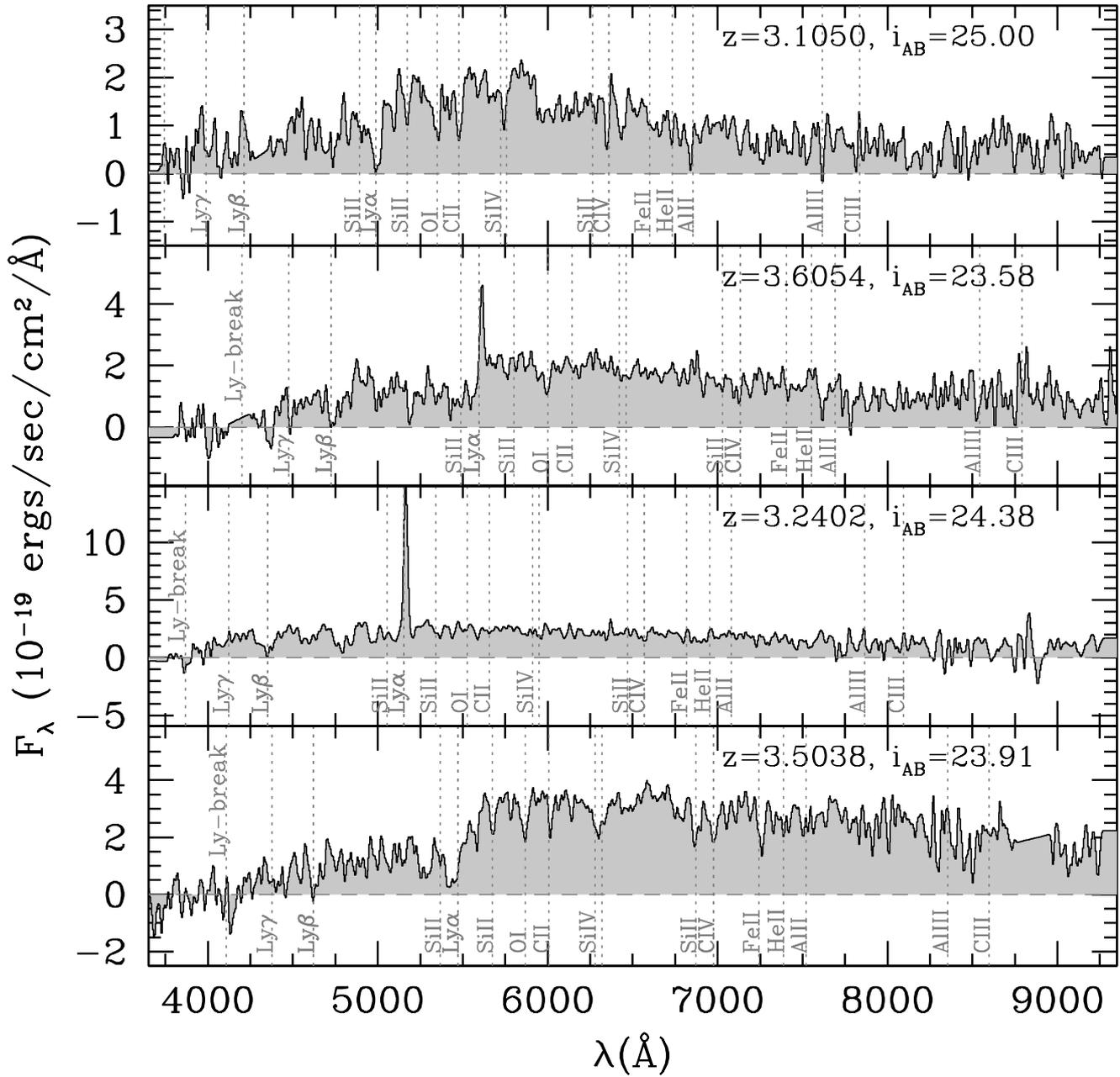}
      \caption{Sample VUDS spectra in the range $3<z<4$.
              }
         \label{spec_z3}
\end{figure*}

\begin{figure*}[h]
   \centering
   \includegraphics[width=\hsize]{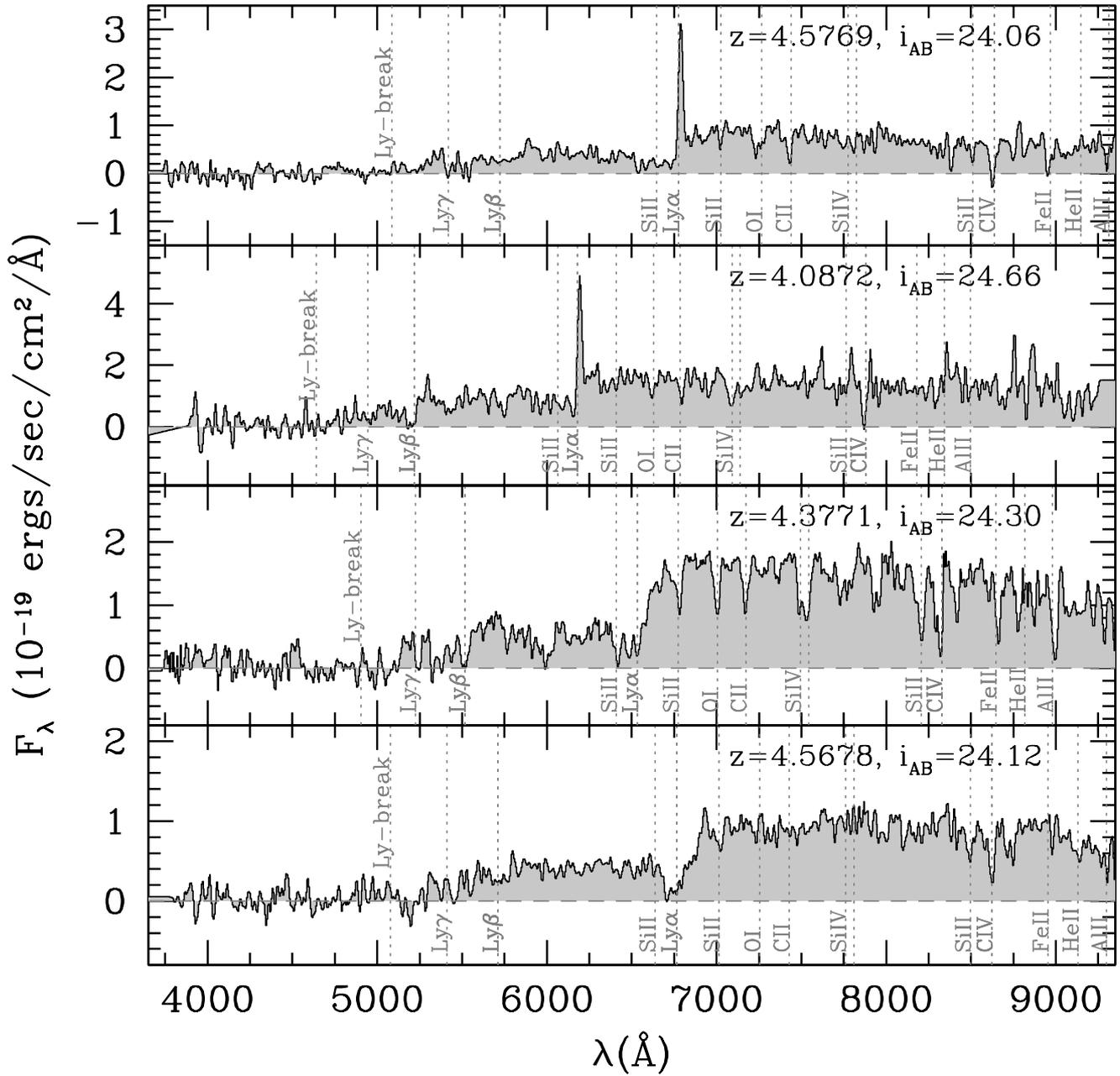}
      \caption{Sample VUDS spectra in the range $4<z<5$.
              }
         \label{spec_z4}
\end{figure*}

\begin{figure*}[h]
   \centering
   \includegraphics[width=\hsize]{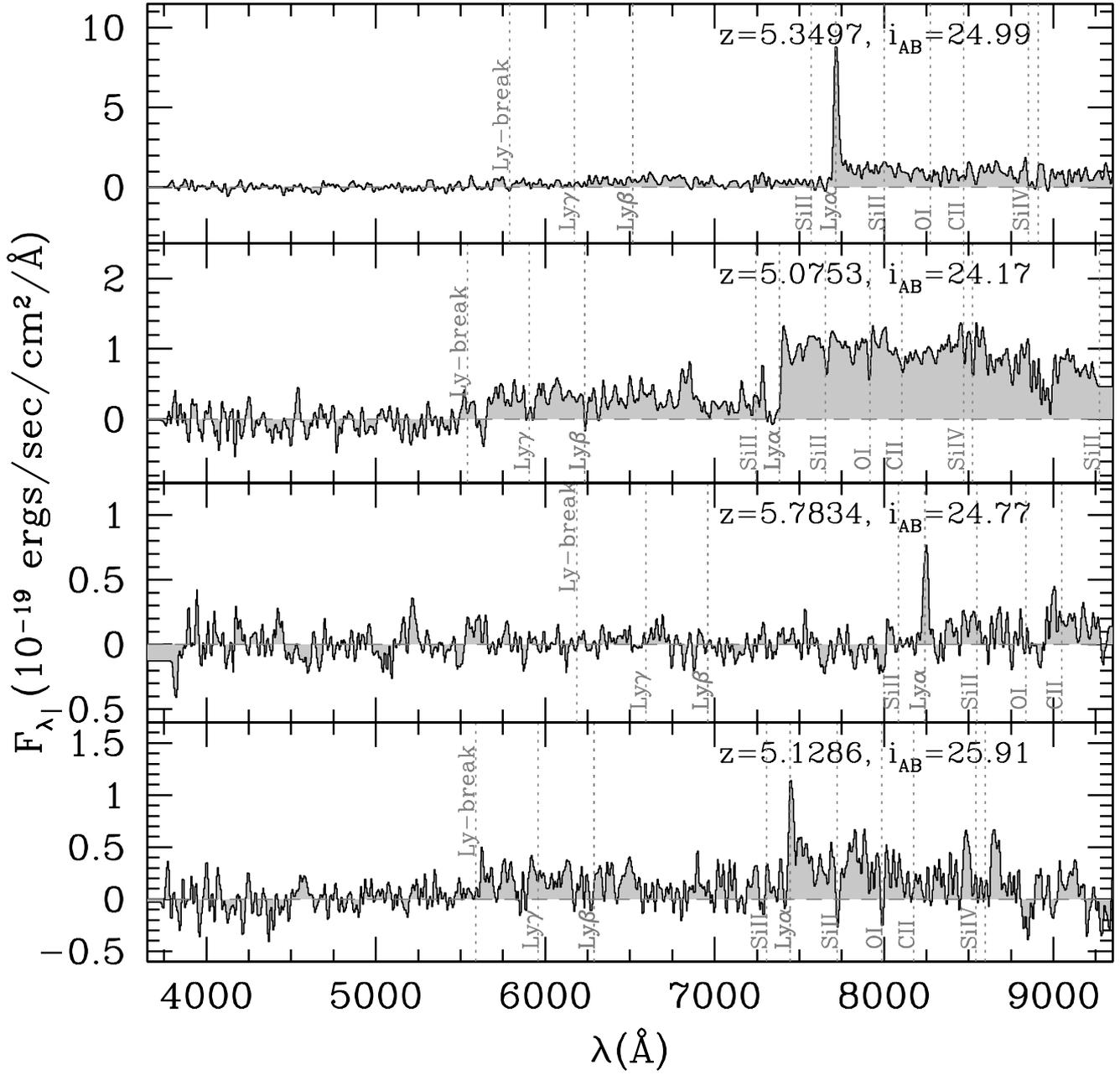}
      \caption{Sample VUDS spectra in the range $5<z<6$.
              }
         \label{spec_z5}
\end{figure*}

We present sample spectra over the redshift range of the survey in Figures \ref{spec_z2} to Figures \ref{spec_z5}.
The signal-to-noise per sampling element ($\sim7$\AA) of the spectra at 1500\AA ~rest-wavelength has a mean 
of $S/N=4.5$, and $\sigma_{S/N}=2.1$, the S/N per spectral resolution element being $\sim2\times$ higher. 
This gives access to a range of spectral features and properties for each individual galaxy. The main
spectral lines identified in individual spectra are Lyman$-\alpha$1215\AA ~(in emission or in absorption), 
OI$\lambda$1303, CII$\lambda$1334, the SiIV-OIV doublet at $\lambda$1394-1403\AA, SiII$\lambda$1527, CIV$\lambda$1549, FeII$\lambda$1608, 
HeII$\lambda$1640, AlII$\lambda$1671, AlIII$\lambda$1856, CIII]$\lambda$1909 
(see the list in Table \ref{spectral_lines}).
Below Lyman$-\alpha$, and depending on the IGM absorption, Lyman$-\beta$, Lyman$-\gamma$ and the Lyman limit at 912\AA ~can be identified.   
The average spectral properties in different redshift ranges are described in section \ref{spec_prop}.

\begin{figure*}[h]
   \includegraphics[width=\hsize]{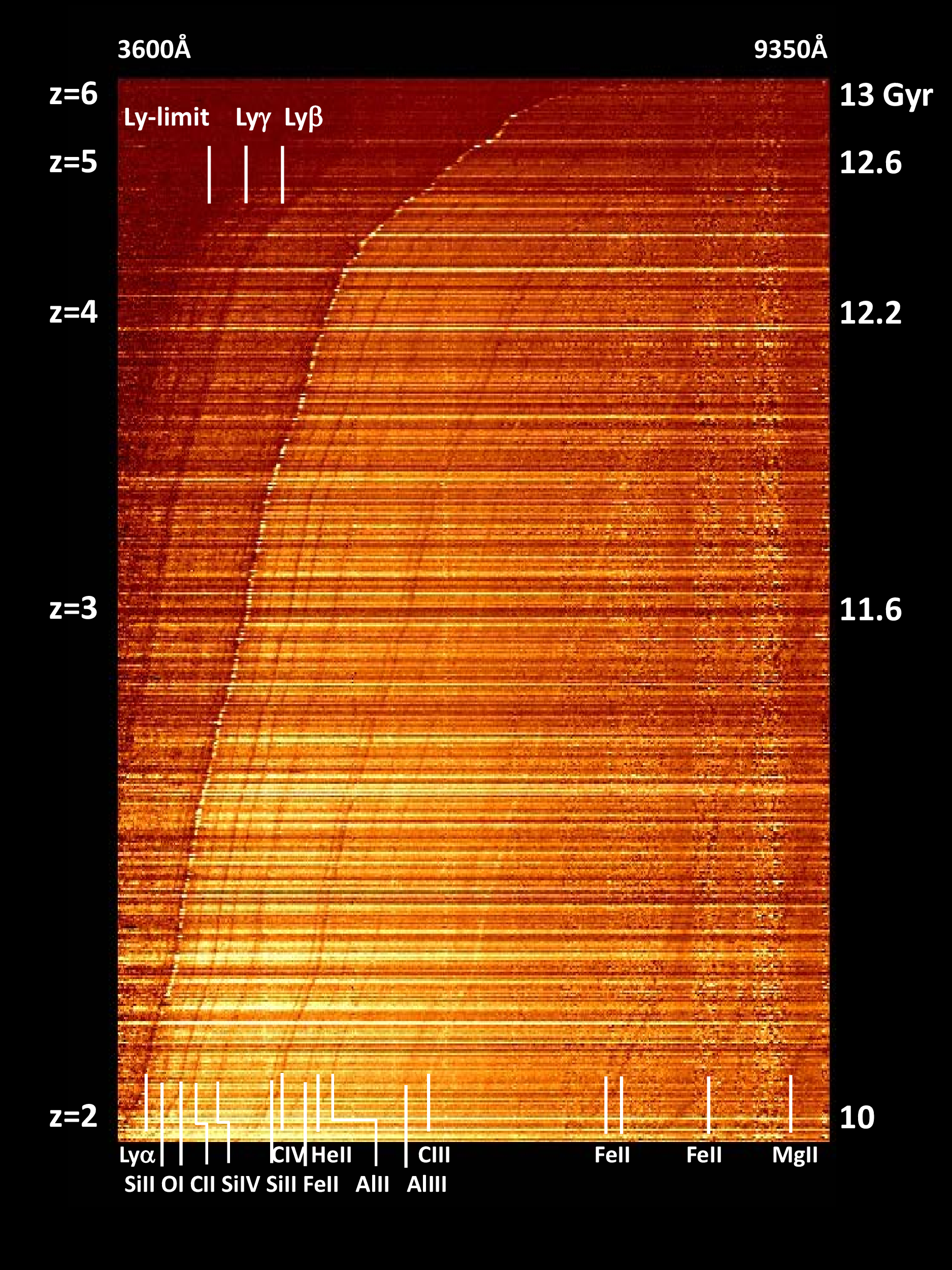}
      \caption{Complete overview of the spectra obtained in the VUDS survey, with the wavelength
               going from 3650 to 9350\AA ~on the X-axis. The redshift is increasing (in a non linear way) 
               as indicated along the left Y-axis, with the corresponding look-back time indicate along the right Y-axis. 
               The image is build with the spectra of all $2<z<4$ VUDS galaxies with flags 3 and 4,
               and all spectra with flags 2, 3 and 4 for $z>4$, ordered one per image line by increasing redshift.
               All the main emission and absorption lines can be readily identified on this image,
               even faint ones, thanks to the increased contrast produced by the continuous display
               of spectra. The main spectral lines are identified on the top left (below Ly$\alpha$) and 
               at the bottom of the plot (above Ly$\alpha$), as listed in Table \ref{spectral_lines}.
               The vertical bands appearing at fixed wavelength in the red
               correspond to increased residual noise after subtraction of the strong atmospheric OH-bands. 
              }
         \label{spec_all}
\end{figure*}

An overview of the VUDS galaxy population over $2<z<\sim6$ is shown in Figure \ref{spec_all}. This figure is built
from all spectra with reliability flags 3 and 4 in $2<z<4$ and all spectra with flags 2, 3 and 4 for $z>4$ assembled in one single image,
one spectrum per image line.
The display is quite striking as the eye is able to follow up some of the weakest spectral features up to the highest redshifts.
The Ly$\alpha$ line is readily visible in absorption or emission all along the redshift range, with the blue
wing of the broad damped Ly$\alpha$ clearly visible in all spectra. Going to the higher redshifts one can note 
the Ly$\beta$, Ly$\gamma$, and Lyman-break (912\AA).

\subsection{Comparison between spectroscopic and photometric redshifts}
\label{sec_select}

\begin{figure*}[h]
   \includegraphics[width=\hsize]{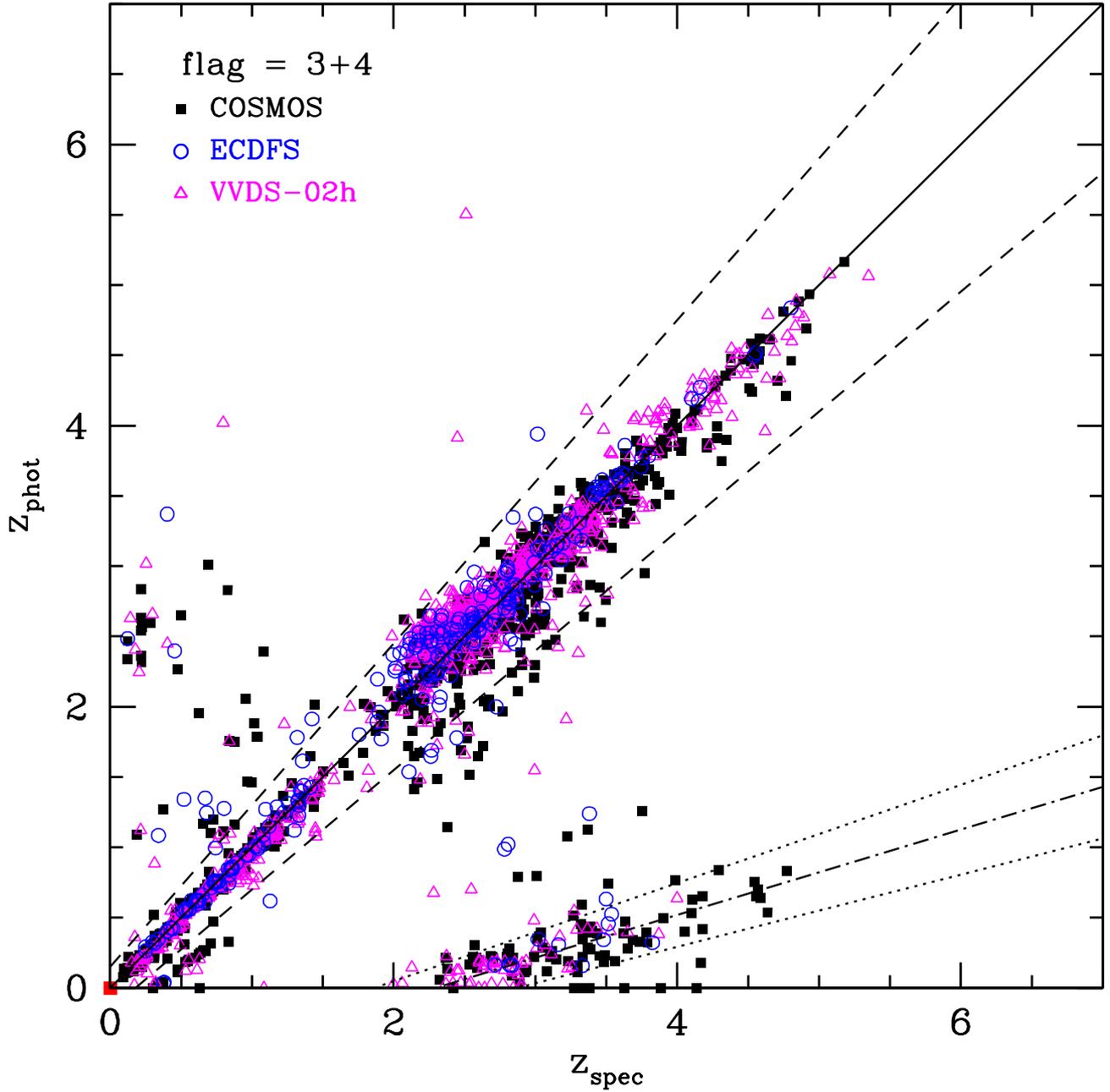}
      \caption{Comparison between VUDS spectroscopic redshifts with spectroscopic flag 3 and 4 (95-100\% reliable)
               and the SED-derived photometric
               redshifts using the Le Phare code. Galaxies in the three different fields are
               identified as filled squares (COSMOS field), open triangles (VVDS-02h),
               and open circles (ECDFS). The 1:1 equality relation is drawn as a continuous line
               with 15\% errors expressed as $0.15 \times (1+z)$ drawn as dashed lines. 
               The known degeneracy of photometric redshifts between a Balmer--4000\AA ~break and a Ly$\alpha$--1215\AA ~break
               is identified by the dot-dash line with 15\% errors as the dotted lines. 
              }
         \label{zphot_34}
\end{figure*}

\begin{figure*}[h]
   \includegraphics[width=\hsize]{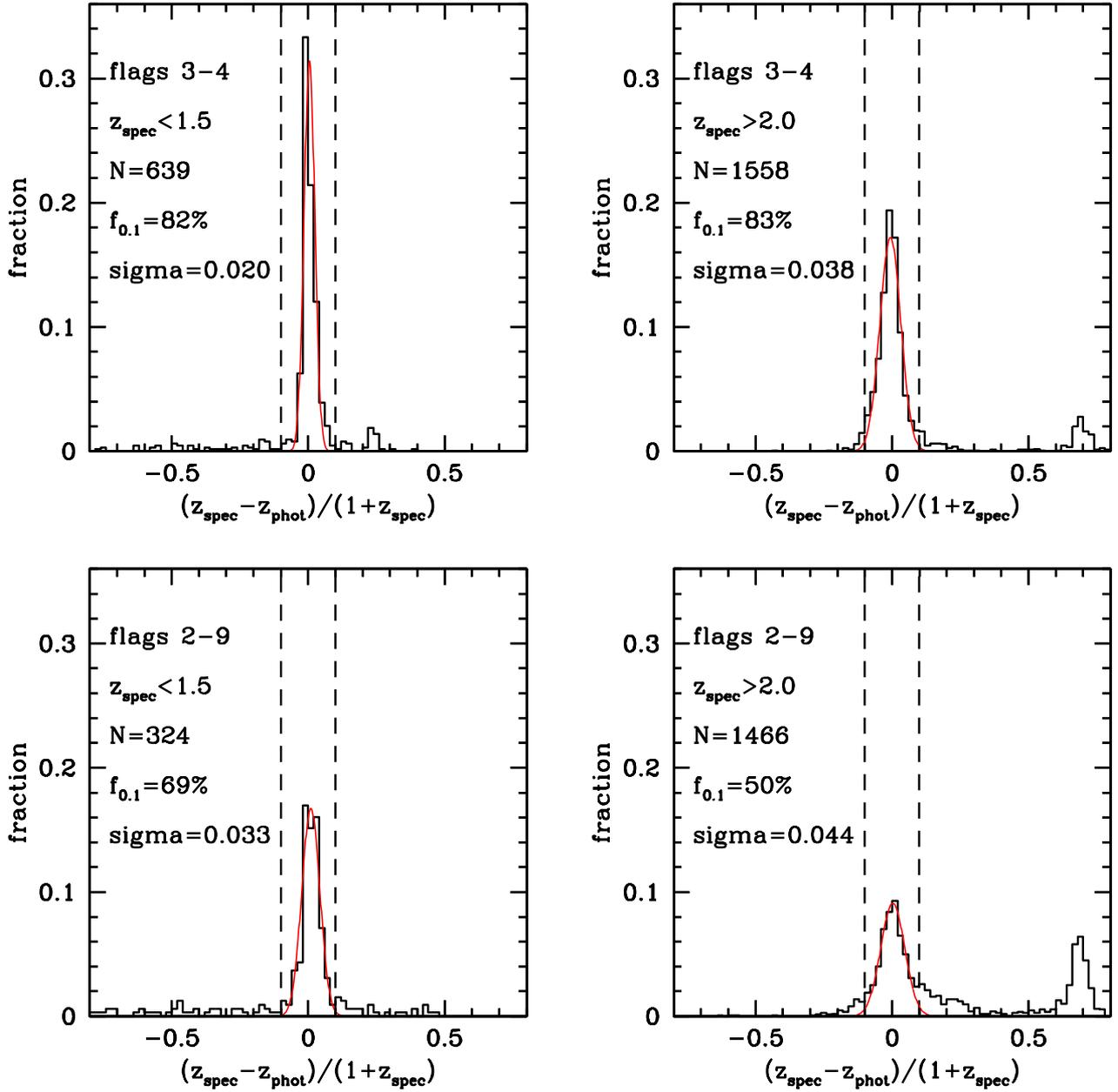}
      \caption{Distribution of the difference between photometric redshifts and VUDS spectroscopic redshifts
               for flags 3+4 (top panels) and flags 2+9 (bottom panels), for redshifts $z<1.5$ (left panels)
               and redshifts $z>2$ (right panels). The number of galaxies is indicated in each panel
               together with the $\sigma$ of the distribution, and the fraction of galaxies
               verifying $\delta z = ( z_{spec}-z_{phot} ) / (1+z_{spec}) \leq 0.1$. The distribution peaked
               at 0 is for the main sample, while a secondary peak appearing beyond $\delta z = 0.5$ is
               produced by the degeneracy between the D4000 and the 1215\AA ~continuum breaks (see text).
              }
         \label{dz}
\end{figure*}

With VUDS targets selected in large part from their photometric redshifts (Table \ref{selection}), we make here
a comparison between the photometric redshifts $z_{phot}$ computed from the multi-wavelength data set and used 
as an input to the target selection, and the spectroscopic redshifts $z_{spec}$. We compare $z_{phot}$ and 
$z_{spec}$ for the VUDS sample with reliable flags 3 and 4 in Figure \ref{zphot_34}. As close to 100\% of the $z_{spec}$(flags 3,4) 
are the "truth", we can directly test the accuracy and degeneracies of the $z_{phot}$. 
We present in Figure \ref{dz} the distribution of
$\delta z = ( z_{spec}-z_{phot} ) / (1+z_{spec})$ for flags 3+4 and 2+9 for $z<1.5$ and z$>2$
separations arbitrarily chosen to distinguish a low redshift regime where the $z_{phot}$ computation heavily
relies on rest-frame visible domain features like the D4000 spectral break, and a high redshift domain where
the computation rests on UV rest-frame features like the Ly$\alpha$ break produced by the intervening IGM and the Lyman-continuum limit. 
The width of the distribution for flags 3+4 below redshift $z=1.5$  is an excellent $\sigma(\delta z/(1+z))=0.02$, particularly
in the COSMOS and ECDFS fields which benefit from medium band photometry. Above 
redshift $z=2$ it is about double which is still excellent but
signals the increasing difficulty to assign accurate photometric redshifts from broad band photometry only. 
It is immediately visible in Figure \ref{zphot_34} that there is a secondary
tight relation with a low $z_{phot}$ corresponding to a high $z_{spec}$, which is easily explained by the degeneracy
between the D4000 continuum break and the Lyman$-\alpha$ break when computing $z_{phot}$, and our use of the secondary
peak of the $z_{phot}$ PDF to select spectroscopic targets. We draw in Figure \ref{zphot_34} the relation
$z_{phot}=(1215/4000) \times (1+z_{spec}) -1$ and $\pm15$\% around this mean. A number of galaxies
are well within these limits, representing 9\% of the flag 3 and 4 sample at $z>2$. 
In all, 95.2\% of our flag 3+4 sample at $z_{spec}>2$ verify either $z_{spec}=z_{phot}$ 
or $z_{phot}=(1215/4000) \times (1+z_{spec}) -1$ within 15\%, meaning that our selection function is particularly effective
in picking--up galaxies at these redshifts, with low 'catastrophic failure' rate,
as further discused in Tasca et al. (in prep.). 

\begin{figure*}[h]
   \includegraphics[width=\hsize]{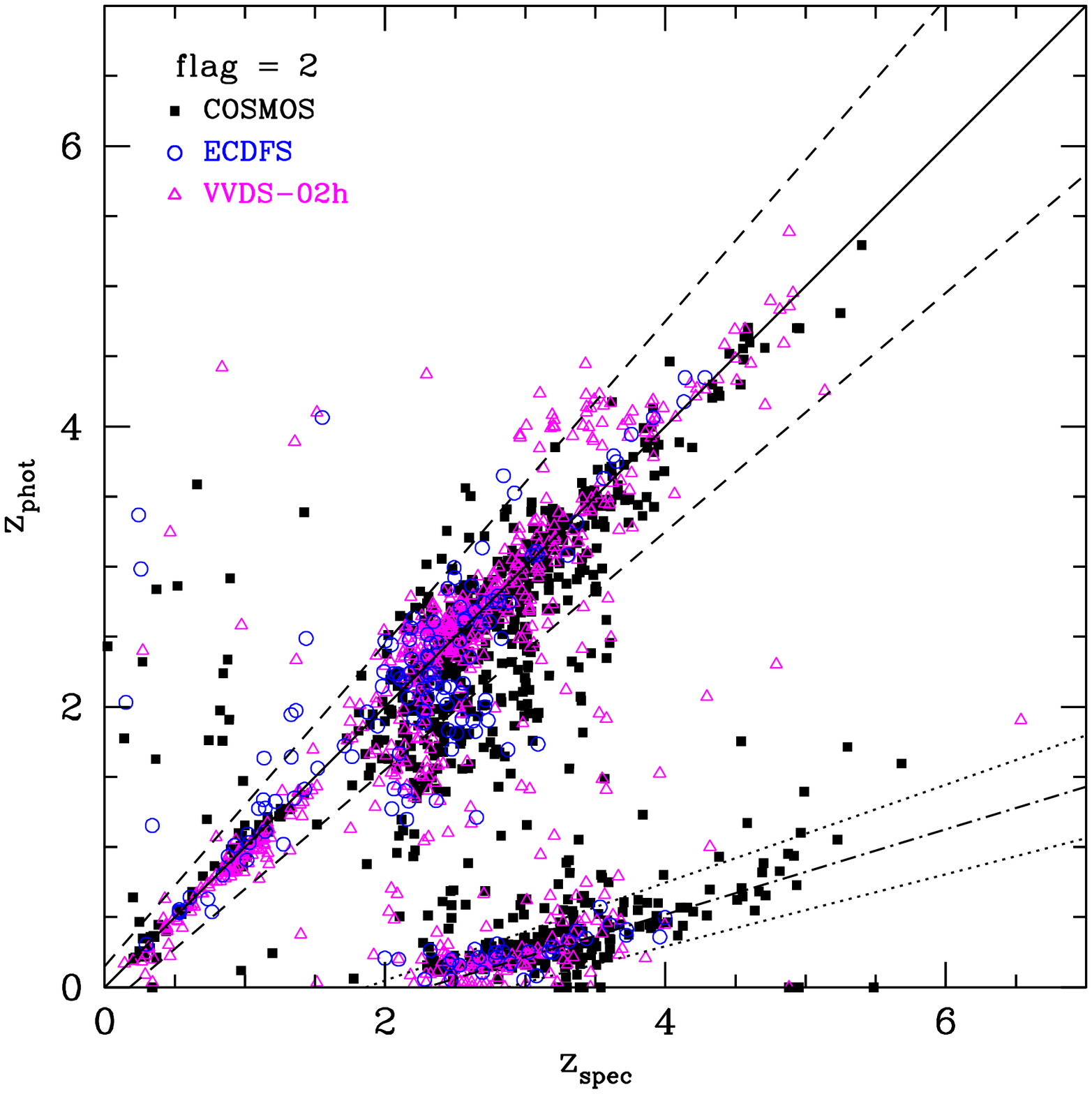}
      \caption{Same as Figure \ref{zphot_34} but for spectroscopic flags 2 ($\simeq85$\% reliable). 
              }
         \label{zphot_2}
\end{figure*}

\begin{figure*}[h]
   \includegraphics[width=\hsize]{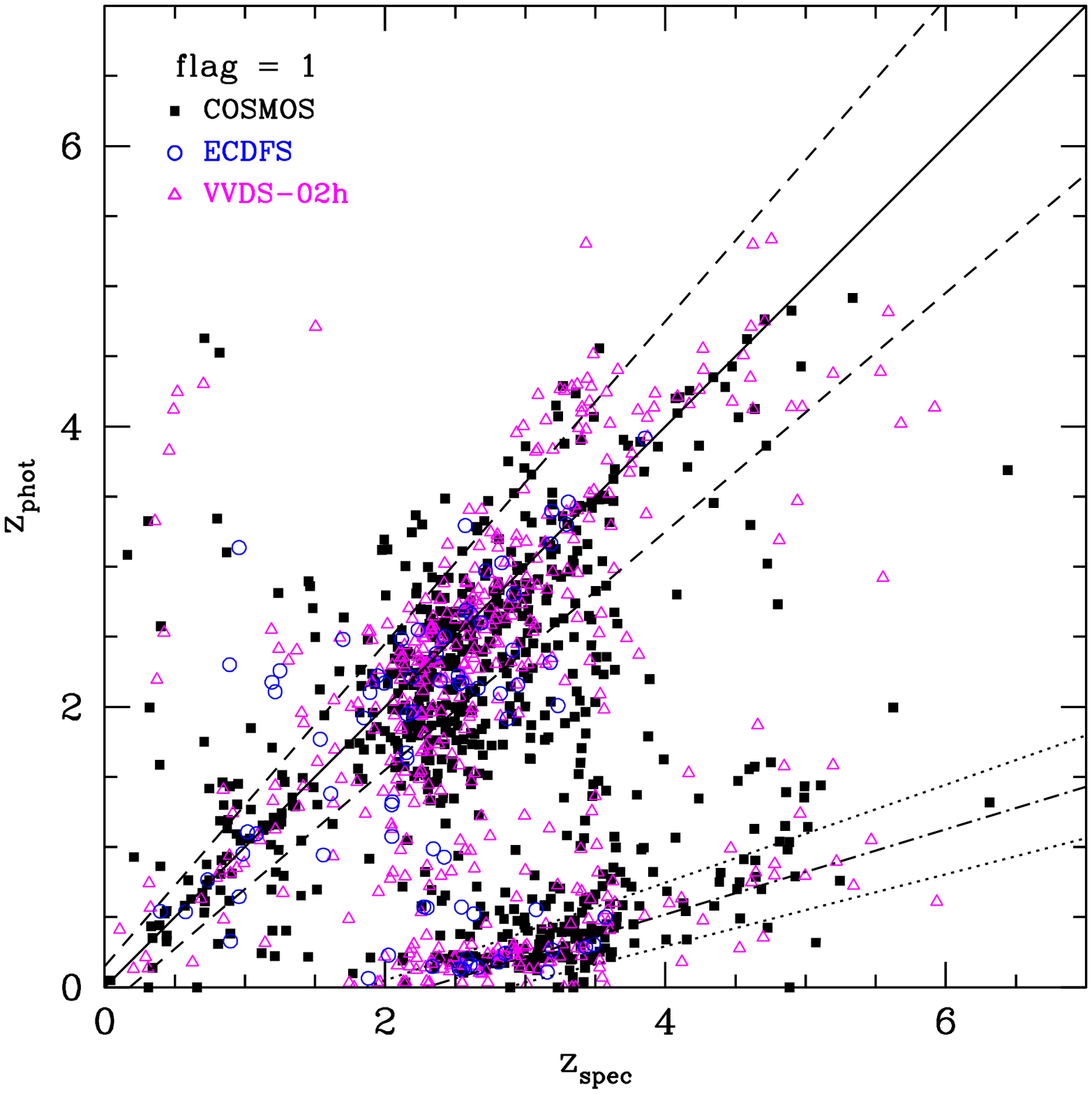}
      \caption{Same as Figure \ref{zphot_34} but for spectroscopic flags 1 ($\simeq50-70$\% reliable). 
              }
         \label{zphot_1}
\end{figure*}

\begin{figure*}[h]
   \includegraphics[width=\hsize]{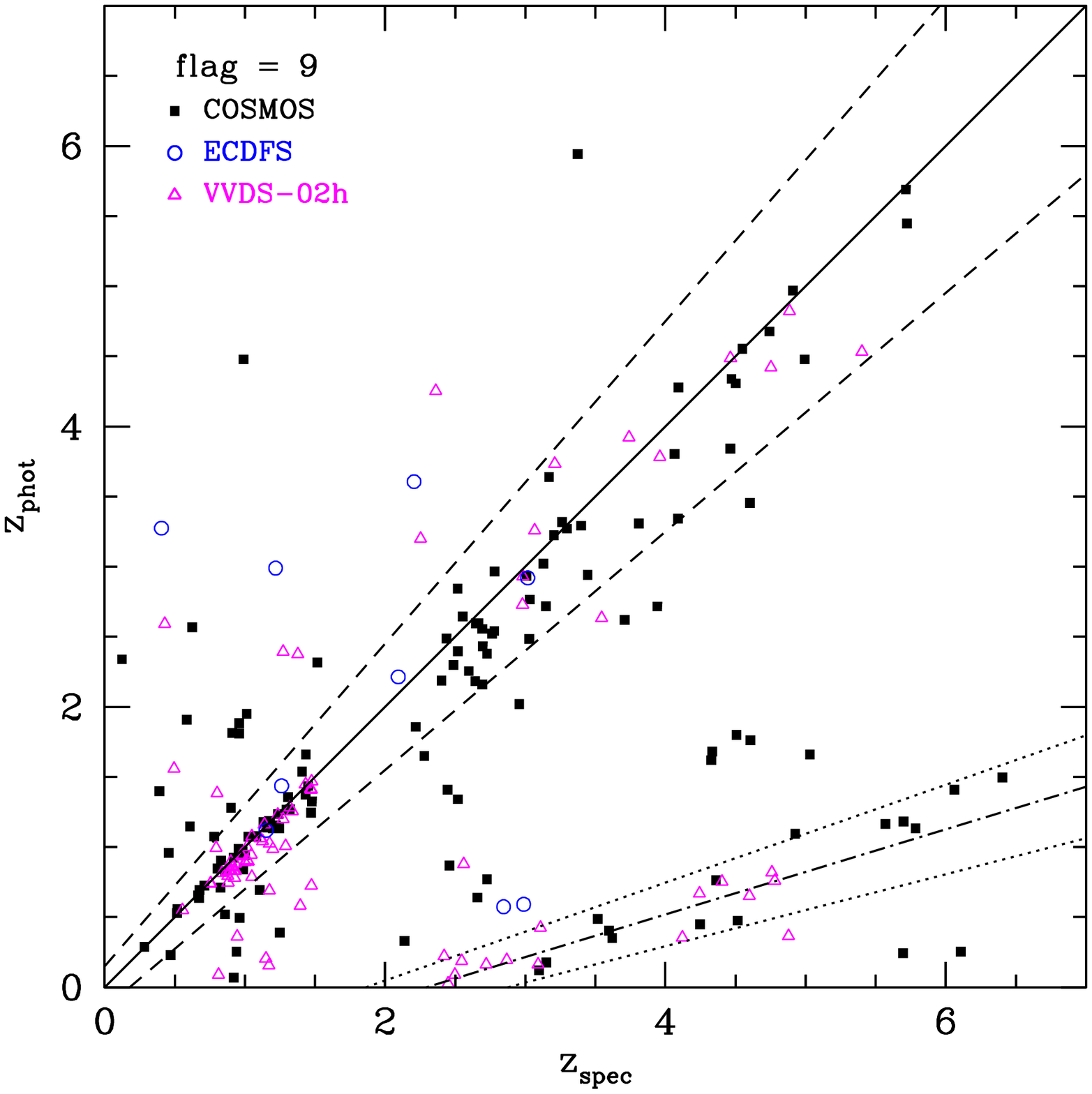}
      \caption{Same as Figure \ref{zphot_34} but for spectroscopic flags 9 ($\simeq 80$\% reliable). 
              }
         \label{zphot_9}
\end{figure*}

We can use the distribution of $(z_{spec},z_{phot})$ for flags 3 and 4 to evaluate the level of agreement between $z_{phot}$
and $z_{spec}$ for other spectroscopic reliability flags 2, 1 and 9.
The comparison between $z_{phot}$ and $z_{spec}$ for reliability flag 2, 1 and 9 are presented in Figure \ref{zphot_2},
\ref{zphot_1} and  \ref{zphot_9}, respectively. These plots are qualitatively similar to Figure \ref{zphot_34}, with the
1:1 and the 4000:1215 relationships well populated. 
For flags 2+9 and $z>2$ we find $\sigma(\delta z) = 0.044$ hence not much different from the flags 3+4 distribution, 
and with about 75\% to 80\% of the objects in this category within 15\% of the 1:1 and 1215:4000 relations, which
fully supports that these redshifts have a high level of reliability.
As noted above, the secondary peak in these distributions for $z>2$ is produced by the degeneracy betweeen 
the D4000 and 1215\AA ~continuum breaks. This peak is more pronounced for flag 2+9, with about 23\% of the objects in this
category within 15\% of the 1215:4000 relation, meaning that the
objects with these reliability flags are more prone to this degeneracy. A possible reason is that
the magnitudes of flag 2+9 objects are on average fainter than the flag 3+4 counterparts.   
Defining the catastrophic failure rate in the selection of $z>2.4$ VUDS galaxies 
as the fraction of galaxies which are outside either $\mid z_{spec}-z_{phot} \mid \leq 0.15 \times (1+z_{spec})$ 
or $z_{phot} - (1215/4000) \times (1+z_{spec}) -1 \leq 0.15$, it is 20\% for flag 2, 37\% for flag 1 and 24\% for flag 9.
Taking into account the intrinsic catastrophic failure rate of 5\% for photometric redshifts as observed for flag 3+4
this gives a qualitative estimate on the reliability level of the different spectroscopy flags.

Based on the excellent match between spectroscopic and photometric redshifts obtained for flag 3 and 4, we
have added a decimal point to the reliability flag as defined in Section \ref{zmeas} translating the level
of agreement between the spectroscopic redshift and the photometric redshift for each galaxy. 
This decimal point may take five different values from 1 for a poor agreement, to 5 for an excellent agreement; 
more specifically:  *.1 means that the spectroscopic and photometric redshifts have a difference 
$dz= \mid z_{spec}-z_{phot} \mid / (1+z_{spec}) \geq 0.5 $,
*.2  is for $0.3 \leq dz < 0.5$, *.3 for $0.2 \leq dz < 0.3$, *.4 for $0.1 \leq dz < 0.2$, and *.5 for $dz < 0.1$.
Adding a *.5 decimal therefore further increases the reliability of the spectroscopic redshift measurement, while a *.1
decimal would rather lower it. However, for the higher spectroscopic reliability flag 3 and 4, a low photometric
decimal *.1 or .2 would rather indicate that the photometric redshift is likely to be incorrect. 
This scheme allows to define a sample depending on the level of robustness required by a particular analysis.

This analysis brings two general comments for studies based on photometric redshifts.
First, if VUDS targets had been selected based only on the primary peak of the $z_{phot}$ PDF, 
$\sim17.5$\% of the sample with $z \geq 2$ would have been lost compared to our selection. 
Second, for studies based only on a $z_{phot}$ sample, which necessarily assign a redshift
using the primary peak of the redshift PDF, 14.2\% of high redshift galaxies would be missed because 
they would be wrongly placed at low redshift instead of $z>2.3$. This average value shows
a variation with redshift, which will be discussed in a forthcoming paper.

\section{General properties of the VUDS sample}
\label{sample}

In this section we report on the main properties of the VUDS sample, giving an overview of the parameter 
space probed by the survey.
As of this writing, the VUDS sample contains 6\,250 objects with a measured redshift, including 6\,003 galaxies, 20 AGNs and 227 stars,
and no redshift measurement could be obtained for 750 objects. These numbers will increase by 15-20\% when data processing
will be completed. 
The projected sky distribution of the VUDS sample follows the layout of the VIMOS pointings, as identified
in Figures \ref{fcosmos} to \ref{fvvds}. With 8 pointings on the COSMOS field, $\sim4150$ objects previously without measurements
have been observed and $\sim 3700$ have now spectroscopic redshift measurements (these numbers are being
revised following the last data processing), covering a total area of $1800$ arcmin$^2$. Over $1125$ arcmin$^2$ in the VVDS-02h
field the data for $\sim 2300$ targets have been processed and $\sim 2100$ objects have now spectroscopic redshifts, and we expect
20\% more when data processing will be completed. In the ECDFS, one pointing has been processed so far with $\sim 550$ objects 
observed and $\sim 500$ with a redshift measurement, and with 2 more pointings to process (one with LR and one with MR grisms) 
we expect this number to increase by about 60\%. 
We use the current sample to discus general properties of the sample below.

\subsection{Sample properties}

The efficiency of the survey target selection can be estimated from the redshift
distribution of the observed sources.  

\begin{figure*}[h]
   \includegraphics[width=\hsize]{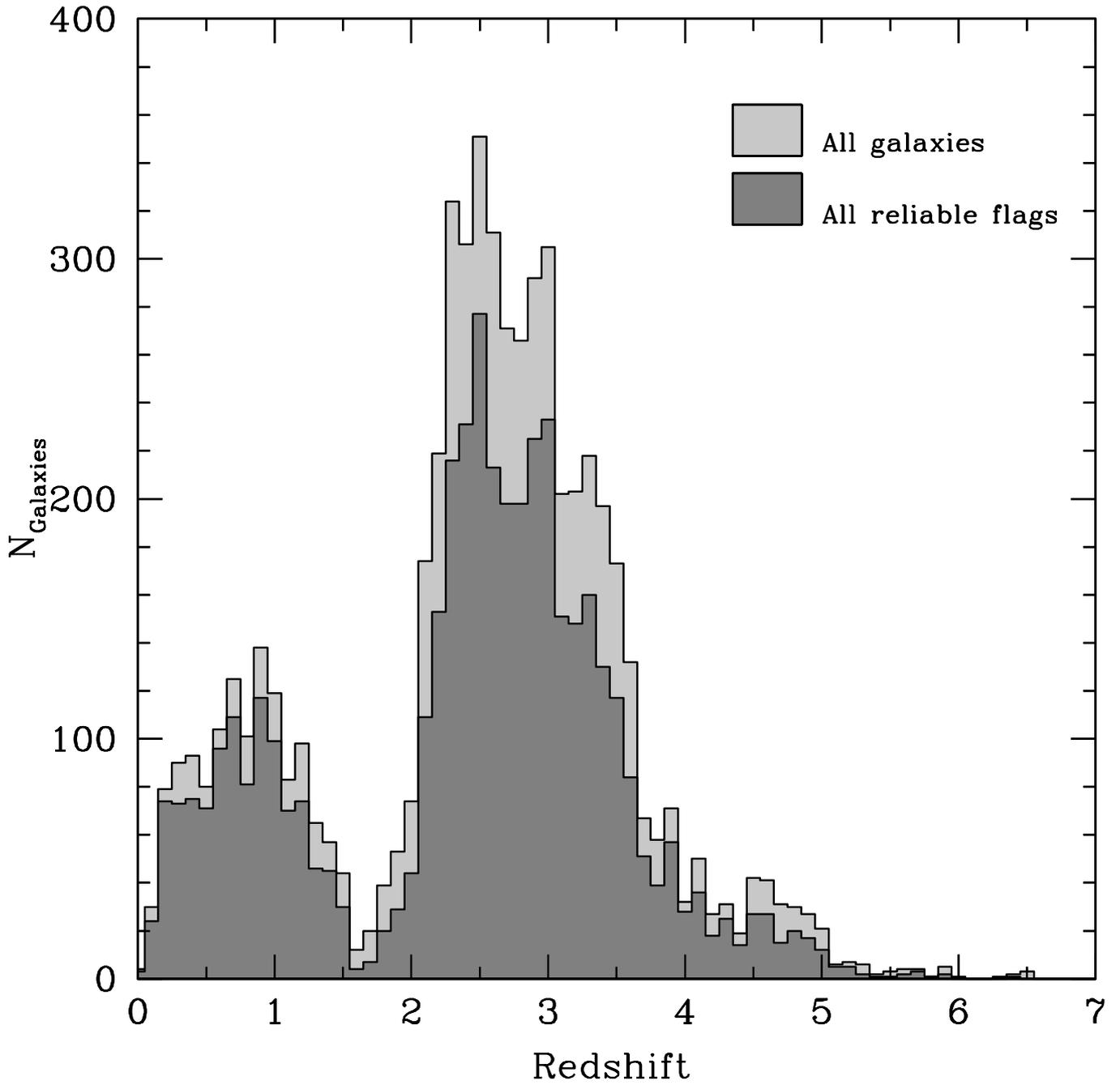}
      \caption{The current redshift distribution from 6057 galaxies already processed in the VUDS survey, for all objects with a redshift measurement (light grey)
               and all objects with a $>80$\% reliable redshift measurement (flags 2, 3, 4, and 9; dark grey). 
              }
         \label{nz}
\end{figure*}

The redshift distribution of the VUDS sample is shown in Figure \ref{nz}.
The N(z) is bimodal, with a high redshift component from $z\sim2$ to 
$z\sim6.5$, and a low redshift component mainly at $z < 1.5$.
The sample above $z=2$ is the sample of interest for the main science 
goals of VUDS; it is representing about 80\% of the total sample, with
currently more than 4500 objects with redshift measured with $z\geq2$,
and, extrapolating for the remaining data to be processed the VUDS 
sample, it will contain $\sim6000$ objects with $2<z<6.5$ in the end.  
VUDS is today the largest sample of galaxies with spectroscopic redshifts in any
of the redshift bins $2<z<3$, $3<z<4$, $4<z<4.7$ or $4.7<z<5.3$, as further discussed in
Section \ref{comparison}.

\begin{figure*}[h]
   \includegraphics[width=\hsize]{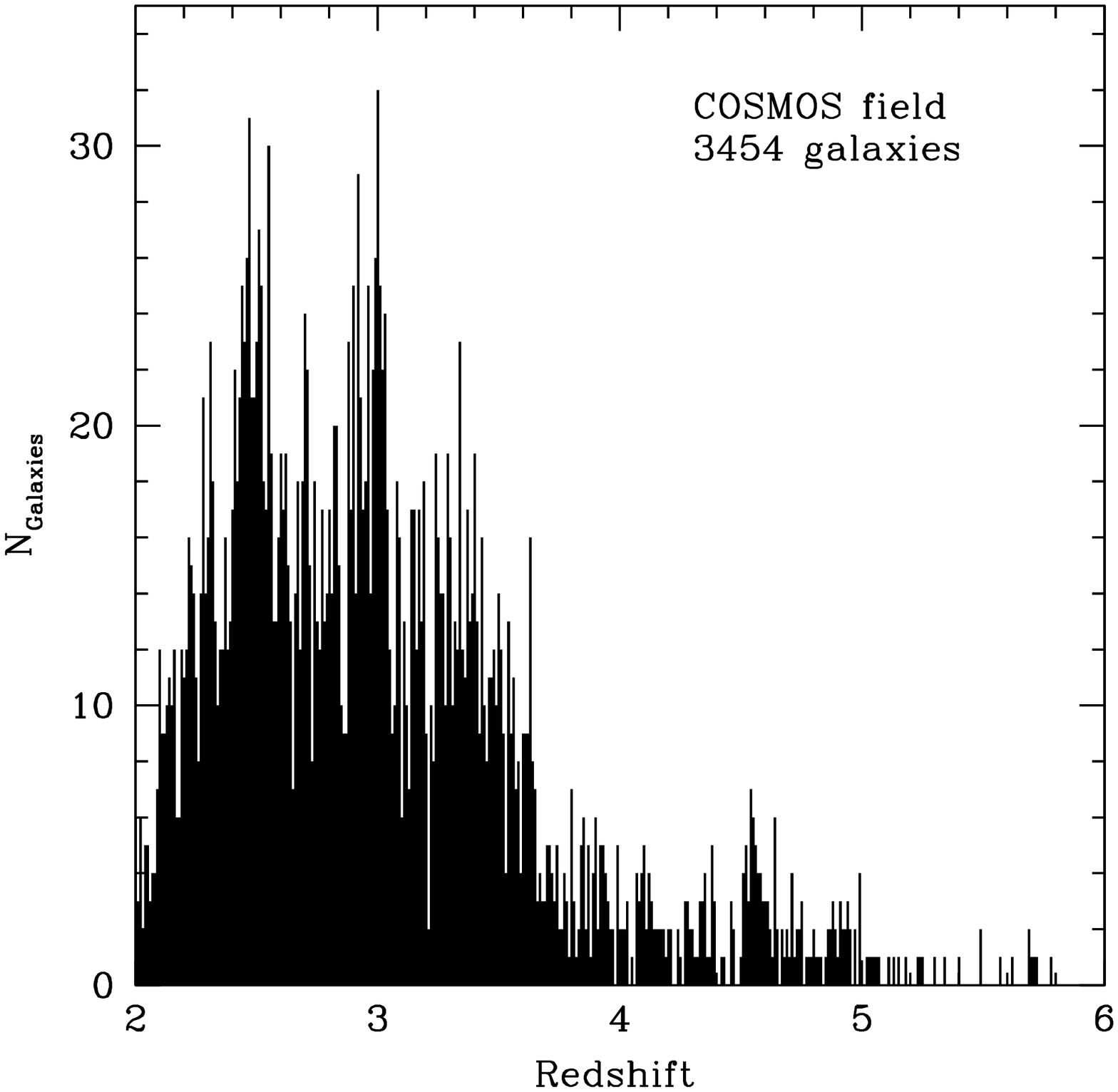}
      \caption{The current redshift distribution from 3495 galaxies already identified in the COSMOS field by the VUDS survey (all objects with a redshift 
               measurement are used). 
              }
         \label{nz_cosmos}
\end{figure*}

\begin{figure*}[h]
   \includegraphics[width=\hsize]{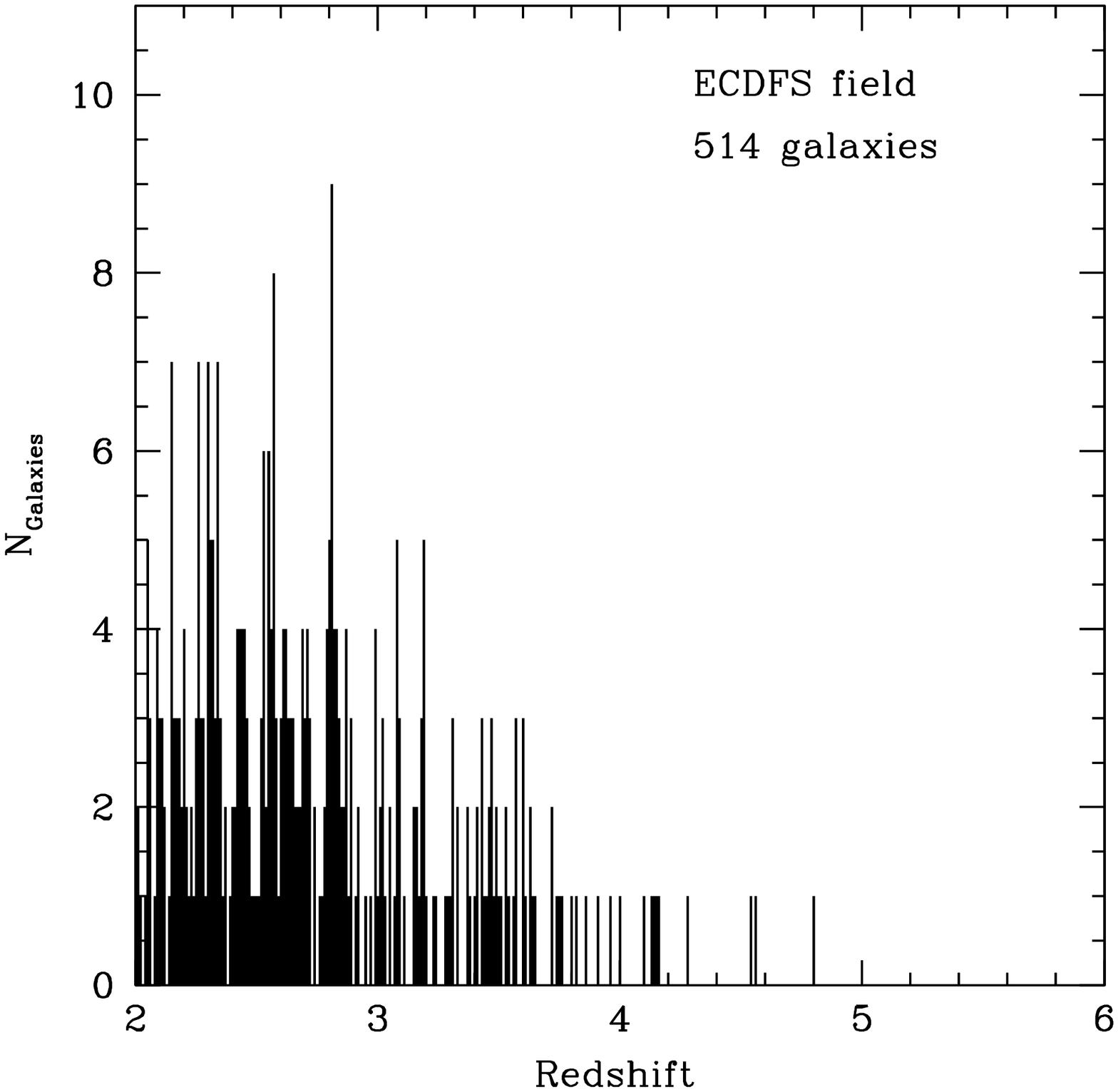}
      \caption{The current redshift distribution from 514 galaxies already identified in the ECDFS field by the VUDS survey (all objects with a redshift 
               measurement are used). 
              }
         \label{nz_ecdfs}
\end{figure*}

\begin{figure*}[h]
   \includegraphics[width=\hsize]{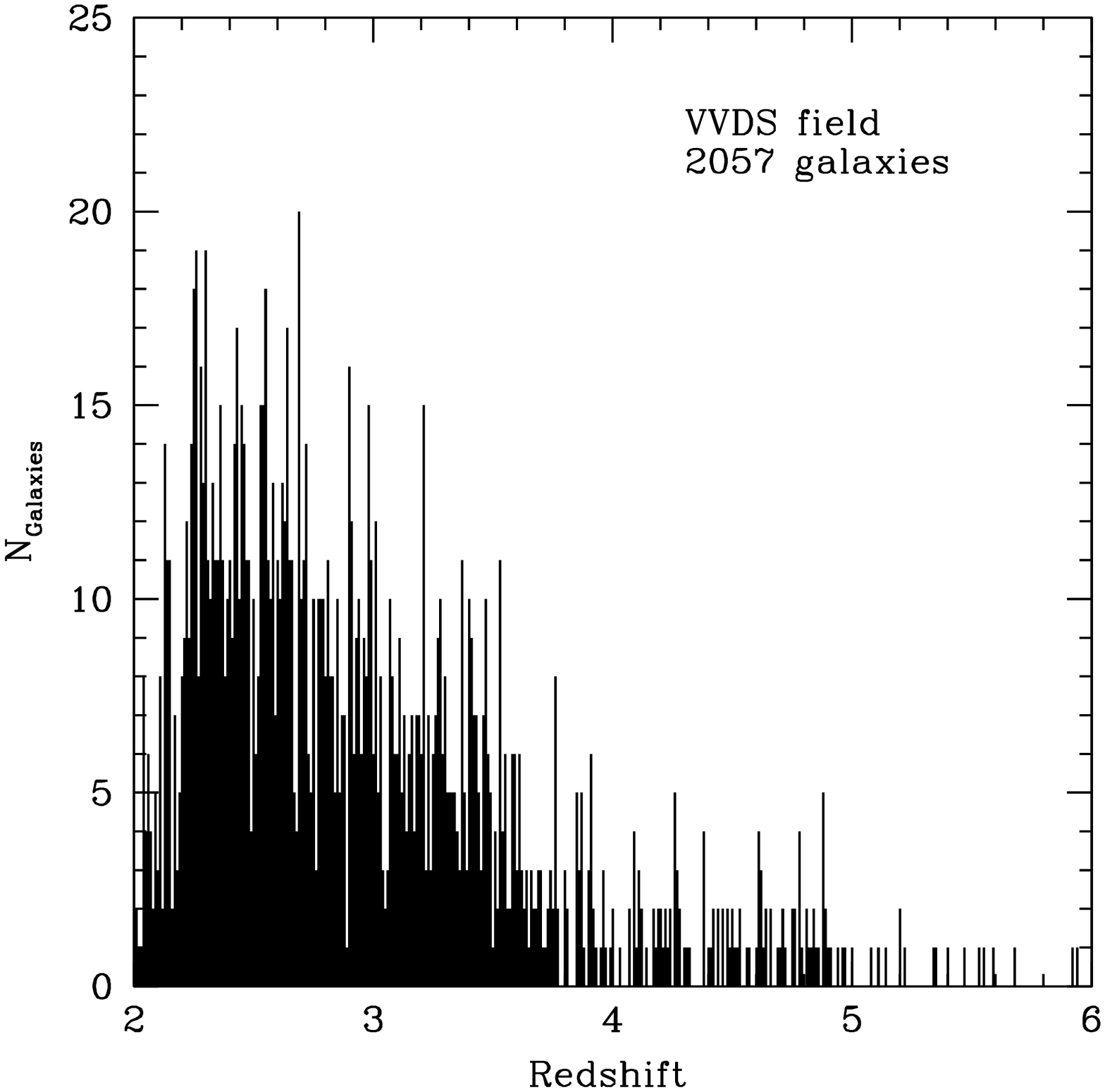}
      \caption{The current redshift distribution from 2059 galaxies already identified in the VVDS-02h field by the VUDS survey (all objects with a redshift 
               measurement are used). 
              }
         \label{nz_vvds}
\end{figure*}

The sample below $z\simeq2$ is made of several sub-samples:
(i) galaxies for which the second peak of the $z_{phot}$ PDF is at $z>2.4$ but
which turned out to be at the lower redshift indicated by the first peak,
(ii) galaxies for which the selection criteria for high redshift failed and which are rather at low redshifts,
(iii) galaxies with $i_{AB} \leq 25$ which have been used as mask fillers and for which the 
N(z) is expected to peak at $z\sim1.5$ (Le F\`evre et al. 2014),
and (iv) some serendipitous sources falling in the slits by chance.
The mean redshift for the sample above $z=2$ is $\bar z=3.0$.
About 10\% of the sample is above $z=4$, and the high redshift tail goes up beyond
redshift $z=6$, the highest reliable redshift so far being $z=6.5363$.
The current redshift distribution for $z\geq2$ in the COSMOS, ECDFS and VVDS-02h fields is shown in Figure \ref{nz_cosmos} to Figure \ref{nz_vvds}.
The redshift bin of $dz=0.01$ enables to show the strong clustering present at all redshifts probed by the survey.
Some of the densest peaks are remarkable examples of clustering in the early universe, as discussed in
Cucciati et al. (submitted) and Lemaux et al. (submitted). 


As our sample is $z_{phot}$ selected, it is interesting to check were the VUDS galaxies are distributed in several
standard color-color diagrams. Following other studies (e.g. Kurk et al. 2013, Le F\`evre et al. 2014),
this {\it a posteriori} analysis gives those
studies using color-color selection an indication of both the efficiency of the selection and the
contamination by galaxies at other redshifts than the redshift of interest in the color space selected.
We examine here the $gzK$ diagram used to select galaxies with $1.4 < z < 2.5$,
and the $ugr$ diagram for galaxies with $2.5 < z < 3.5$.
Other color-color diagrams like the $gri$ or $riz$ for $3.5 < z < 4.5$ and $4.5<z<5.5$, will be investigated in a forthcoming paper.

\begin{figure*}[h]
   \includegraphics[width=\hsize]{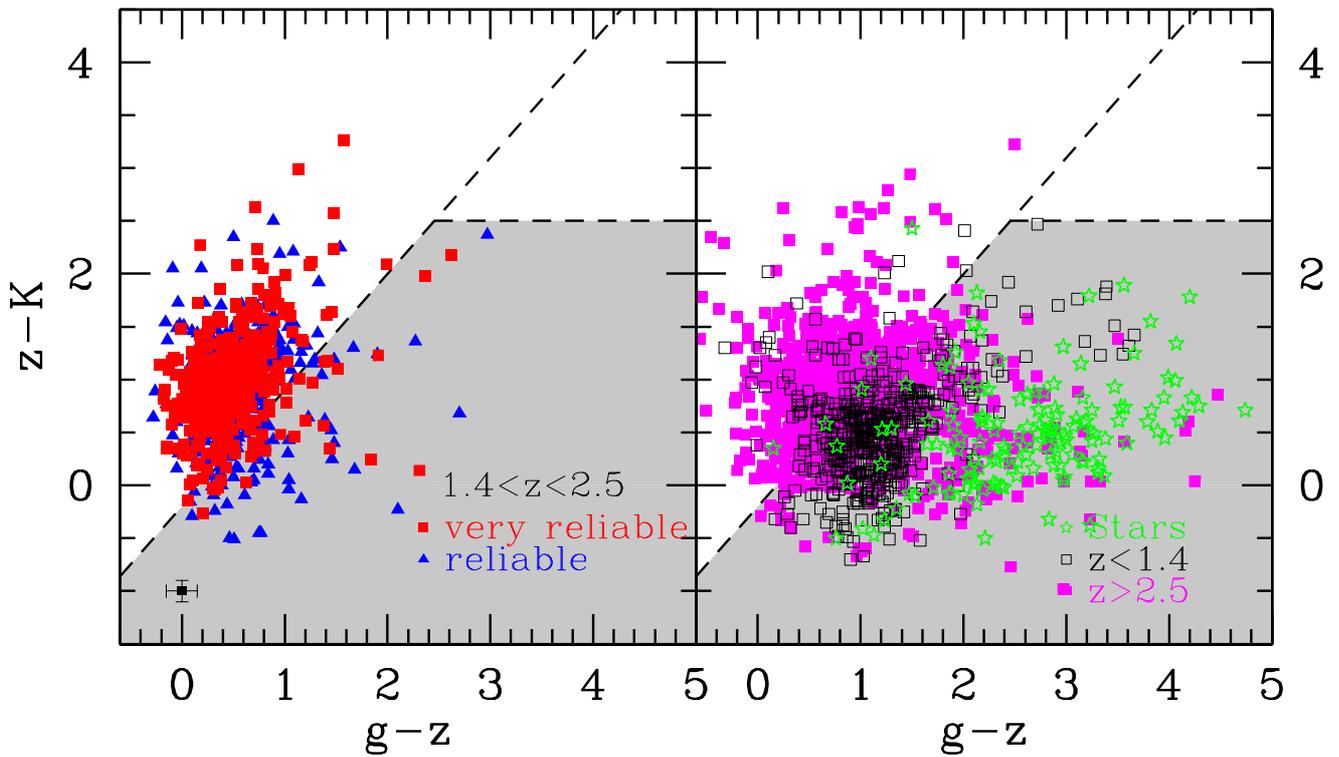}
      \caption{(left panel) (g-z) vs. (z-K) color-color diagram for VUDS galaxies with $1.4 < z < 2.5$ (flag 3+4: red squares, flag 2: blue triangles).
               (right panel) same for galaxies either with $z<1.4$ (open squares) or $z>2.5$ (magenta filled squares). Star are represented by green
               starred symbols. The $gzK$ selection would have selected 91\% of the VUDS galaxies with $1.4 < z < 2.5$, but with a high
               level of contamination as 58\% of galaxies in the selection area of the $gzK$ are outside the redshift range $1.4 < z < 2.5$,
               coming mostly from galaxies at $z>2.5$.  
              }
         \label{gzk}
\end{figure*}

The BzK selection is based on the identification of the Balmer and D4000 break crossing the $z$ band for $1.4 < z < 2.5$
(Daddi et al. 2004). We present the $gzK$ diagram in Figure \ref{gzk}, with the color-color area adjusted
to take into account the different wavelength coverage of the $g$ filter compared to the B filter (Bielby et al. 2012, Le F\`evre et al. 2014).
The $gzK$ criteria are efficient to select 92\% of the VUDS galaxies with $K \leq 24$. However, we note that the level of contamination
of a $gzK$ selected sample down to $K_{AB}=24$ would be quite high as 58\% of galaxies in the selection area of the $gzK$
diagram selecting the $1.4 < z < 2.5$ redshift range would be outside this range at $z<1.4$ or $z>2.5$ (right panel of Figure \ref{gzk});
at $K_{AB}=22$ we find this contamination to be lower, at the $\sim30$\% level. This trend
is similar to that found by Le F\`evre et al. (2014), who identified a magnitude-dependent contamination level.
As the VUDS data is selected with $z_{phot}>2.4$ the contamination level is further enhanced compared to the
pure $i-band$ magnitude selection of the VVDS.

\begin{figure*}[h]
   \includegraphics[width=\hsize]{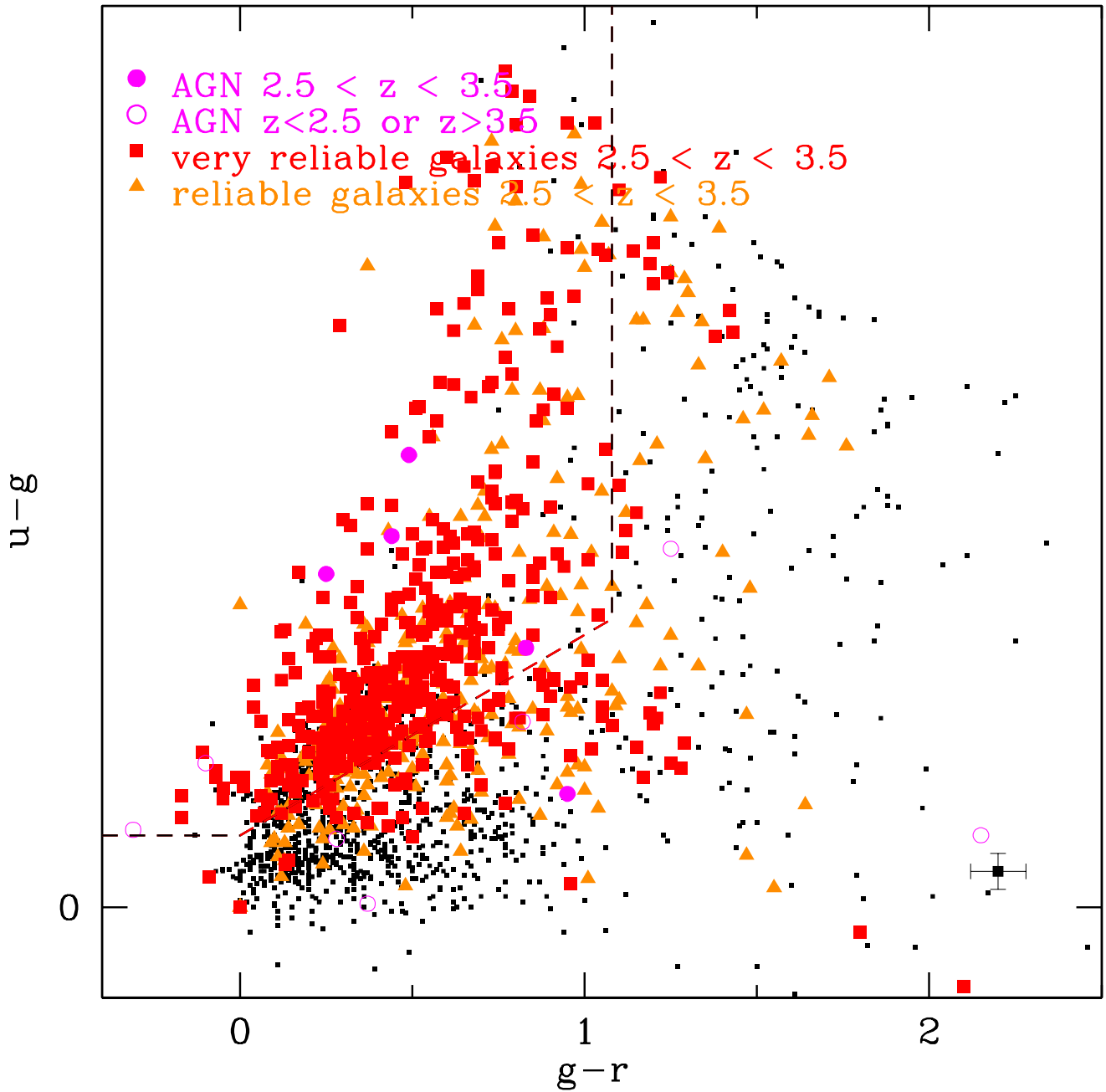}
      \caption{(left panel) (u-g) vs. (g-r) color-color diagram for VUDS galaxies with $2.5 < z < 3.5$ (flag 3+4: red squares, flag 2: blue triangles).
               (right panel) same for galaxies either with $z<2.5$ (open squares) or $z>3.5$ (magenta filled squares). 
               The $ugr$ selection would have selected 80\% of the VUDS galaxies with $2.5 < z < 3.5$, but with a high
               level of contamination as 40\% of galaxies in the selection area of the $ugr$ are outside the redshift range $2.5 < z < 3.5$.  
              }
         \label{ugr}
\end{figure*}

Going to higher redshifts, we show the distribution of VUDS galaxies with $2.5<z<3.5$ in the $ugr$  color-color diagram in Figure \ref{ugr}.
Here again a large fraction $\sim80$\% of galaxies appear in the expected locus of the color-color diagram. 
The contamination by galaxies at $z<2.5$ or $z>3.5$ is quite high at about 40\%.
We note that we did not attempt to optimize the redshift range on the basis of the exact shape of the photometric bands used
for the CFHTLS photometry used in this plot, but we rather elected to show the distribution of a large population.

This a posteriori color-color analysis shows that the VUDS sample is behaving as generaly expected for galaxies
at these redshifts. The photometric redshift selection allows to identify galaxies beyond the classical color-color
locus of the $gzK$ and $ugr$ diagrams. We also point out the strong contamination present in color-color selected samples from galaxies
at other redshifts outside the targeted redshift range.

\subsection{Absolute magnitudes, stellar masses and star formation rates}

\begin{figure*}[h]
   \includegraphics[width=\hsize]{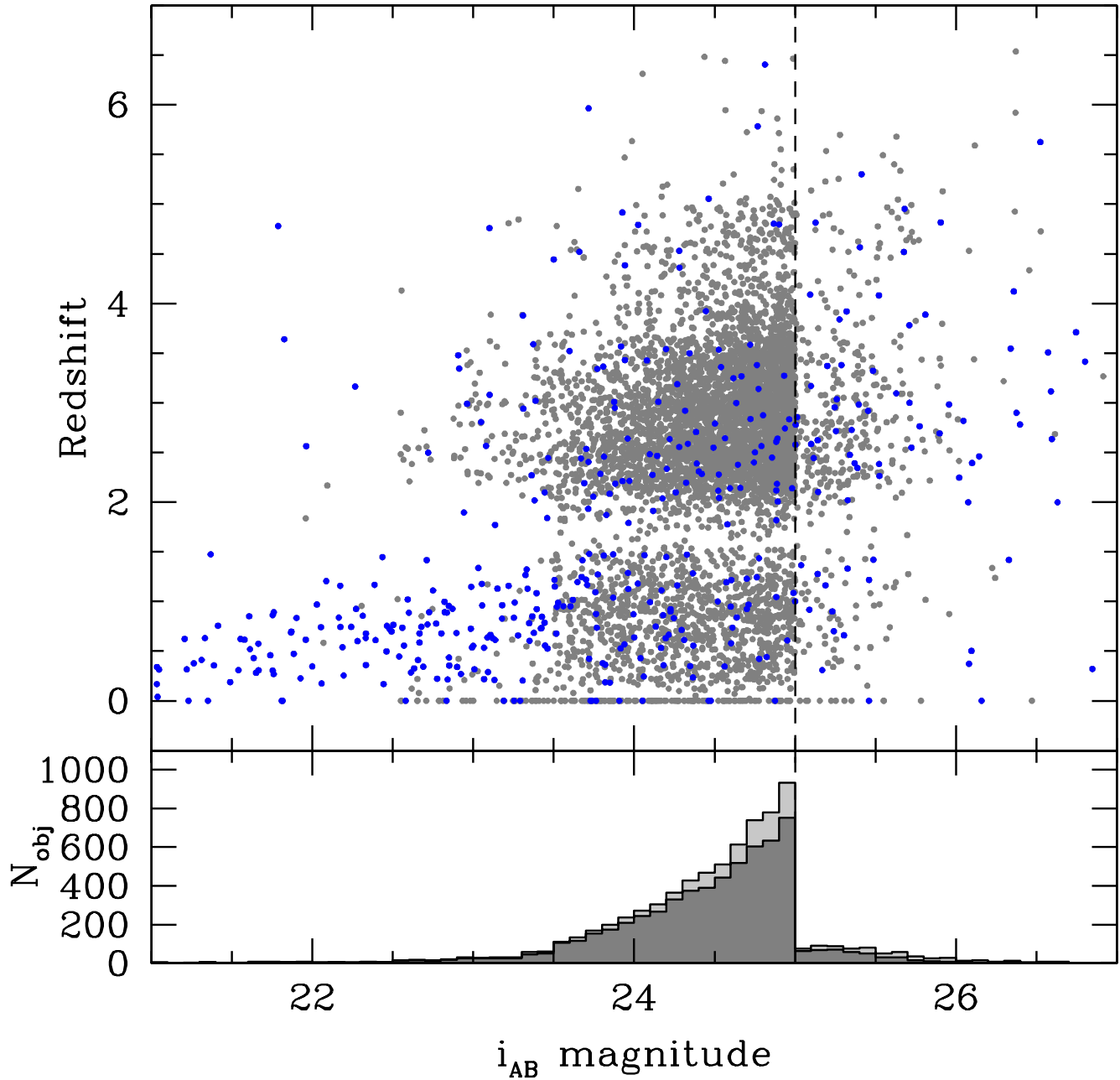}
      \caption{(top panel) $i_{AB}$ magnitude - redshift distribution for the full VUDS sample (grey points)
      and for serendipitous objects in the slits (blue points). (bottom panel) Distribution of $i_{AB}$ magnitudes in the VUDS survey for
      all objects observed (light grey), and for those with a redshift measurement (any non 0 flag, dark grey).
      The $i_{AB}=25$ imposed on the sample selected by photometric redshifts is indicated. The fainter objects are pre-selected
      from one of the other color or SED  criteria.
              }
         \label{magi_z}
\end{figure*}

The distribution of apparent magnitudes with redshift is shown in Figure \ref{magi_z}. In all $\sim90$\% of the distribution
is within $23 \leq i_{AB} \leq 25$, while $\sim10$\% have $25 \leq i_{AB} \leq 27$. 

\begin{figure*}[h]
   \includegraphics[width=\hsize]{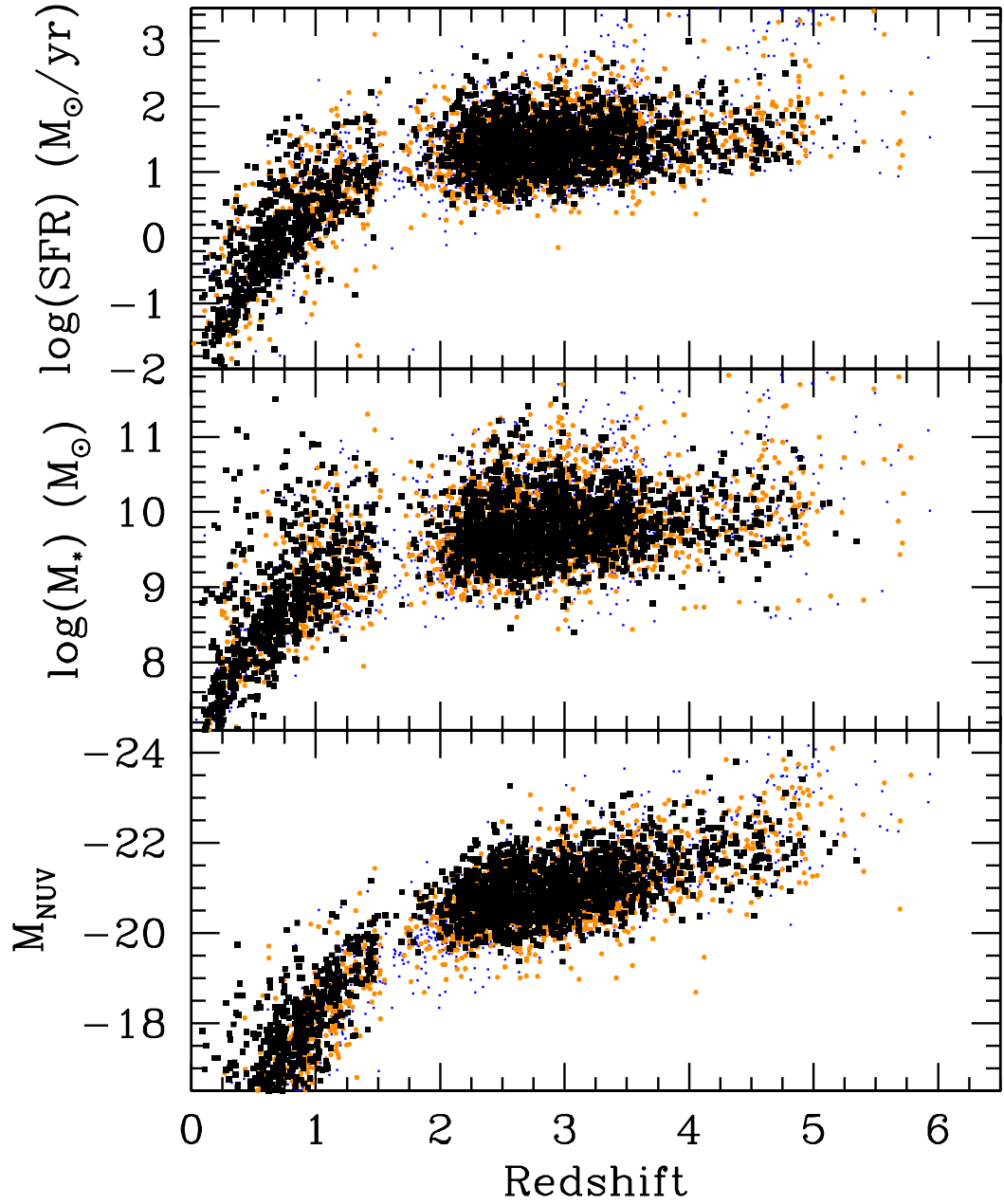}
      \caption{The distribution of absolute U-band magnitudes (bottom), stellar masses (center), and star formation rate (top). 
               Stellar masses and star formation rates are derived from template SED fitting at the spectroscopic redshift (see text). 
               Black squares are for galaxies with reliability flags 3 and 4, orange filled circles are for flag 2, and blue dots are for flag 1.
              }
         \label{u_mass_sfr}
\end{figure*}

The VUDS sample covers a large range of galaxy physical properties. Using the VUDS spectroscopic redshifts
we perform SED fitting on the multi-wavelength photometry using the code Le Phare (Ilbert et al. 2006), as described in Ilbert et al. (2013).
We summarize here a few key points, but we refer to the recent description of the fitting process
in Ilbert et al. (2013) for a detailed account and associated limitations.
Galaxy luminosities are transformed into stellar mass using the best fit synthetic template, from a list of templates
built from Bruzual \& Charlot (2003) stellar population synthesis models with 3 metallicities 
($Z=0.004$, $Z=0.008$, and solar $Z=0.02$), and exponentially declining and delayed SFR with 9 different $\tau$
values from 0.1 to 30 Gyr. We use a Calzetti (2000) extinction
law, and emission lines are added to the synthetic spectra as described in Ilbert et al. (2009).
The output of this fitting process includes among other parameters: absolute magnitudes integrated into standard bands,
stellar masses, star formation rates, and extinction. 
The distributions in absolute NUV (2300\AA ~rest) magnitude, stellar masses, and star formation rate of the sample are presented in Figure \ref{u_mass_sfr}.
At redshifts $z \sim 3-4$ the NUV absolute magnitude ranges from $NUV=-20.5$ to $NUV=-23$,
the stellar mass from $10^9$M$_{\sun}$ to $10^{11}$M$_{\sun}$, and the star formation rate from below $1$M$_{\sun}$/yr up to
several hundred M$_{\sun}$/yr. 

The VUDS survey therefore covers a large parameters space both in the
observed properties and the physical properties of the sample galaxies.

\subsection{Average spectral properties: stacked spectra}
\label{spec_prop}

Stacked spectra have been produced from previous surveys, like Shapley et al. (2003) for $z\sim3$,
or for B, V and R dropout galaxies (Vanzella et al. 2009). The large statistics and wavelength coverage
of VUDS offers the opportunity to produce composite spectra over a large redshift coverage. 
The average spectral properties of galaxies over the redshift range $2 \leq z_{spec} \leq 6.5$ are derived
from stacking VUDS spectra in different redshift bins. 
For each redshift bin, the average spectra are produced using the \texttt{odcombine} task in IRAF, 
averaging spectra after scaling to the same median continuum value, i.e. luminosity weighted,
and weighting spectra to their mean flux in the same rest-frame wavelength range.
Average spectra using sigma clipping, removing at each wavelength pixel those 
with a value $1.5-3$ times the $1\sigma$ value, have been compared to the straight weighted average 
described above, and very little difference have been observed when the number of spectra is large ($>50$).
For smaller samples, sigma clipping helps improving the S/N of the stacks by removing e.g. the
left-over signatures of the sky subtraction process.
  
\begin{figure*}[h]
   \includegraphics[width=\hsize]{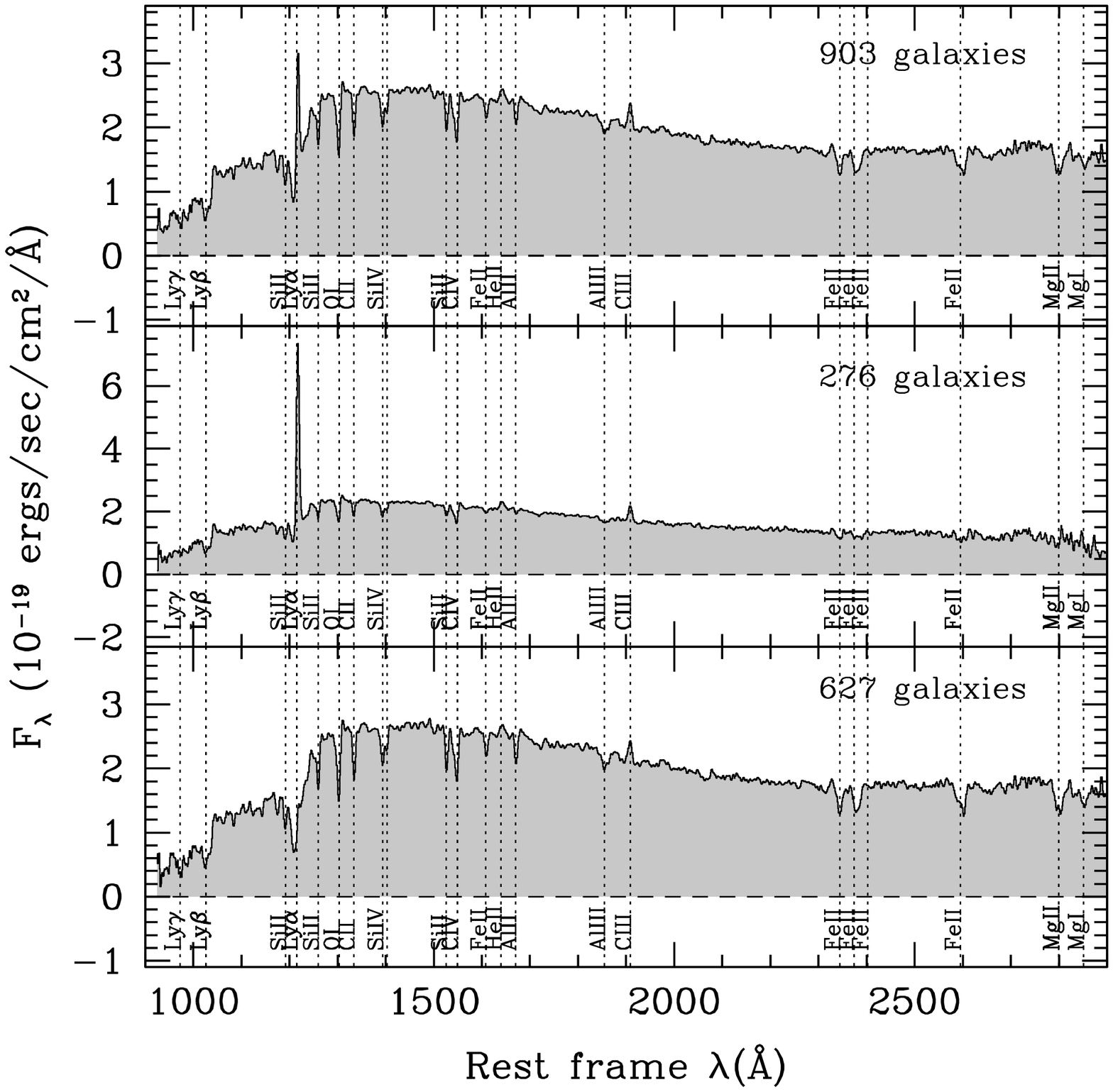}
       \caption{Average rest-frame spectra ($F_{\lambda}$) of galaxies with flags 3 and 4 in VUDS with $2 \leq z \leq 3$:
	{\it (Top)}: stack of all galaxy spectra;
	{\it (Middle)}: stack of galaxies with Ly-$\alpha$ in emission;
	{\it (Bottom)}: stack of galaxies with Ly-$\alpha$ in absorption.
              }
         \label{avg_spec_z2.5}
\end{figure*}

\begin{figure*}[h]
   \includegraphics[width=\hsize]{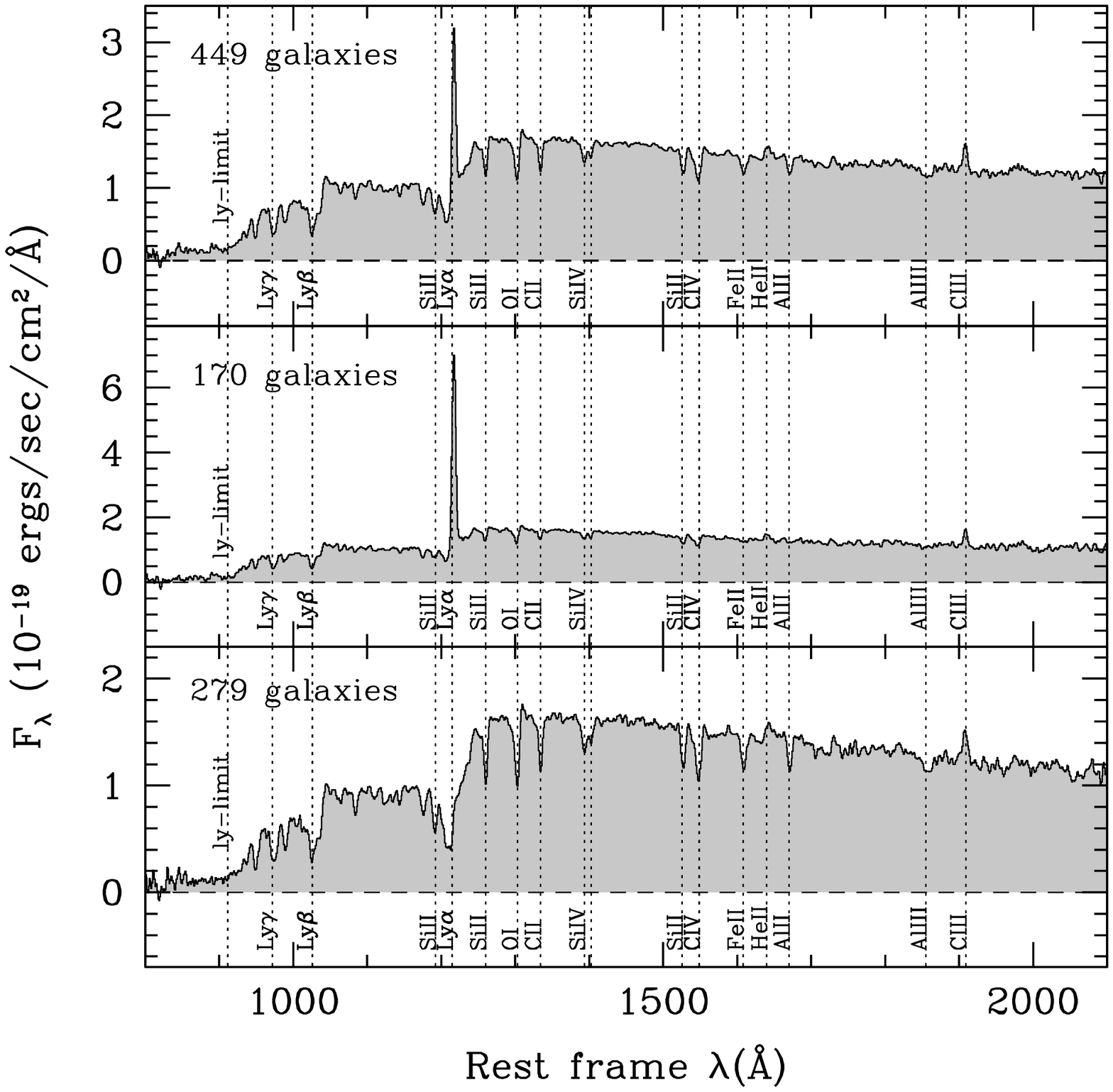}
       \caption{Average rest-frame spectra ($F_{\lambda}$) of galaxies with flags 3 and 4 in VUDS with $3 \leq z \leq 4$:
	{\it (Top)}: stack of all galaxy spectra;
	{\it (Middle)}: stack of galaxies with Ly-$\alpha$ in emission;
	{\it (Bottom)}: stack of galaxies with Ly-$\alpha$ in absorption.
              }
         \label{avg_spec_z3.5}
\end{figure*}

\begin{figure*}[h]
   \includegraphics[width=\hsize]{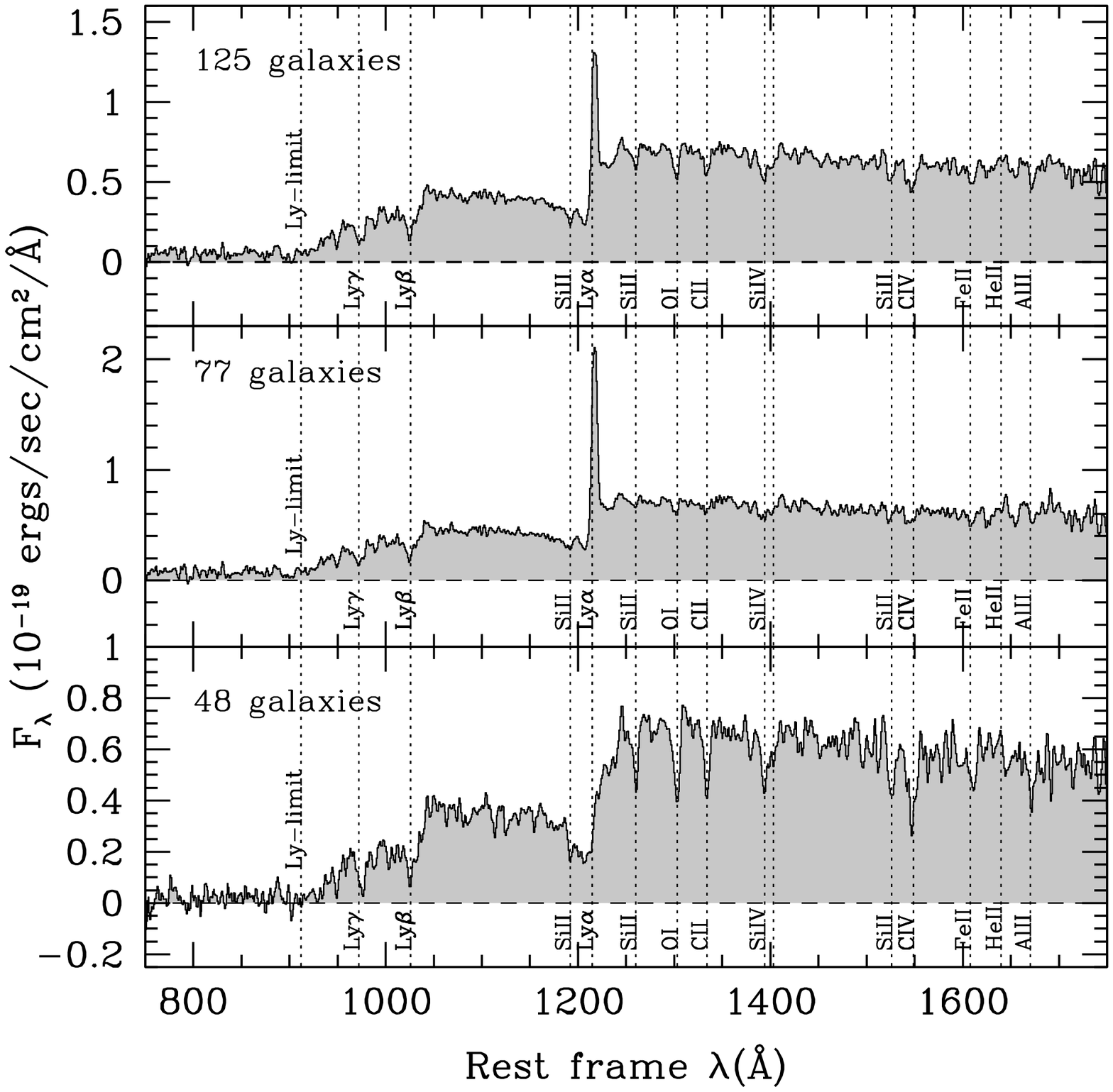}
       \caption{Average rest-frame spectra ($F_{\lambda}$) of galaxies with flags 2, 3 and 4 in VUDS with $4 \leq z \leq 4.7$:
	{\it (Top)}: stack of all galaxy spectra;
	{\it (Middle)}: stack of galaxies with Ly-$\alpha$ in emission;
	{\it (Bottom)}: stack of galaxies with Ly-$\alpha$ in absorption.
              }
         \label{avg_spec_z4.4}
\end{figure*}

\begin{figure*}[h]
   \includegraphics[width=\hsize]{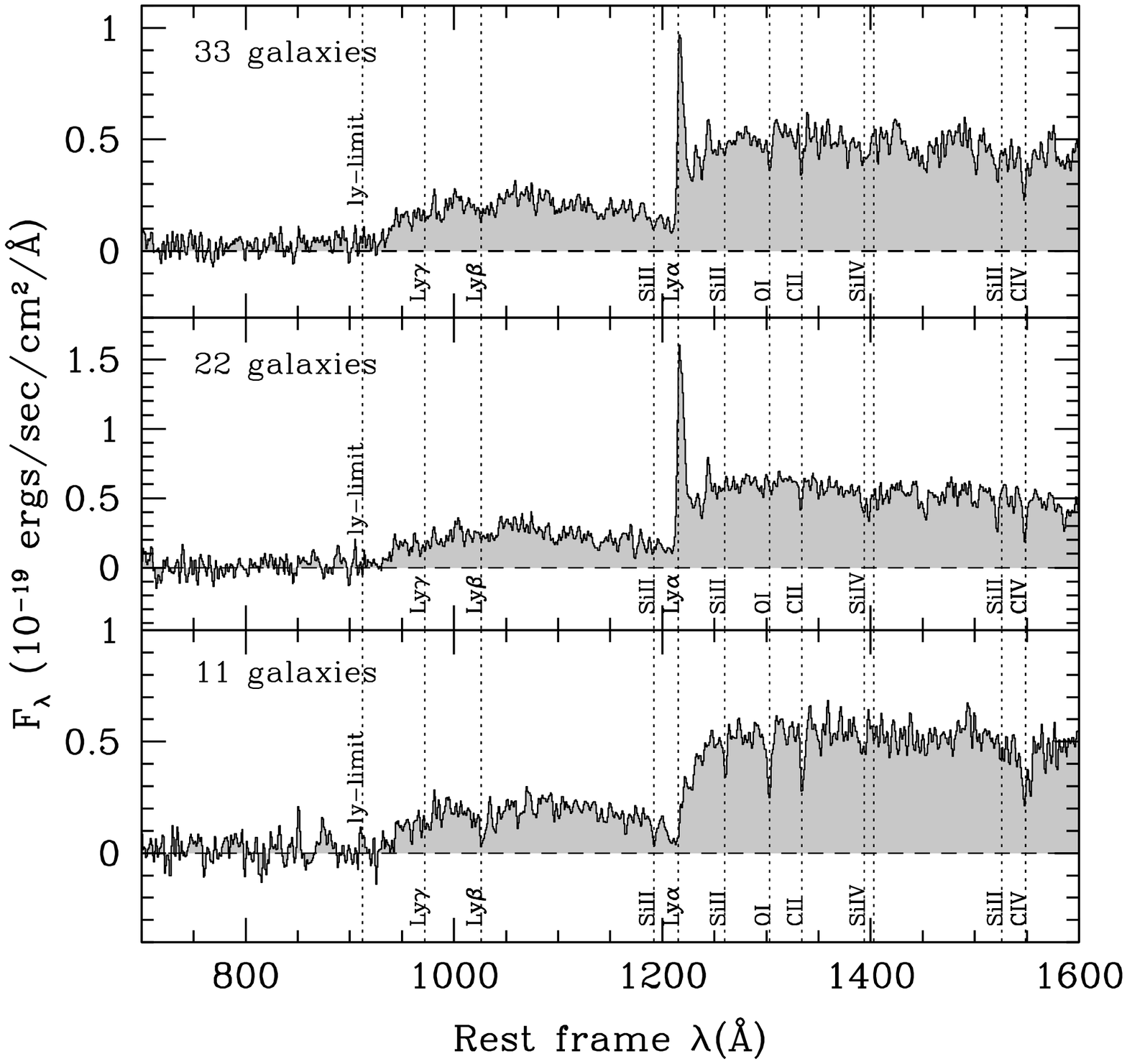}
       \caption{ 
Average rest-frame spectra ($F_{\lambda}$) of galaxies with flags 2, 3 and 4 in VUDS with $4.7 \leq z \leq 5.3$:
	{\it (Top)}: stack of all galaxy spectra;
	{\it (Middle)}: stack of galaxies with Ly-$\alpha$ in emission;
	{\it (Bottom)}: stack of galaxies with Ly-$\alpha$ in absorption.
              }
         \label{avg_spec_z5}
\end{figure*}

\begin{figure*}[h]
   \includegraphics[width=\hsize]{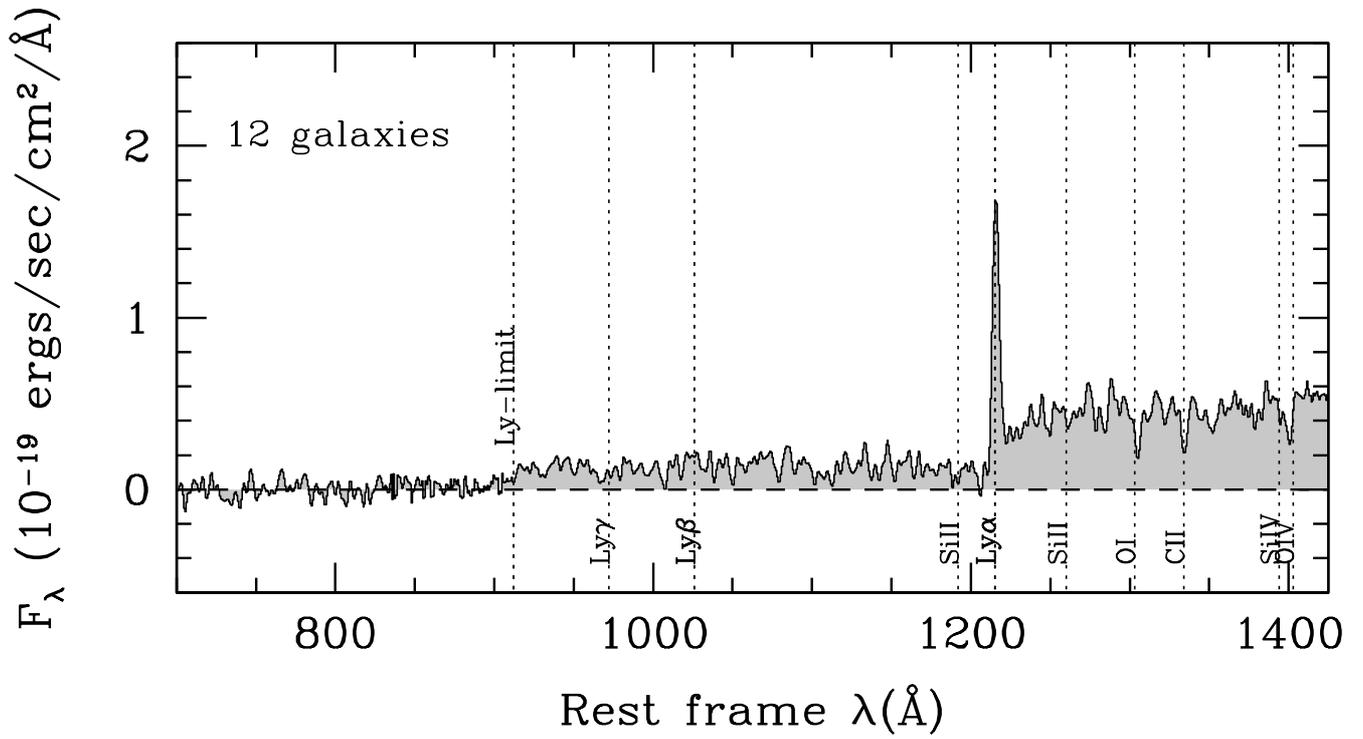}
       \caption{Average rest-frame spectra ($F_{\lambda}$) of galaxies with flags 2, 3 and 4 in VUDS with $5.3 \leq z \leq 6.5$.
              }
         \label{avg_spec_z6}
\end{figure*}

The average spectra of all galaxies for several redshift bins are shown in Figures \ref{avg_spec_z2.5}
to \ref{avg_spec_z6}, together with the average spectra of those with and without Ly-$\alpha$ emission.
We discuss some of the key features below.

We find that the fraction of galaxies with any trace of Ly$\alpha$ emission 
is strongly changing with redshift with $30.6\pm1.8$\% in $2 \leq z \leq 3$, 
$38.3\pm2.9$\% in $3 \leq z \leq 4$,  
$61.6\pm7.0$\% in $4 \leq z \leq 4.7$, 
and $66.6\pm14.2$\% in $4.7 \leq z \leq 5.3$.  
The detailed properties of the Ly$\alpha$ emitting fraction and implications are examined in details in the accompaning paper
by Cassata et al. (submitted).

Besides Ly$\alpha$ a number of spectral features are noteworthy, as indicated in Table \ref{spectral_lines}. 
The main absorption features redder than Ly$\alpha$ include the SiII$\lambda$1260, OI$\lambda$1303, CII$\lambda$1334, 
SiIV$\lambda$1394/1403, SiII$\lambda$1527, CIV$\lambda$1549 lines.
In emission, weak CIII]$\lambda$1909 is quite common, and one can identify S-shape absorption-emission features for e.g. 
SiII$\lambda$1260, OI$\lambda$1303 or CIV$\lambda$1549, indicative of strong outflows. The HeII$\lambda$1640 emission is ubiquitous, as seen on all spectra where
this line is in the wavelength range (Figures \ref{avg_spec_z2.5} to \ref{avg_spec_z4.4}). This line, when several hundreds
of km/s in width, indicates the presence of strong winds around Wolf-Rayet stars. If narrow it may indicate some other processes
like the presence of a population of low-metallicity stars with properties akin to Population III (Cassata et al. 2013).

Below the Ly$\alpha$ line, the main features identified in galaxies at these redshifts are
SiII$\lambda$1192, and then the Ly$\beta$ and Ly$\gamma$ lines, followed by the Lyman-limit producing
a continuum break at 912\AA. 
It is interesting to note that the flux below 912\AA ~in our stacked spectra is not zero as would
be expected if the lyman continuum escape fraction was 0\%. 
We observe a significant detection of flux below the Lyman-limit in the stacked spectra  of galaxies at $3<z<4$, $4<z<4.7$ and $4.7<z<5.3$.
The ratio of continuum flux density in the [1400,1500]\AA ~range over the flux in [800,900]\AA ~range below the 912\AA ~Lyman-limit is 
$f_{1500}/f_{900}=32\pm3$, $39\pm5$, and $33\pm10$,
in these three redshift bins respectively. Although these measurements are based on a much larger sample
covering a larger redshift range, these values are comparable to the observed value of 
$58 \pm 18$ reported by Shapley et al. (2006) at $z\sim3$. 
While it is tempting to interpret this in terms of the Lyman-continuum escape fraction,
we note that at the faint magnitudes we are observing, the flux observed below 912\AA ~in stacked spectra could also be coming in part from 
objects more or less contaminated by lower redshift interlopers producing observed flux below that of the Lyman-limit at the 
rest-frame of the distant source (e.g. Vanzella et al. 2010). The {\it observed} non-zero flux below 912\AA ~has consequences on the
selection of high redshift galaxies based on the Lyman-break technique, as expected colors of high redshift galaxies are 
affected by apriori hypotheses, e.g. on the $f_{1500}/f_{900}$ ratio. This and the corrected Lyman escape fraction  will be analyzed in forthcoming papers. 

Our stacked spectra beautifully show the evolution of the 'stair-case' pattern of IGM absorption
as a function of redshift produced by continuum blanketing from the Lyman series of galaxies along the line of sight  in the 
volume probed (Madau 1995). 
The comparison of the observed mean and distribution of IGM absorption properties as a function of redshift in VUDS to the
models of Madau (1995) and Meiksin (2010) will be  discussed in Thomas et al. (in preparation). 

A detailed analysis of the spectral properties of stacked spectra and individual galaxies in the VUDS 
sample will be presented in forthcoming papers.

\section{Comparison with other spectroscopic surveys at high redshifts}
\label{comparison}

Large spectroscopic surveys at $z>2$ published in the literature are understandably 
relatively limited compared to surveys at $z<1.5$ because of the faintness of the sources
and requirements on deep imaging to perform pre-selection.

\begin{figure*}[h]
   \includegraphics[width=18cm]{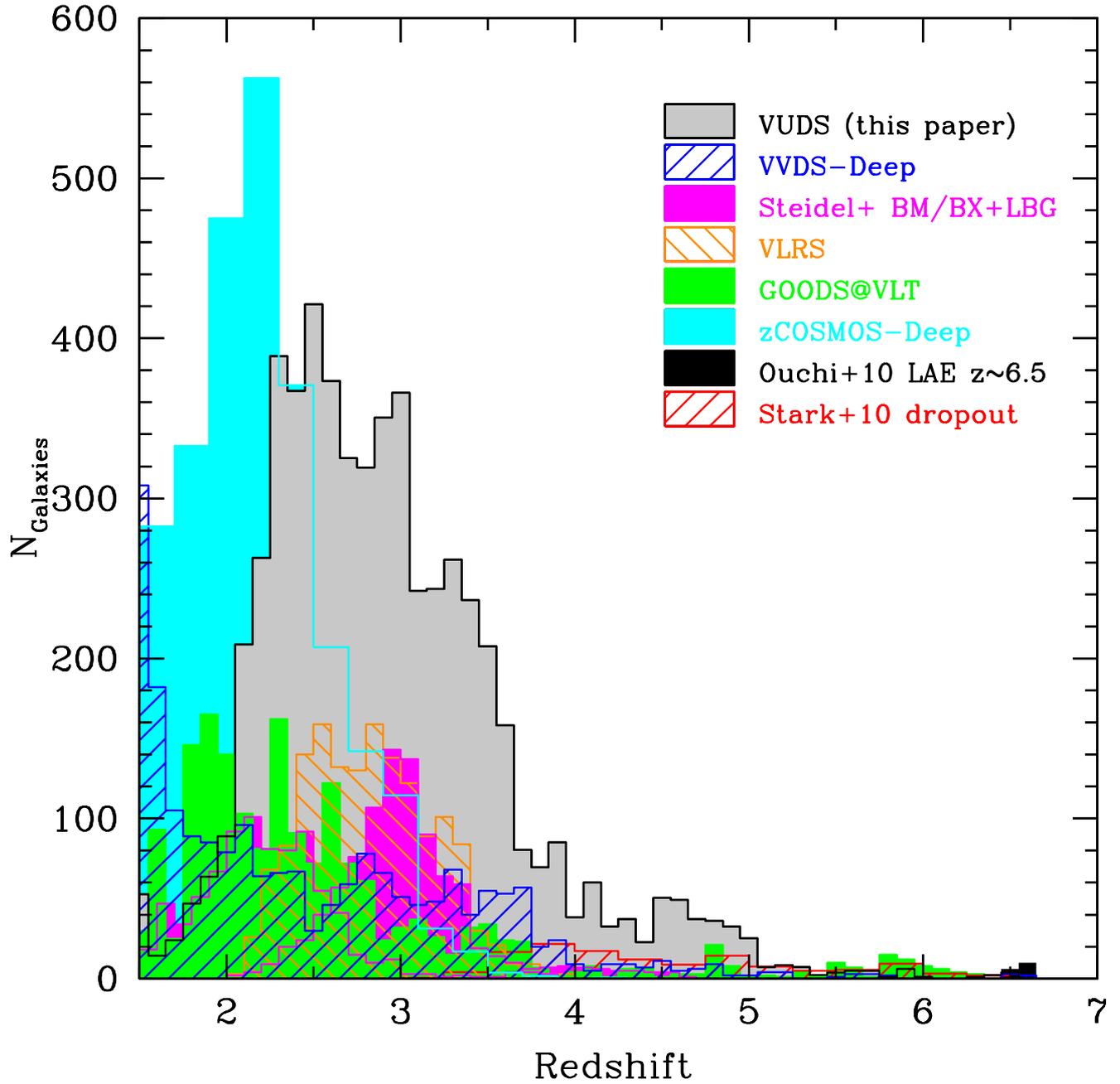}
       \caption{Comparison of the redshift distribution of the VUDS survey with other published spectroscopic surveys at $z>2$ 
                (as listed in Table \ref{comp_surveys}). The VUDS
                counts from existing measurements 
                have been scaled by 1.2 to account for the data not yet processed at the time of this writing.
              }
         \label{nz_surveys}
\end{figure*}

We compile a list, likely not exhaustive, of the main surveys in the redshift range $2<z<7$ in Table \ref{comp_surveys},
and we show the corresponding redshift distribution of these surveys in Figure \ref{nz_surveys}.
The largest numbers of galaxies can be found in the range $2<z<3$ with the LBG-selected surveys of Steidel et al. (1999, 2003, 2004), the 
zCOSMOS-Deep survey (Lilly et al. 2007, Lilly et al. in prep.), the VLRS (Bielby et al. 2013), and the VVDS-Deep/Ultra-Deep (Le F\`evre et al. 2013).
In the redshift range $2.5<z<3.5$, these surveys accumulate $\sim2700$ galaxies with spectroscopic redshifts.
The VUDS survey contributes $\sim2800$ spectroscopic redshifts, therefore as much as all these other surveys combined, but 
in a single  well-controled survey.

Beyond redshift $z\sim3.5$ both the number of spectroscopic surveys and the number of galaxies with measured spectroscopic
redshifts dramatically fall down. Spectroscopic campains on the GOODS-South area with VIMOS and FORS2 on the VLT have 
produced about 114 galaxies with redshifts $z\sim3-6$. The Keck-DEIMOS surveys of Stark et al. 2010  contribute more than 
300 galaxies in the range $3<z<6.5$. 
VUDS brings more than 800 new galaxies with spectroscopic redshifts in the range $3.5<z<5$, 
or about twice the number of all other surveys combined as can be seen in Figure \ref{nz_surveys}. 

At the highest redshifts $z>5$, we enter the game of single redshift confirmations following-up on galaxies
identified as {\it dropout}, i.e. with a sharp change in flux between two adjacent photometric bands (e.g.
Stark et al. 2010, Curtis-Lake et al. 2012), or following narrow-band 
Lyman Alpha Emitter candidates (e.g. Ouchi et al. 2010). These different
surveys are bringing several tens of spectroscopic identifications beyond $z=5$. 
VUDS contributes about 60 new sources with spectroscopic redshifts $z>5$. This sample is
being consolidated, particularly at $z>6$, and will be the subject of future studies. 

\begin{figure*}[h]
   \includegraphics[width=18cm]{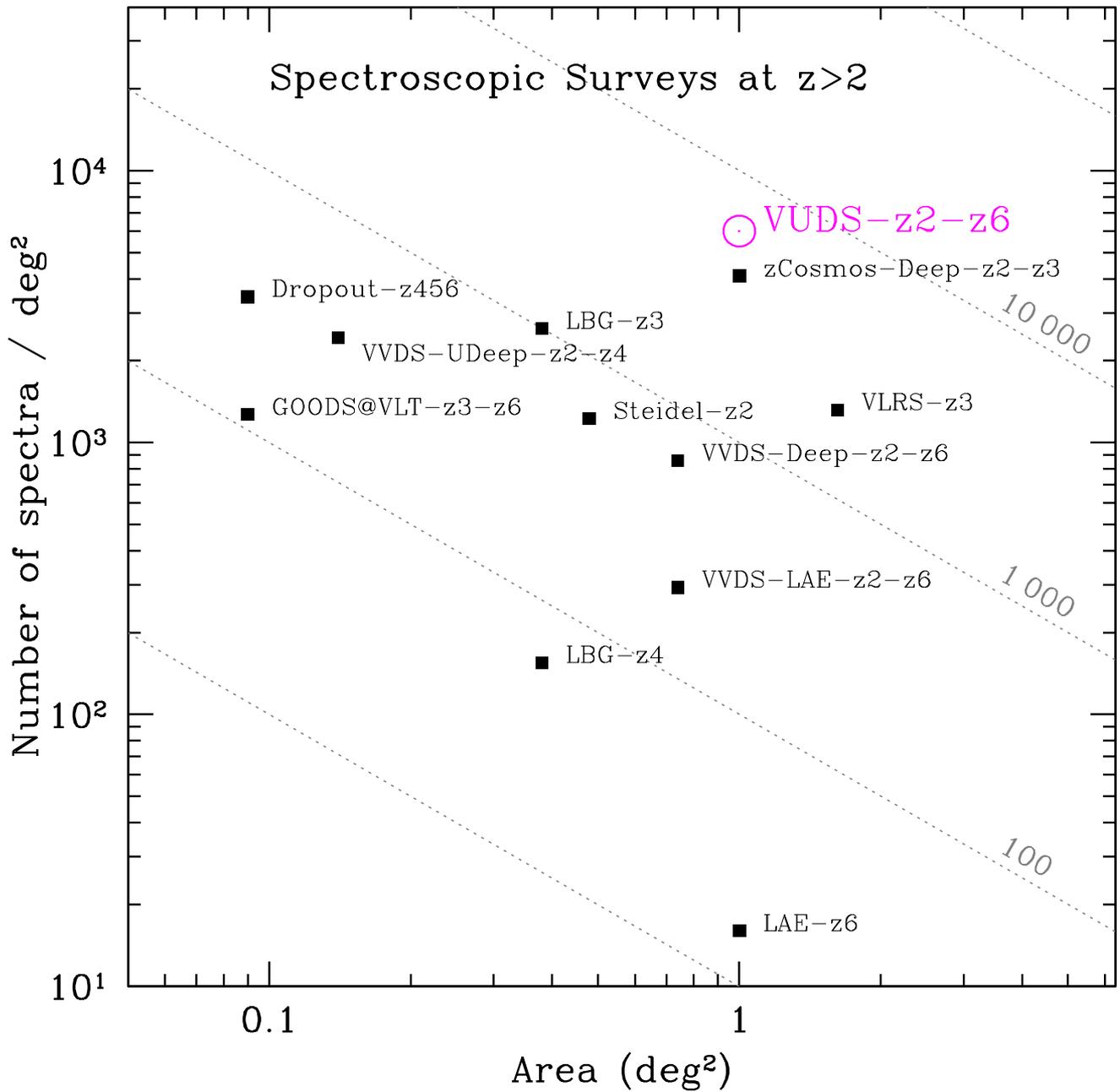}
       \caption{Comparison of the number density of spectra in the VUDS survey with other published spectroscopic surveys at $z>2$  
                (as listed in Table \ref{comp_surveys}). The VUDS
                counts have been scaled by 1.2 to account for the data not yet processed at the time of this writing. 
                Dashed lines are for surveys with a total number of spectra as indicated.
              }
         \label{surveys_dens}
\end{figure*}

An important element to keep in perspective is the area which has been sampled by a survey.
As discussed in Moster et al. (2011), the fluctuations in number density of objects observed
in deep surveys result from the cosmic variance in the distribution of galaxies in
large scale structures. Surveys with one square degree at $z\sim3$ will be subject
to an uncertainty of about 10\% from cosmic variance, in addition to the uncertainties
related to the number of objects in a survey. For fields of the GOODS size ($\sim 150$arcmin$^2$), 
the cosmic variance at $z\sim3$ is expected to be in excess of 50\% for galaxies with stellar masses $M_*>10^{10.5}$M$_{\sun}$.  
We compare in Figure \ref{surveys_dens} the number of spectra per square degree vs. the area for different surveys, 
a high redshift version of that presented in Baldry et al. (2010) and updated by
Le F\`evre et al. (2013).  While at $z\sim3$, several surveys
have surveyed about 1 square degree (Steidel et al. 2003, Lilly et al. 2007, Bieby et al. 2013, or Le F\`evre et al. 2013), 
the highest redshift surveys have surveyed only about 0.1 deg$^2$ (e.g. Vanzella et al. 2009, Stark et al. 2010). 
In this context, VUDS has selected galaxies at any redshift $z>2$ in three different fields for a total of $\sim1$deg$^2$.

\section{Summary}
\label{summary}

The VIMOS Ultra Deep Survey (VUDS) is a deep spectroscopic redshift survey aiming to study the early phases of galaxy assembly at $2<z<6.5$ from a sample
of $\sim10\,000$  galaxies observed with the VIMOS multi-slit spectrograph at the ESO-VLT. 
The survey target selection is based on photometric redshifts derived from extensive multi-wavelength data, combined to color
and color-color selection as well as magnitude-selection. Most of the sample is limited down to $i_{AB}=25$, but galaxies are observed
as faint as  $i_{AB}=27$. 

The combination of a wide wavelength coverage from 3650\AA ~to 9350\AA, and exposure times of $\simeq14$h, lead to a
spectroscopic success rate in redshift measurement of about 91\%  (74\% for spectroscopic reliability flags 2 to 9) down to $i_{AB}=25$.
The comparison of photometric redshifts to the VUDS spectroscopic redshifts shows that the VUDS strategy minimizes the
loss of galaxy populations compared to more restrictive selection criteria. 

We report on the general properties of the sample based on $\sim80$\% of the data which has already been processed.
The redshift distribution of the current sample at $z \geq 2$  peaks at a mean $z=3$, and extends beyond $z=6$. 
A secondary sample at $z<2$ is the result of the selection function and provides interesting very low intrinsic luminosity galaxies.
The average spectral properties of galaxies with $z>2$ are discussed based on high S/N stacks of VUDS spectra in several increasing redshift bins.
Galaxies with and without Ly$\alpha$ in emission are found at any redshift, but it is found that the fraction 
of galaxies with Ly$\alpha$ in emission increases with redshift. This is quantified in
the accompaning paper by Cassata et al. (submitted). Using stacked spectra we find that there is observed flux below the 912\AA ~Lyman limit
in all the redshift ranges explored, the origin of this is being investigated and will be the subject of future papers.

Following an early measurement of the merger rate at $z\sim3$ (Tasca et al. 2013), several papers presenting results
from the VUDS are submitted together with this paper (Cucciati et al. submitted, Cassata et al. submitted, Lemaux et al. submitted, 
Amorin et al. submitted). A number of other analyses are in progress.

VUDS is the first deep spectroscopic survey covering such a large redshift range with such a large sample of galaxies
with confirmed spectroscopic redshifts. It is ideally suited for detailed studies of the galaxy population at early 
times $2<z<6$. When the full data set will be completed, it is foreseen to make future VUDS data releases publicly available.

\begin{acknowledgements}
We thank ESO staff for their continuous support for the VUDS survey, particularly the Paranal staff conducting the observations and Marina Rejkuba and the ESO user support group in Garching.
This work is supported by funding from the European Research Council Advanced Grant ERC-2010-AdG-268107-EARLY and by INAF Grants PRIN 2010, PRIN 2012 and PICS 2013.
AC, OC, MT and VS acknowledge the grant MIUR PRIN 2010--2011.
DM gratefully acknowledges LAM hospitality during the initial phases of the project.
This work is based on data products made available at the CESAM data center, Laboratoire d'Astrophysique de Marseille.
This work partly uses observations obtained with MegaPrime/MegaCam, a joint project of CFHT and CEA/DAPNIA, at the Canada-France-Hawaii Telescope (CFHT) which is operated by the National Research Council (NRC) of Canada, the Institut National des Sciences de l'Univers of the Centre National de la Recherche Scientifique (CNRS) of France, and the University of Hawaii. This work is based in part on data products produced at TERAPIX and the Canadian Astronomy Data Centre as part of the Canada-France-Hawaii Telescope Legacy Survey, a collaborative project of NRC and CNRS.
\end{acknowledgements}

\clearpage


\begin{table*}[h]
\begin{center}
      \caption[]{VUDS fields}
      \[
        \begin{array}{lllccrr}
           \hline \hline
            \noalign{\smallskip}
            Field      &  \alpha_{2000} & \delta_{2000} & b  & l   &  $Area$ & Depth \\
            \noalign{\smallskip}
            \hline
            \noalign{\smallskip}
            $COSMOS$   &  10h00m04.0s & +02\deg12\arcmin40\arcsec &  42.1 & 236.8 & 1800 $ arcmin$^2 & i_{AB}\simeq25 \\
            $ECDFS$    &  03h32m28.0s & -27\deg48\arcmin30\arcsec & -54.0 & 223.5 &  675 $ arcmin$^2 & i_{AB}\simeq25 \\
            $VVDS-02h$ &  02h26m00.0s & -04\deg30\arcmin00\arcsec & -57.5 & 172.0 & 1125 $ arcmin$^2 & i_{AB}\simeq25 \\
            \noalign{\smallskip}
            \hline
         \end{array}
      \]
\label{fields}
\end{center}
\end{table*}

   \begin{table*}[h]
      \caption[]{VUDS spectroscopic target selection criteria}
      \[
         \begin{array}{llcc}
           \hline \hline
            \noalign{\smallskip}
            Criterion      &  $Value$ & $Limiting magnitude$ & $Fraction of targets$  \\ 
            \noalign{\smallskip}
            \hline
            \noalign{\smallskip}
            $Photometric redshift, 1st PDF peak$ & z_{phot} + 1\sigma \geq 2.4  &  22.5 \leq i_{AB} \leq 25 & 58.3\% \\
            $Photometric redshift, 2nd PDF peak$ & z_{phot} \geq 2.4            &  22.5 \leq i_{AB} \leq 25 & 19.3\% \\
            $LBG ugr$                            & 2.7<z<3.5                    &  i_{AB} \leq 27           & 1.9\%  \\
            $LBG gri$                            & 3.5<z<4.5                    &  i_{AB} \leq 27           & 1.9\%  \\
            $LBG riz$                            & 4.5<z<5.5                    &  i_{AB} \leq 27           & 3.6\%  \\
            $Break SED-based$                    &  -                           &  K_{AB} \leq 24           & 3.5\% \\
            $Magnitude-selected$                 &  -                           &  23\leq i_{AB}\leq25      & 11.5\% \\
\noalign{\smallskip}
            \hline
         \end{array}
     \]
\label{selection}
   \end{table*}

   \begin{table*}[h]
      \caption[]{VUDS: observed VIMOS fields }
      \[
         \begin{array}{p{0.2\linewidth}llc}
           \hline \hline
            \noalign{\smallskip}
            Pointing      &  \alpha_{2000} & \delta_{2000} & $Grism$ \\ 
                 &  &&  \\ 
            \noalign{\smallskip}
            \hline
            \noalign{\smallskip}

            COSMOS-P01 &  09h59m02.39s & +01\deg54\arcmin35.9\arcsec &  $LRBLUE \& LRRED$  \\ 
            COSMOS-P02 &  10h00m04.08s & +01\deg54\arcmin35.9\arcsec &  $LRBLUE \& LRRED$  \\ 
            COSMOS-P03 &  10h01m05.76s & +01\deg54\arcmin35.9\arcsec &  $LRBLUE \& LRRED$  \\ 
            COSMOS-P04 &  09h59m02.39s & +02\deg12\arcmin41.4\arcsec &  $LRBLUE \& LRRED$  \\ 
            COSMOS-P05 &  10h00m04.08s & +02\deg12\arcmin41.4\arcsec &  $LRBLUE \& LRRED$  \\ 
            COSMOS-P06 &  10h01m05.76s & +02\deg12\arcmin41.4\arcsec &  $LRBLUE \& LRRED$  \\ 
            COSMOS-P07 &  10h00m04.08s & +02\deg30\arcmin46.7\arcsec &  $LRBLUE \& LRRED$  \\ 
            COSMOS-P08 &  10h01m05.76s & +02\deg30\arcmin46.7\arcsec &  $LRBLUE \& LRRED$  \\ 
                       &               &                             &         \\ 
	    ECDFS-P01  &  03h32m25.99s & -27\deg41\arcmin59.9\arcsec &  $LRBLUE \& LRRED$  \\ 
	    ECDFS-P02  &  03h32m34.00s & -27\deg53\arcmin59.9\arcsec &  $LRBLUE \& LRRED$  \\ 
	    ECDFS-P03$^a$  &  03h32m15.00s & -27\deg49\arcmin59.9\arcsec &  $MR$ \\ 
                       &               &                             &         \\ 
            VVDS02-P01 &  02h26m44.51s & -04\deg16\arcmin42.8\arcsec &  $LRBLUE \& LRRED$  \\ 
            VVDS02-P02 &  02h25m40.34s & -04\deg16\arcmin42.8\arcsec &  $LRBLUE \& LRRED$  \\ 
            VVDS02-P03 &  02h26m44.51s & -04\deg34\arcmin50.3\arcsec &  $LRBLUE \& LRRED$  \\ 
            VVDS02-P04 &  02h25m40.34s & -04\deg34\arcmin50.3\arcsec &  $LRBLUE \& LRRED$  \\ 
            VVDS02-P05 &  02h24m36.14s & -04\deg44\arcmin57.9\arcsec &  $LRBLUE \& LRRED$  \\ 
\noalign{\smallskip}
            \hline
         \end{array}
      \]
 \begin{list}{}{}
 \item[$^{\mathrm{a}}$] The ECDFS-P03 has been observed with a VIMOS PA=70deg on the sky.
 \end{list}
\label{obs}
   \end{table*}


\begin{table*}[h]
      \caption[]{Main spectral features in VUDS spectra}
         \label{spectral_lines}
\centering          
\begin{tabular}{l c c  }     
\hline\hline     
            \noalign{\smallskip}
            Spectral line     &  $\lambda_{rest}$ (\AA) & Line type \\
            \noalign{\smallskip}
            \hline
            \noalign{\smallskip}
	    Lyman$-$limit      &   912.0              & Continuum break \\
	    Lyman$-\gamma$     &   972.0              & HI absorption      \\
	    Lyman$-\beta$      &  1025.2              & HI absorption      \\
            SiII$\lambda$1192          &  1192.0              & ISM, blend 1190+1193 \\
	    Lyman$-\alpha$     &  1215.7              & HI emission \& absorption   \\
	    SiII$\lambda$1260          &  1260.4              & ISM              \\
	    OI+SiII-1303       &  1303.2              & ISM, blend       \\
	    CII$\lambda$1334           &  1334.5              & ISM              \\
	    SiIV$\lambda$1394          &  1393.8              & ISM              \\
	    SiIV$\lambda$1403          &  1402.8              & ISM              \\
	    SiII$\lambda$1527          &  1526.7              & ISM              \\
	    CIV$\lambda$1549           &  1549.1              & ISM, blend 1548.2+1550.8  \\
	    FeII$\lambda$1608          &  1608.5              & ISM              \\
	    HeII$\lambda$1640          &  1640.0              & Nebular          \\
	    AlII$\lambda$1671          &  1670.8              & ISM              \\
	    FeII$\lambda$1855          &  1854.7              & ISM             \\
            FeII$\lambda$1863          &  1862.8              & ISM             \\ 
	    CIII]$\lambda$1909         &  1908.7              & Nebular, blend 1907+1909      \\
            FeII$\lambda$2344          &  2343.5              & ISM             \\
            FeII$\lambda$2371          &  2370.5              & ISM             \\
            FeII$\lambda$2402          &  2402.6              & ISM             \\
            FeII$\lambda$2594          &  2593.7              & ISM             \\
            MgII$\lambda$2796          &  2796                & ISM             \\
            \noalign{\smallskip}
            \hline
 \end{tabular}
 \end{table*}

   \begin{table*}
      \caption[]{Comparison of the VUDS survey with other spectroscopic redshift surveys at $z>2$ in the literature, by order of increasing mean redshift } 
      \[
         \begin{array}{lrlrrclll}
           \hline \hline
            \noalign{\smallskip}
Survey         & Area    & Depth      & N_{obj}    &  N_{obj}     & z_{range} & z_{mean} & Selection & Reference \\ 
               & deg^2   & i_{AB} ~eq. & in ~Survey  &  at ~z\geq2   &           &          &           &           \\
            \noalign{\smallskip}
            \hline
            \noalign{\smallskip}

$VVDS-Deep$    &    0.74 &  24.00 & 11\,601    & 634 &   0-5      & 0.92  &  17.5 \leq I_{AB} \leq 24.0  & $Le F\`evre et al. 2013$ \\
$VVDS-UDeep$   &    0.14 &  24.75 & 941        & 341 &   0-4.5    & 1.38  &  23.0 \leq i_{AB} \leq 24.75 & $Le F\`evre et al. 2014$ \\
$Steidel-z2$   &    0.48 & R=25.5 &    851     & 588 &  1.4-2.5   & 2.0  &  BM-BX                         & $Steidel et al. 2004$ \\
$zCosmos-Deep$ &      1  &  23.75 & \sim7\,500 & \sim4\,100 &1.5-3  &  2.1  & B_{AB} \leq 25 + $color$    & $Lilly et al. 2007$, ~$Lilly et al. in prep.$  \\
$VLRS$        &    1.62 &  24.7  & 2\,135     & 2\,135 & 2-3.5   &  2.8 &  23<R<25                    & $Bielby et al. 2013$ \\
$LBG-z3$       &    0.38 &  24.8  & 1\,000     & &   2.7-3.5  &  3.2  & R_{AB}<25.5 + $color$             & $Steidel et al. 2003$ \\
$GOODS$        &    0.09 &  z=26  & 887        & 114 &  3         &  3.5 &  (i_{775}-z_{850} ) > 0.6, z_{850} < 26 & $Vanzella et al. 2009$ and ref. therein \\
$VVDS-LAE$     &    0.74 &    -   & 217        & 217 &   0-6.7    & 3.5  &  23.0 \leq i_{AB} \leq 24.75 & $Cassata et al. 2011$ \\
$LBG-z4$       &    0.38 &  25.0  & 300        & 300 &   3.5-4.5  &  4.0  & I_{AB}<25  $color$                & $Steidel et al. 1999$ \\
$Dropout-z456$ &    0.09 &   -    & 310        & 310 &   3-6.5    &    4.5  &  $Dropout, Ly$\alpha ~$break$ & $Stark et al. 2010$\\
$LAE-z6$       &      1  & \sim27 & 16         & 16  &   \sim6.5  &   6.5 &  $LAE narrow band$         & $Ouchi et al. 2010$\\ 
            \noalign{\smallskip}
            \hline
            \noalign{\smallskip}
{\bf VUDS}   &      1  &  25-27   & \sim10\,000^a  & \sim6\,000^a &   2-6.7    &  3.7  & i_{AB}<25 + $photo-z$          & $This paper$  \\ 
\noalign{\smallskip}
\hline
\noalign{\smallskip}
         \end{array}
      \]
 \begin{list}{}{}
 \item[$^{\mathrm{a}}$] Adjusted to take into account new data not yet processed
 \end{list}
\label{comp_surveys}
   \end{table*}


\clearpage

\clearpage

\clearpage

\clearpage

\newpage

\newpage

\newpage

\newpage

\newpage

\newpage

\begin{thebibliography}{}

   \bibitem[2009]{aba} Abazajian et al., 2009, \apjs, 182, 543  




   \bibitem[2007]{arnouts07} Arnouts, S., et al., 2007, \aap, 476, 137  

   \bibitem[Appenzeller et al. 1998]{appen} Appenzeller, I., et al. 1998, The Messenger, 94, 1  

   \bibitem[2010]{baldry10} Baldry, I., et al., 2010, MNRAS, 404, 86  

   \bibitem[1996]{Sex} Bertin, E., Arnouts, S.,  1996, \aaps, 117, 393 

   \bibitem[Bielby et al. 2012]{bielby} Bielby, R., et al., 2012, \aap, 545, 23    

   \bibitem[Bielby et al. 2013]{bielby13} Bielby, R., et al., 2013, \mnras, 430, 425 


   \bibitem[2003]{bondi} Bondi, M., et al., 2003, \aap, 403, 857   


   \bibitem[Bouwens et al. 2009]{bowens09} Bouwens, R. et al., 2009, \apj, 705, 936  

   \bibitem[Bruzual \& Charlot 2003]{bc03} Bruzual, G., \& Charlot, S., 2003, \mnras, 344, 1000 


   \bibitem[Calzetti et al. (2000)]{calzetti} Calzetti, D., et al., 2000, \apj, 533, 682  


   \bibitem[Capak et al. (2011)]{capak11} Capak, P., et al.,  2011, Nature, 470, 233  

   \bibitem[Caputi et al. 2012]{caputi12} Caputi, K., et al., 2012, \apj, 750, 20  

   \bibitem[Cardamone et al. 2010]{cardamone10} Cardamone, C., et al., 2010, \apjs, 189, 270  

   \bibitem[2011]{cassata11} Cassata, P., Le F\`evre, O., et al., 2011, \aap,  525, 143  

   \bibitem[2013]{cassata13} Cassata, P., Le F\`evre, O., et al., 2013, \aap, 556, 68  

   \bibitem[2002]{cimatti} Cimatti, A., et al., 2002, \aap, 381, 68     

   \bibitem[2008]{cimatti08} Cimatti, A., et al., 2008, \aap, 482, 21     



   \bibitem[Colless et al. (2001)]{colless} Colless, M., et al., 2001, \mnras, 328, 1039   



   \bibitem[Cooper et al. 2012]{cooper12} Cooper, M.C., et al., 2012, \mnras, 425, 2116  

   \bibitem[Cucciati et al. 2012]{cucciati12} Cucciati, O., et al., 2012, \aap, 539, 31   



   \bibitem[Cuillandre et al. 2012]{cuillandre12} Cuillandre, J.C., et al., 2012, Proc. SPIE 8448, Observatory Operations: Strategies, Processes, and Systems IV, 84480  

   \bibitem[Curtis-Lake et al. 2012]{curtis12} Curtis-Lake, E., et al., 2012, \mnras, 422, 1425  


   \bibitem[2004]{daddi} Daddi, E., et al., 2004, \apj, 617, 746   

   \bibitem[Davis et al., 2003]{davis} Davis, M., et al.,  2003, SPIE, 4834, 161 







   \bibitem[Ellis et al. (2013)]{ellis13} Ellis, R.S., et al. 2013, \apj, 763, L7   



   \bibitem[Faber et al. 2003]{faber03} Faber, S. M., et al. 2003, Proc. SPIE, 4841, 1657  






   \bibitem[Garilli et al. (2010)]{garilli2} Garilli, B., et al., 2010, PASP, 122, 827  
  
   \bibitem[Giacconi et al. (2002)]{giacconi} Giacconi, R., et al., 2002,  \apjs, 139, 369   

   \bibitem[Grogin et al. 2011]{grogin11} Grogin, N., et al., 2011, \apjs, 197, 35   

   \bibitem[Guo et al. 2011]{guo11} Guo, Q., et al., 2011, \mnras, 413, 101    


   \bibitem[Guzzo et al. (2013)]{guzzo13} Guzzo, L., et al., 2014, arXiv:1303.2623  

   \bibitem[Hammersley et al. 2010]{Hammersley} Hammersley, P. et al., 2010, The Messenger, 142, 8 




   \bibitem[Ilbert et al. (2006)]{ilbert06} Ilbert, O., et al., 2006, \aap, 457, 841   

   \bibitem[Ilbert et al. (2009)]{ilbert09} Ilbert, O., et al., 2009, \apj, 690, 1236  

   \bibitem[Ilbert et al. (2013)]{ilbert13} Ilbert, O., et al., 2013, \aap, 556, 55  



   \bibitem[Kashikawa et al. 2011]{kashi}  Kashikawa, N., et al., 2011, \apj, 734, 119  

   \bibitem[Koekemoer et al. (2007)]{koek} Koekemoer, A., et al., 2007, \apjs, 172, 196 

   \bibitem[Koekemoer et al. (2011)]{koek11} Koekemoer, A., et al., 2011, \apjs, 197, 36 




   \bibitem[Le F\`evre et al. 1995]{olf95} Le F\`evre, O., Crampton, D., Lilly, S.J., Hammer, F., Tresse, L., 1995, \apj, 455, 60  
 

   \bibitem[Le F\`evre et al. 2003]{olf03} Le F\`evre, O., et al., 2003, SPIE, 4841, 1670   


   \bibitem[Le F\`evre et al. (2004)]{olfima} Le F\`evre, O., et al., 2004, \aap, 417, 839  

   \bibitem[Le F\`evre et al. (2005a)]{olf2} Le F\`evre, O., et al., 2005a, \aap, 439, 845   

   \bibitem[Le F\`evre et al. (2005b)]{olf1} Le F\`evre, O., et al., 2005b, Nature, 437, 519  

   \bibitem[Le F\`evre et al. (2013)]{olf13} Le F\`evre, O., et al., 2013, \aap, 559, 14  

   \bibitem[Le F\`evre et al. (2014)]{olf14} Le F\`evre, O., et al., 2014, \aap, arXiv:1307.6518  

   \bibitem[Lemaux et al. (2014)]{lemaux} Lemaux, B., et al., 2014, \aap, in press, arXiv:1311.5228  



   \bibitem[Lilly et al. (1995)]{lilly1} Lilly, S.J., Le F\`evre, O., Crampton, D., Hammer, F., 1995a, \apj, 455, 50  
  

   \bibitem[Lilly et al. (1996)]{lilly3} Lilly, S. J., Le F\`evre, O., Hammer, F., Crampton, D., 1996, \apj, 460, 1  

   \bibitem[Lilly et al. (2007)]{lilly2} Lilly, S.J., Le F\`evre, O., et al., 2007, \apjs, 172, 70  

   \bibitem[2003]{lonsdale} Lonsdale, C.C., et al., 2003, \pasp, 115, 897   



   \bibitem[Madau 1995]{madau95} Madau, P., 1995, \apj, 441, 18   

   \bibitem[Madau et al. (1996)]{mad} Madau, P., Ferguson, H.C.,  Dickinson, M.,  Giavalisco, M.,
    Steidel, C.C.,  Fruchter, A., 1996, \mnras, 283, 1388                   

   \bibitem[Madau Dickinson 2014]{madau14} Madau, P., \& Dickinson, P., 2014, ARAA, in press (arXiv:1403.0007)    

   \bibitem[Mauduit et al. 2012]{Mauduit12} Mauduit et al., 2012, \pasp, 124, 714  
  

   \bibitem[McCracken et al. 2012]{hjmcc12} McCracken, H.J., et al., 2012, \aap,   


   \bibitem[Mo et al. 2010]{mo} Mo, H., van den Bosch, F., White, S., Galaxy Formation and Evolution, Cambridge University Press, 2010  

   \bibitem[Moster et al. 2011]{moster1} Moster, B.P., Somerville, R.S., Newman, J.A., Rix, H-W., 2011, \apj, 731, 113
  


   \bibitem[Oke et al. 1995]{oke} Oke, J.B., 1995, \pasp, 107, 3750   

   \bibitem[Oliver et al. (2012)]{hermes} Oliver, S., et al., 2012, \mnras, 424, 1614   


   \bibitem[Ouchi et al. (2008)]{ouchi08} Ouchi, M., et al.,  2008, \apjs, 176, 301  

   \bibitem[Patat et al. 2011]{Patat} Patat, N., et al., 2011, \aap, 527, 91  

   \bibitem[2004]{pierre} Pierre, M., et al., 2004, JCAP, 09, 011   

   \bibitem[Popesso et al. (2009)]{popesso} Popesso, et al.,  2009, \aap, 494, 443   




   \bibitem[Rix et al. 2004]{rix04} Rix, H.W., et al., 2004, 2004, \apjs, 152, 163  

   \bibitem[Sanders et al. 2007]{sanders07} Sanders, D., et al., 2007, \apjs, 172, 86  


   \bibitem[Schlegel et al. 1998]{Schlegel} Schlegel, D., et al., 1998, \apj, 500, 525  

   \bibitem[2005]{sco05} Scodeggio, M., et al., 2005, PASP, 117, 1284    

   \bibitem[2009]{sco09} Scodeggio, M., et al., 2009, The Messenger, 135, 13    

   \bibitem[Scoville et al. (2007)]{scov} Scoville, N., et al., 2007, \apjs, 172, 1  


   \bibitem[Shapley et al. 2006]{shapley06} Shapley, A., et al., 2006, \apj, 651, 688  


   \bibitem[Shimasaku et al. 2006]{Shimasaku} Shimasaku, K., 2006, \pasj, 5, 313  

   \bibitem[Springel et al. 2008]{springel08} Springel, V. et al., 2008, \mnras, 391, 1685  

   \bibitem[Stark et al. 2010]{stark10} Stark, D., et al., 2010, \mnras, 408, 1628  

   \bibitem[Steidel et al. 1996]{steidel96} Steidel, C.C., Giavalisco, M., Pettini, M., Dickinson, M., Adelberger, K.L., 1996, \apj, 462, 17    

   \bibitem[Steidel et al. (1999)]{steidel99} Steidel, C.C., Adelberger, K.L., Giavalisco, M., Dickinson, M., Pettini, M., 
   1999, \apj, 519, 1                     


   \bibitem[2003]{steidel03} Steidel, C.C., et al., 2003, \apj, 592, 728   

   \bibitem[2004]{steidel04} Steidel, C.C., et al., 2004, \apj, 604, 534   

   \bibitem[Taniguchi et al. 2005]{Tani05} Taniguchi, Y., et al., 2005, \pasj, 57, 165  

   \bibitem[Taniguchi et al. 2007]{Tani07} Taniguchi, Y., et al., 2007, \apjs, 172, 9  

   \bibitem[Tasca et al. 2014]{Tasca14} Tasca, L., et al., 2014, A\&A, in press, arXiv:1303.4400  


   \bibitem[2007]{tresse} Tresse, L., et al., 2007, \aap, 472, 403  



   \bibitem[Vanzella et al. (2008)]{vanzella08} Vanzella, E., et al.,  2008, \aap, 478, 83  

   \bibitem[Vanzella et al. (2009)]{vanzella09} Vanzella, E., et al.,  2009, \apj, 695, 1163  

   \bibitem[Vanzella et al. (2010)]{vanzella10} Vanzella, E., et al.,  2010, \mnras, 404, 1672 


   
   \bibitem[Wiklind et al. 2008]{wiklind} Wiklind, T. et al., 2008, \apj, 676, 781   

   \bibitem[Williams et al. 2000]{williams} Williams, R., et al.,  2000, \aj, 120, 2735  

   \bibitem[Windhorst et al. 2011]{windhorst} Windhorst, R., et al., 2011, \apjs, 193, 27  




\end{thebibliography}
\end{document}